\documentclass[10pt,journal,compsoc]{IEEEtran}
\ifCLASSOPTIONcompsoc
  \usepackage[nocompress]{cite}
\else
  \usepackage{cite}
\fi
\ifCLASSINFOpdf
  \usepackage[pdftex]{graphicx}
\else
  \usepackage[dvips]{graphicx}
  \graphicspath{{./figs/}}
  \DeclareGraphicsExtensions{.eps}
\fi
\usepackage{amsmath}

\usepackage[ruled]{algorithm2e} 

\SetAlFnt{\small}
\SetAlCapFnt{\small}
\SetAlCapNameFnt{\small}
\SetAlCapHSkip{0pt}

\usepackage{subfigure}
\usepackage{xcolor}
\usepackage{overpic}
\usepackage{multirow}
\usepackage{algpseudocode}
\usepackage{amsfonts}

\newcommand{\ud} {\mathrm{d}}
\newcommand{\mbx}{\mathbf{x}}
\newcommand{\mby}{\mathbf{y}}
\newcommand{\mbz}{\mathbf{z}}

\newcommand{\mbp}{\mathbf{p}}
\newcommand{\mbomega}{\boldsymbol{\omega}}

\newcommand{\cfbox}[2]{%
    \colorlet{currentcolor}{.}%
    {\color{#1}%
    \fbox{\color{currentcolor}#2}}%
}

\begin{document}
%
\title{Rendering Discrete Participating Media with Geometrical Optics Approximation}
%
%
%
%

\author{Jie~Guo, Bingyang~Hu, Yanjun~Chen, Yuanqi~Li, Yanwen~Guo and Ling-Qi~Yan
\IEEEcompsocitemizethanks{
\IEEEcompsocthanksitem Jie~Guo, Bingyang Hu, Yanjun Chen, Yuanqi Li and Yanwen Guo are with the State Key Lab for Novel Software Technology, Nanjing University, Nanjing, Jiangsu 210023, P.R. China.\protect\\
E-mail: guojie@nju.edu.cn, fhymyang@gmail.com, cujooyer@gmail.com, dz1833015@smail.nju.edu.cn and ywguo@nju.edu.cn
\IEEEcompsocthanksitem Ling-Qi Yan is with Department of Computer Science, UC Santa Barbara, United States.\protect\\
E-mail: lingqi@cs.ucsb.edu}
}

%
%

\markboth{IEEE Transactions on Visualization and Computer Graphics, In Submission.}%
{Shell \MakeLowercase{\textit{et al.}}: Discrete Participating Media}
%



\IEEEtitleabstractindextext{%
\begin{abstract}
We consider the scattering of light in participating media composed of sparsely and randomly distributed discrete particles. The particle size is expected to range from the scale of the wavelength to the scale several orders of magnitude greater than the wavelength, and the appearance shows distinct graininess as opposed to the smooth appearance of continuous media. One fundamental issue in physically-based synthesizing this appearance is to determine necessary optical properties in every local region. Since these optical properties vary spatially, we resort to geometrical optics approximation (GOA), a highly efficient alternative to rigorous Lorenz-Mie theory, to quantitatively represent the scattering of a single particle. This enables us to quickly compute bulk optical properties according to any particle size distribution. Then, we propose a practical Monte Carlo rendering solution to solve the transfer of energy in discrete participating media. Results show that for the first time our proposed framework can simulate a wide range of discrete participating media with different levels of graininess and converges to continuous media as the particle concentration increases.

\end{abstract}

}

\maketitle

\IEEEdisplaynontitleabstractindextext

%
\IEEEpeerreviewmaketitle

\IEEEraisesectionheading{\section{Introduction}\label{sec:introduction}}
Rendering participating media is a long-standing problem in computer graphics, with much effort devoted to solving this problem plausibly and efficiently \cite{Cerezo2005,novak18monte}. From the physics point of view, radiative transfer is rather complicated and should be rigorously derived from Maxwell's electromagnetic theory. To make the simulation tractable, some compromises are made in physically-based rendering over the past decades. Two main assumptions are \emph{independent scattering} \cite{Chandrasekhar1960} and \emph{local continuity} (or \emph{statistical homogeneity} \cite{Bohren1983}). The first assumption of independent scattering means that the particles forming the medium are far apart from each other and are mutually unaffected. Recently, this assumption has been relaxed in computer graphics via incorporating spatial correlations between scatterers into radiative transfer frameworks \cite{Jarabo:2018:RTF:3197517.3201282,Bitterli:2018:RTF:3272127.3275103,Eugene18,deon2018reciprocal,Guo:2019:FGF:3306346.3323031}, leading to non-exponential attenuation of light.

The second assumption of local continuity implies that the medium is homogeneous and compact in each differential volume even if macroscopic heterogeneity exists. Under this circumstance, the light is not sensitive to the discrete spatial distribution of the scatterers, but only to their local average properties. Consequently, light scattering phenomena take place at any point of the medium, resulting in locally smooth renderings. However, many participating media are composed of separate particles distributed randomly within a given volume. The scattering will happen only at the particle positions \cite{Hulst1981,Bohren1983}. These facts indicate that the assumption of continuous media only holds when the particle size is much smaller as compared with the resolution of the sensor (e.g., human eyes) and the quantity is sufficiently large \cite{Arvo93transferequations}. Otherwise, individual grains can be observed when zoomed in. To break through this constraint, the graininess of the medium should be taken into consideration.

Several recent studies in computer graphics \cite{Moon:2007:RDR:2383847.2383878,Meng:2015:MMR:2809654.2766949,Muller:2016:ERH:2980179.2982429} have noticed the graininess in rendering granular materials. They typically rely on explicit geometries and precomputed transport functions to capture the appearance of discernible grains. As these approaches are designed for a certain amount of very large particles under geometric optics, the scattering behaviors of particles forming the media are rather limited. For instance, diffraction, which dominates light scattering from small particles, is often ignored. In this paper, we attempt to put forward a more general approach to model and render discrete participating media with a large amount of scatterers whose particle size distributions (PSDs) range widely. These media are omnipresent in natural and artificial environments, such as flying dusts, blowing snows, powder suspension, and air bubbles in liquid.

Unlike conventional continuous media, these discrete media have spatially-varying optical properties (e.g., the extinction coefficient) that cannot be determined in advance but should be evaluated on-the-fly. They are closely related to the scattering behavior of each individual particle and the fluctuation of PSDs. To derive necessary optical properties for any local region of a given discrete participating medium in a physically-based manner, one can resort to Lorenz-Mie theory \cite{Lorenz1890,Mie1908} which offers substantial realism in rendering participating media \cite{Frisvad:2007:CSP:1275808.1276452}. However, numerical evaluation of the Lorenz-Mie coefficients is known to be difficult and time-consuming when the particle size becomes large \cite{Glantschnig:81}. To ameliorate this issue, we introduce geometrical optics approximation (GOA) \cite{Glantschnig:81,Ungut:81} and use it to simplify the computation of light scattering when the particle is sufficiently large. We show how to perform a smooth transition between Lorenz-Mie theory and GOA in computing the optical properties, enabling both high accuracy and low computational cost.

Due to the variations in local PSDs, the optical properties exhibit multi-scale effects with respect to the scene configuration. We derive a novel \emph{multi-scale volumetric rendering equation (VRE)} and propose a practical Monte Carlo rendering solution to solve it. Our solution only relies on the position and the radius of each particle distributed randomly according to some PSDs, avoiding cumbersome geometric modeling and lengthy precomputation. Experimental results verify that the proposed solution is able to capture the distinct grainy appearance of discrete participating media and guarantee temporal coherence in animation. We also show that it converges to continuous media in the limit of particle concentration.

In summary, the main contributions of this paper are:
\begin{itemize}
  \item a general and physically-based framework for modeling and rendering discrete participating media, considering diffraction, polarization, a wide range of PSDs, etc,
  \item the use of GOA for efficient and accurate evaluation of the multi-scale bulk optical properties at any local region of a medium, and
  \item a new Monte Carlo rendering solution that captures both low-frequency haziness and high-frequency graininess in discrete participating media.
\end{itemize}

\section{Related Work}

\subsection{Participating Media Rendering}
Rendering participating media is a challenging but important problem, which requires efficiently solving the VRE \cite{Chandrasekhar1960,Cerezo2005} by means of Monte Carlo path integration \cite{Lafortune:1996:Rendering,veach97,Pauly:2000:Metropolis,novak18monte,doi:10.1111/cgf.13342}, photon density estimation \cite{Jensen:1998:Efficient,Jarosz:2008:Beam,Jarosz:2011:CTV:1899404.1899409,jarosz11progressive,hachisuka13starpm,Bitterli:2017:Points,deng19photon}, or a combination of both \cite{Krivanek:2014:UPB:2601097.2601219}. In our current framework, we choose Monte Carlo path integration by virtue of its elegant simplicity, generality, and accuracy. This technique operates by stochastically constructing a large number of light paths between sensors and emitters to simulate the light transport in the scene. To facilitate the query of particles along paths, we augment each ray with a cylinder, in a way similar to the photon beam \cite{Jarosz:2008:Beam,Jarosz:2011:CTV:1899404.1899409,Krivanek:2014:UPB:2601097.2601219}. Multiple importance sampling \cite{10.1145/218380.218498,10.1145/3306346.3323025} algorithms are beneficial for reducing the large variance caused by Monte Carlo sampling.

\subsection{Detailed Volumetric Modeling}
Since the original VRE is only a rough approximation to the real radiative transport in participating media, the range of appearance that can be faithfully simulated is limited. Recent trend in computer graphics tries to capture more details in the volume by relaxing the assumptions in the original VRE, enriching the range of achievable appearances. For example, to account for angular anisotropy, the VRE is extended with local directional dependency based on the microflake model \cite{Jakob:2010:RTF:1778765.1778790,Heitz:2015:SMD:2809654.2766988,sre.20161210}. It is also possible to extend the VRE to simulate the effects of spatial correlations \cite{Jarabo:2018:RTF:3197517.3201282,Bitterli:2018:RTF:3272127.3275103,Guo:2019:FGF:3306346.3323031}, yielding non-exponential attenuation of light. Notably, these methods still assume the media to be statistically homogeneous within each differential volume \cite{Bohren1983}, ignoring any sub-pixel details.

To handle participating media with complex 3D structures, volumetric representations of explicit geometries have been widely used. By capturing the geometric and optical properties of a fabric down to the fiber level, micro-appearance models, described using high-resolution volumes, offer state-of-the-art renderings for fabrics and textiles \cite{Zhao:2011:BVA:2010324.1964939,Zhao:2012:SSP:2185520.2185571,10.1145/2818648,10.1145/2897824.2925932,doi:10.1111/cgf.13222}. Unfortunately, these methods are highly data-intensive and plagued by heavy computation. To improve the performance while maintaining good accuracy, some downsampling strategies \cite{Zhao:2016:DSP:2980179.2980228,loubet:hal-01702000} are developed. Current rendering solutions for granular materials are also based on explicit geometries and pre-captured optical properties of each individual grain \cite{Moon:2007:RDR:2383847.2383878,Meng:2015:MMR:2809654.2766949,Muller:2016:ERH:2980179.2982429}. They generally employ shell tracing to make large jumps inside media. Even so, the computational cost is still high. The goal of this work is to develop a general framework for handling participating media with graininess which also allows rapid computation and convenient usage.

\subsection{Glittery Surface Simulation}
Our work is also closely related to the simulation of glints on surfaces. Yan et al. \cite{Yan:2014:RGH:2601097.2601155,Yan:2016:PDE:2897824.2925915} suggested using explicit high-resolution normal maps to model sub-pixel surface details and successfully simulated spatially-varying glints with a patch-based normal distribution function. Subsequent work adopted a wave optics model to achieve more accurate results with noticeable color effects \cite{10.1145/3197517.3201351}. There are also other methods focusing specifically on capturing spatially-varying highlights from scratched surfaces, under either geometric optics \cite{Merillou2001,doi:10.1111/j.1467-8659.2004.00767.x,10.1145/2897824.2925945} or wave optics \cite{10.1145/3130800.3130840}. To ease the burden of computation and storage, Kuznetsov et al. \cite{10.1145/3355089.3356525} proposed to learn high-frequency angular patterns from existing examples, using a generative adversarial network (GAN). Jakob et al. \cite{Jakob:2014:DSM:2601097.2601186} addressed the problem of glittery surface simulation using a purely procedural approach which requires far less storage and supports on-the-fly point queries. This approach has been extended to incorporate iridescence \cite{10.1111:cgf.13476} and allow fast global illumination \cite{10.1111:cgf.13547}.

\subsection{Lorenz-Mie Theory and GOA}
Lorenz-Mie theory \cite{Lorenz1890,Mie1908} develops a rigorous solution to the problem of light scattering by spherical particles. It was introduced to the graphics community by Rushmeier \cite{Rushmeier:1995} to accurately simulate the physics of light transport in participating media. Later, Callet \cite{Callet:96} used this theory to model pigmented materials consisting of pigmented particles in a transparent solvent. Atmospheric phenomena, such as halos and rainbows, are especially favored by this theory \cite{Jackel1997,DBLP:conf/cgi/NishitaD01,Riley:2004:ERA:2383533.2383584,Laven:03}. Frisvad et al. \cite{Frisvad:2007:CSP:1275808.1276452} generalized the original Lorenz-Mie theory and used it to compute the appearance of materials with different mixed particle concentrations. Though accurate, this theory is computationally expensive. We show that GOA \cite{Glantschnig:81,Ungut:81,Hovenac:91,Zhou:03,WU200754,YU2008340,YU20091178} is much more efficient than Lorenz-Mie theory in computing optical properties of individual particles in various media, especially when the particle size is large. Moreover, the computation of GOA can be made in non-ideal situations such as absorbing particles \cite{YU20091178} and non-spherical particles \cite{Hovenac:91,HE20121467,LU201990,sadeghi11physically}. Compared to Lorenz-Mie theory, GOA is less explored in computer graphics. We choose GOA in our framework, taking advantage of its high performance.

\section{Light Scattering by A Single Particle}
We first study light scattering by a single particle. We suppose that the particle is approximately spherical and has a set of physical properties including its radius $r$ and the refractive index $\eta_\mathrm{p}$. Currently, we assume that particles forming the medium have the same composition and only their sizes vary. In this case, the refractive index $\eta_\mathrm{p}$ is fixed. Supposing that the host medium has the refractive index $\eta_\mathrm{m}$, we can define the relative refractive index of the particle as $\eta = \eta_\mathrm{p}/\eta_\mathrm{m}$. The size of a spherical particle may also be expressed in terms of the dimensionless size parameter $\alpha = kr = 2\pi\eta_\mathrm{m} r / \lambda$, where $k$ is the wave number defined by $k=2\pi\eta_\mathrm{m}/\lambda$ and $\lambda$ is the wavelength of light in the medium.


To describe the scattering, we need two scattering amplitude functions: $S_1(\theta,\varphi, r)$ and $S_2(\theta,\varphi, r)$, where $\theta$ is the scattering angle and $\varphi$ is the azimuth angle. The subscripts $1$ and $2$ denote perpendicular and parallel polarizations, respectively. For spherical particles, $S_1$ and $S_2$ are invariant with respect to $\varphi$, but they change depending on the radius $r$. For unpolarized light, these two functions define the phase function of a single particle as \cite{Bohren1983}
\begin{equation}
  f_{\mathrm{p}}(\theta,r) = \frac{|S_1(\theta,r)|^2+|S_2(\theta,r)|^2}{2|k|^2 C_\mathrm{s}(r)}
\end{equation}
which is properly normalized by the scattering cross section $C_\mathrm{s}$:
\begin{equation}
  C_\mathrm{s}(r) = \int_0^{2\pi} \int_0^{\pi} \frac{|S_1(\theta,r)|^2+|S_2(\theta,r)|^2}{2|k|^2} \sin \theta \ud \theta \ud \phi.
\end{equation}
Another important property of the particle is the extinction cross section $C_\mathrm{t}$ which is evaluated by
\begin{equation}\label{eq:ecs}
  C_\mathrm{t}(r) = 4\pi \mathrm{Re}\left\{\frac{S(0,r)}{|k|^2}\right\}
\end{equation}
with $S(0,r)=S_1(0,r)=S_2(0,r)$. The notation $\mathrm{Re}$ takes the real part of a complex number. For particles with absorption, the absorption cross section is given by $C_\mathrm{a}(r) = C_\mathrm{t}(r)-C_\mathrm{s}(r)$.

As seen, once the scattering amplitude functions $S_1(\theta,r)$ and $S_2(\theta,r)$ are available, we can easily find the scattering, extinction and absorption cross sections as well as the phase function of the particle. For light scattering of an electromagnetic wave from a homogeneous spherical particle, exact solutions of the two scattering amplitude functions are given by Lorenz-Mie theory~\cite{Lorenz1890,Mie1908}. Its accuracy has been validated against real measurements in various literature \cite{10.1145/2508363.2508377,mam.20161247}. Please refer to Appendix A for more details.

As a rigorous and general electromagnetic treatment of light scattering by spherical particles, Lorenz-Mie theory can precisely handle a wide range of particle sizes.
However, as the particle size increases, numerical calculations of the Lorenz-Mie coefficients become very tedious and time-consuming, due to the fact that the number of terms to be computed in the series for $S_1(\theta,r)$ and $S_2(\theta,r)$ is proportional to the size parameter $\alpha$ \footnote{It is suggested that an appropriate number of terms to sum is $\lceil|\alpha|+4.3|\alpha|^{1/3}+1 \rceil$ \cite{Cachorro91newimprovements}.} \cite{Glantschnig:81,Frisvad:2007:CSP:1275808.1276452}. For this reason, simpler approximate expressions should be developed to reduce the computational complexity. In the case that the particle size is large with respect to the wavelength of the illuminating light, geometrical optics approximation (GOA) \cite{Glantschnig:81,Ungut:81,Hovenac:91,Zhou:03,WU200754,YU2008340,YU20091178} provides a simplified but also good solution.


%
%

\subsection{Geometrical Optics Approximation}\label{sec:goa}

\begin{figure}[t]
  \centering
  \includegraphics[width=0.9\linewidth]{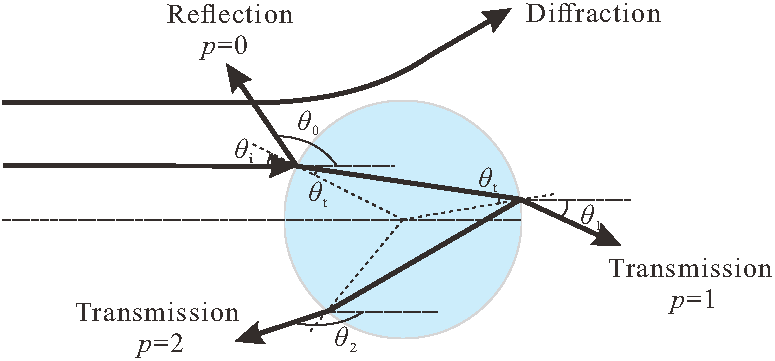}\\
  \caption{Light scattering by a spherical particle in the GOA picture. }\label{fig:goa}
\end{figure}

Within the framework of GOA, light scattering is calculated by a superposition of classical diffraction, geometrical reflection and transmission. The diffraction is independent of the particle's composition (i.e., the refractive index). Its amplitude functions for the forward direction are readily described by the Fraunhofer diffraction as \cite{Glantschnig:81}
\begin{equation}
  S_{\mathrm{D},1}(\theta,r) = S_{\mathrm{D},2}(\theta,r) = \alpha^2 \frac{J_1(\alpha\sin \theta)}{\alpha \sin \theta}
\end{equation}
where $J_1$ is the first-order Bessel function.

Leaving out diffraction, a light ray hitting a spherical particle at an incident angle $\theta_\mathrm{i}$ is partially reflected and partially refracted depending on the properties of the interface, as sketched in Fig. \ref{fig:goa}. The refracted ray may undergo a number of internal reflections before leaving the particle. For each emerging ray, we use an integer $p$ to denote the number of chords it makes inside the spherical particle. Obviously, the externally reflected ray has $p=0$ while the other rays are transmitted with $p-1$ internal reflections. The angle of deflection $\theta_p$ between the $p$th emerging ray and the direction of the incident ray is given by
\begin{equation}
  \theta_p = 2p \theta_\mathrm{t}-2\theta_\mathrm{i}-(p-1)\pi
\end{equation}
with $\sin \theta_\mathrm{i} = \eta \sin \theta_\mathrm{t}$ according to Snell's law. The scattering angle $\theta$ is further determined by the deflection angle $\theta_p$ as
\begin{equation}
  \theta = q(\theta_p - 2\pi l)
\end{equation}
where $q\in\{1,-1\}$ \footnote{$q=1$ indicates that the incident ray hits the particle on the upper hemisphere and $q=-1$ for the lower.} and $l$ is an integer ensuring that the scattering angle $\theta$ is well defined in the range between $0$ and $\pi$.

Clearly, reflected and transmitted rays depend on the shape and composition of the particle. Their scattering amplitudes for each polarization are derived as \cite{Hulst1981,Ungut:81}
\begin{equation}\label{eq:amp}
  S_j^{(p)}(\theta,r) = \alpha \epsilon_j(\theta_\mathrm{i})\sqrt{\frac{\sin2\theta_\mathrm{i}}{2\sin\theta|\ud \theta_p/\ud \theta_\mathrm{i}|}}e^{\mathrm{i}\phi} \mbox{\quad} j=1,2.
\end{equation}
Here, the fraction $\epsilon_j(\theta_\mathrm{i})$, which is due to the reflection and/or refraction for an emergent ray of order $p$, is defined as
\begin{equation}\label{eq:fraction}
  \epsilon_j(\theta_\mathrm{i}) = \left\{ \begin{array}{ll}
                                        R_j(\theta_\mathrm{i}) & \quad p=0\\
                                        (1-R_j(\theta_\mathrm{i})^2)(-R_j(\theta_\mathrm{i}))^{p-1} & \quad p>0
                                        \end{array}
                                        \right.
\end{equation}
with $R_j(\theta_\mathrm{i})$ being the Fresnel reflection coefficients. The phase difference $\phi = \phi_\mathrm{p} + \phi_\mathrm{f}$ includes $\phi_\mathrm{p}$ due to the length of optical path:
\begin{equation}
  \phi_\mathrm{p} = 2\alpha(\cos \theta_\mathrm{i} - p \eta \cos \theta_\mathrm{t})
\end{equation}
and $\phi_\mathrm{f}$ due to focal line:
\begin{equation}
  \phi_\mathrm{f} = \frac{\pi}{2}\left(1+p-2l-\frac{s}{2}-\frac{q}{2}\right)
\end{equation}
with $s=\mathrm{sgn}(\ud \theta_p/\ud\theta_\mathrm{i})=\mathrm{sgn}(2p \tan \theta_\mathrm{t}/ \tan \theta_\mathrm{i}-2)$.

Putting together $S_{\mathrm{D},j}(\theta,r)$ and $S_j^{(p)}(\theta, r)$, we are able to get the total amplitude functions of GOA as
\begin{equation}\label{eq:tot_amp}
  S_j(\theta, r) = \left\{ \begin{array}{ll}
                                        \sum_{p=0}^{\infty} S_j^{(p)}(\theta, r) + S_{\mathrm{D},j}(\theta,r)  & \theta\in[0,\frac{\pi}{2})\\
                                        \sum_{p=0}^{\infty} S_j^{(p)}(\theta, r)  & \theta\in[\frac{\pi}{2},\pi]
                                        \end{array}
                                        \right.
\end{equation}
with $j=1,2$. These expressions can be evaluated quite efficiently.

In GOA, analytical expression of $C_\mathrm{t}(r)$ can be derived as (see the derivation in Appendix B)
\begin{equation}\label{eq:ct}
  C_\mathrm{t}(r) = 2\pi r^2 + \frac{2\pi r}{|k|} \sum_{p\in \mathcal{P}}  \frac{\epsilon_j(0)}{|p/\eta-1|}\cos(\phi_\mathrm{p} + \phi_\mathrm{f})
\end{equation}
where $\mathcal{P}=\{1,3,5,\cdots\}$. This expression is fast to evaluate and well captures the ripple structures \cite{1977ApOpt}.


For absorbing particles, we certainly have $C_\mathrm{a}(r)=C_\mathrm{t}(r)- C_\mathrm{s}(r)>0$. Within GOA, the absorption cross section $C_\mathrm{a}(r)$ is faithfully approximated by \cite{Bohren1983}
\begin{equation}\label{eq:ca}
  C_\mathrm{a}(r) = \frac{16\pi^2 r^3 \eta_\mathrm{i}}{3\lambda\eta_\mathrm{r}}\left[\eta_\mathrm{r}^3-(\eta_\mathrm{r}^2-1)^{\frac{3}{2}}\right]
\end{equation}
in which $\eta_\mathrm{r}$ and $\eta_\mathrm{i}$ are the real and imaginary parts of $\eta$, respectively. The scattering amplitude functions $S_1(\theta,r)$ and $S_2(\theta,r)$ are also slightly different. The details are provided in Appendix C.

\subsection{Discussions on $p$}
There is also an infinite summation in computing $S_1$ and $S_2$ of GOA. However, unlike that in Lorenz-Mie theory, the number of terms needed is independent of the particle size, and a small $p$ suffices in most cases. As shown in Fig. \ref{fig:goa_p}, when calculating $(|S_1|+|S_2|)/2$ with $p=3$ in GOA, we get an almost identical curve with that of $p=100$, irrespective of the particle size. This is because higher-order reflections ($p>3$) carry much less energy and have very little impact on the scattered light intensities. Regarding this, $p$ is safely set to 3 in what follows. Please see more discussions in Appendix E.
\begin{figure}[t]
\centering
\subfigure[$r=1~\mathrm{{\mu}m}$]{
\begin{minipage}[b]{0.48\linewidth}
\begin{overpic}[width=1.0\linewidth]{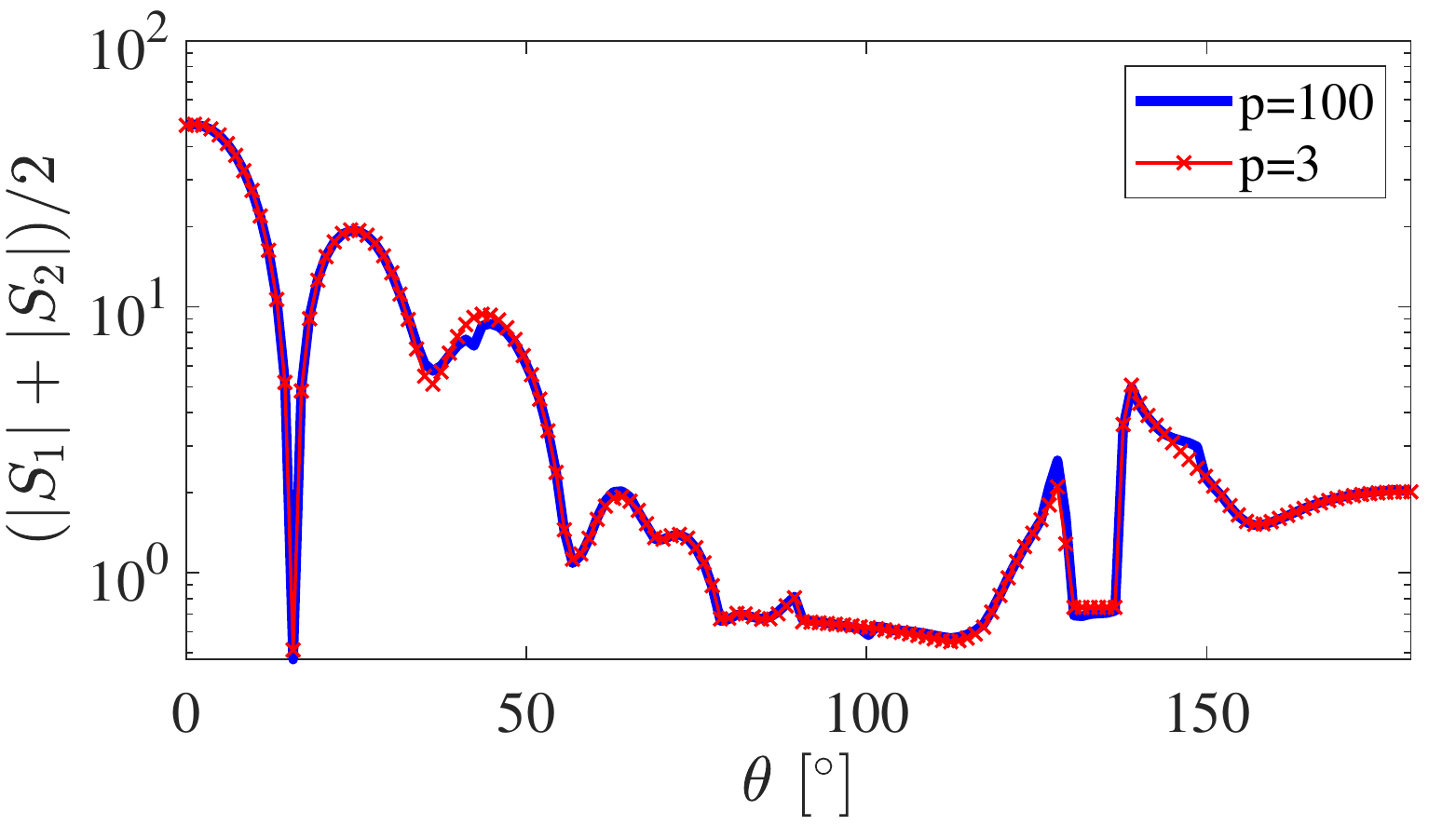}
\end{overpic}
\end{minipage}
}
\hspace{-0.12in}
\subfigure[$r=100~\mathrm{{\mu}m}$]{
\begin{minipage}[b]{0.48\linewidth}
\begin{overpic}[width=1.0\linewidth]{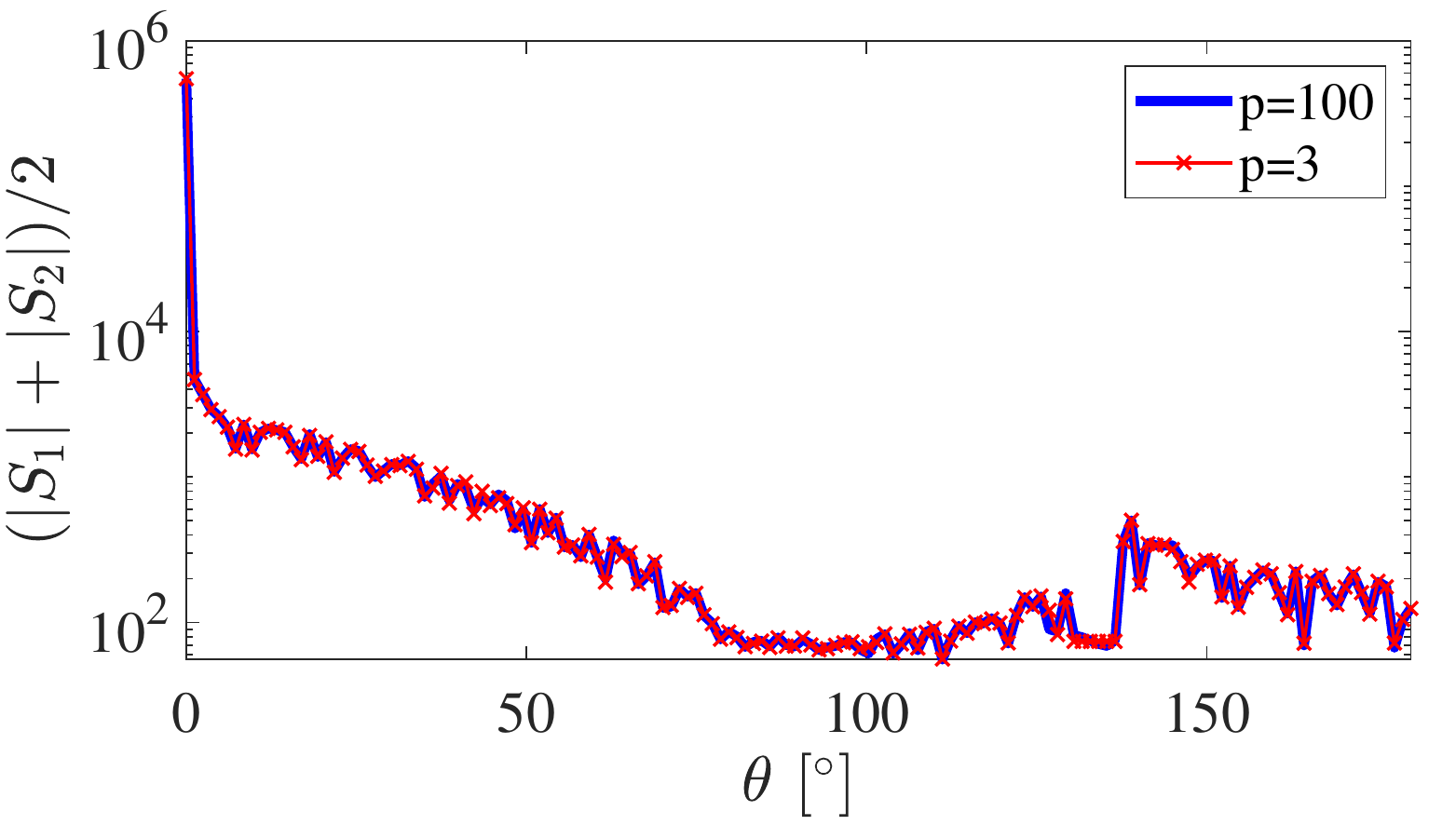}
\end{overpic}
\end{minipage}
}
\caption{\label{fig:goa_p} Impact of $p$ over $(|S_1|+|S_2|)/2$ in GOA. Here, $\eta = 1.33$ and $\lambda = 0.6~\mathrm{{\mu}m}$.}
\end{figure}

\begin{figure}
\centering
\subfigure{
\begin{minipage}{0.48\linewidth}
\begin{overpic}[width=1.0\linewidth]{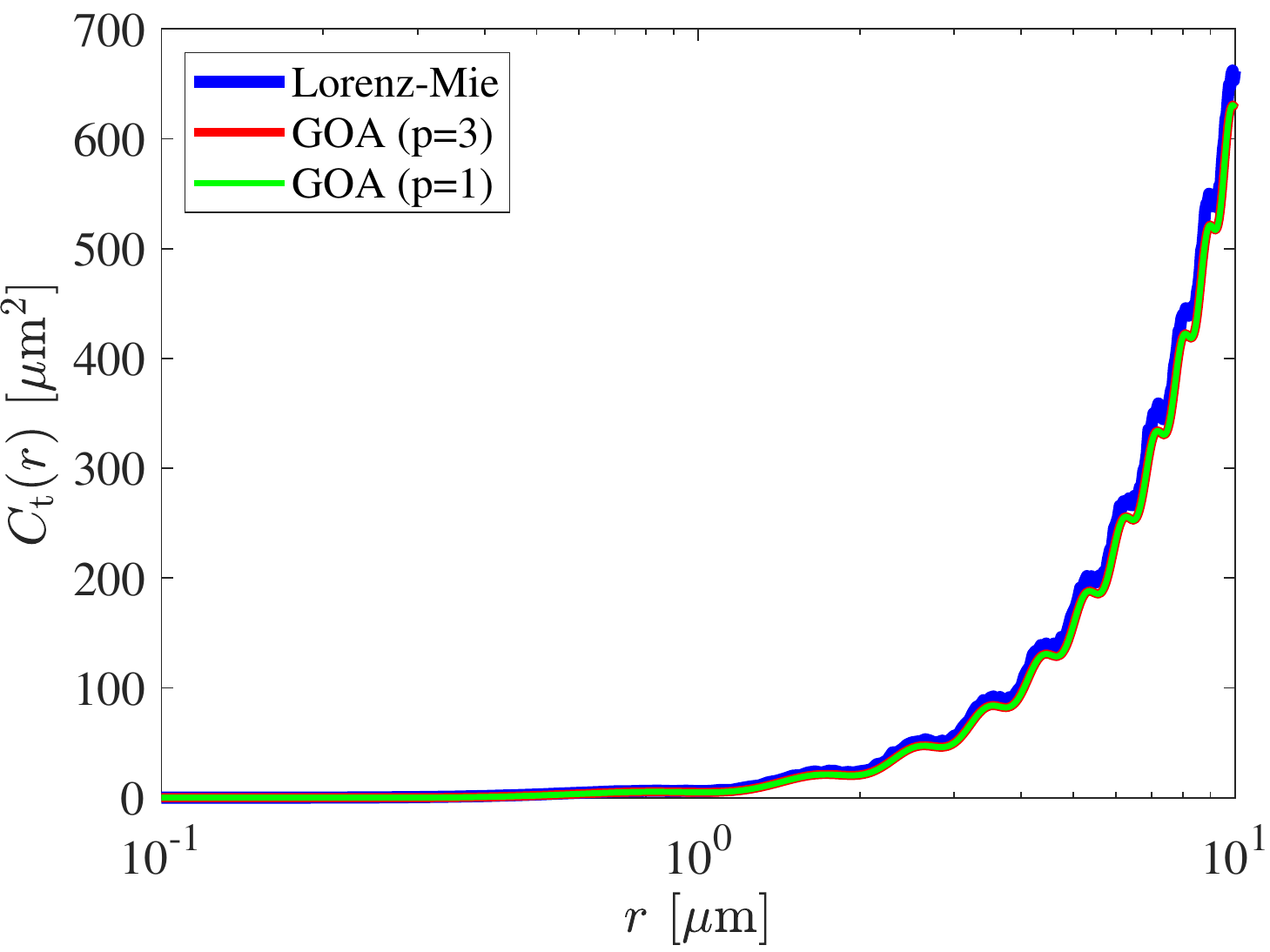}
\end{overpic}
\end{minipage}
}
\hspace{-0.08in}
\subfigure{
\begin{minipage}{0.48\linewidth}
\begin{overpic}[width=1.0\linewidth]{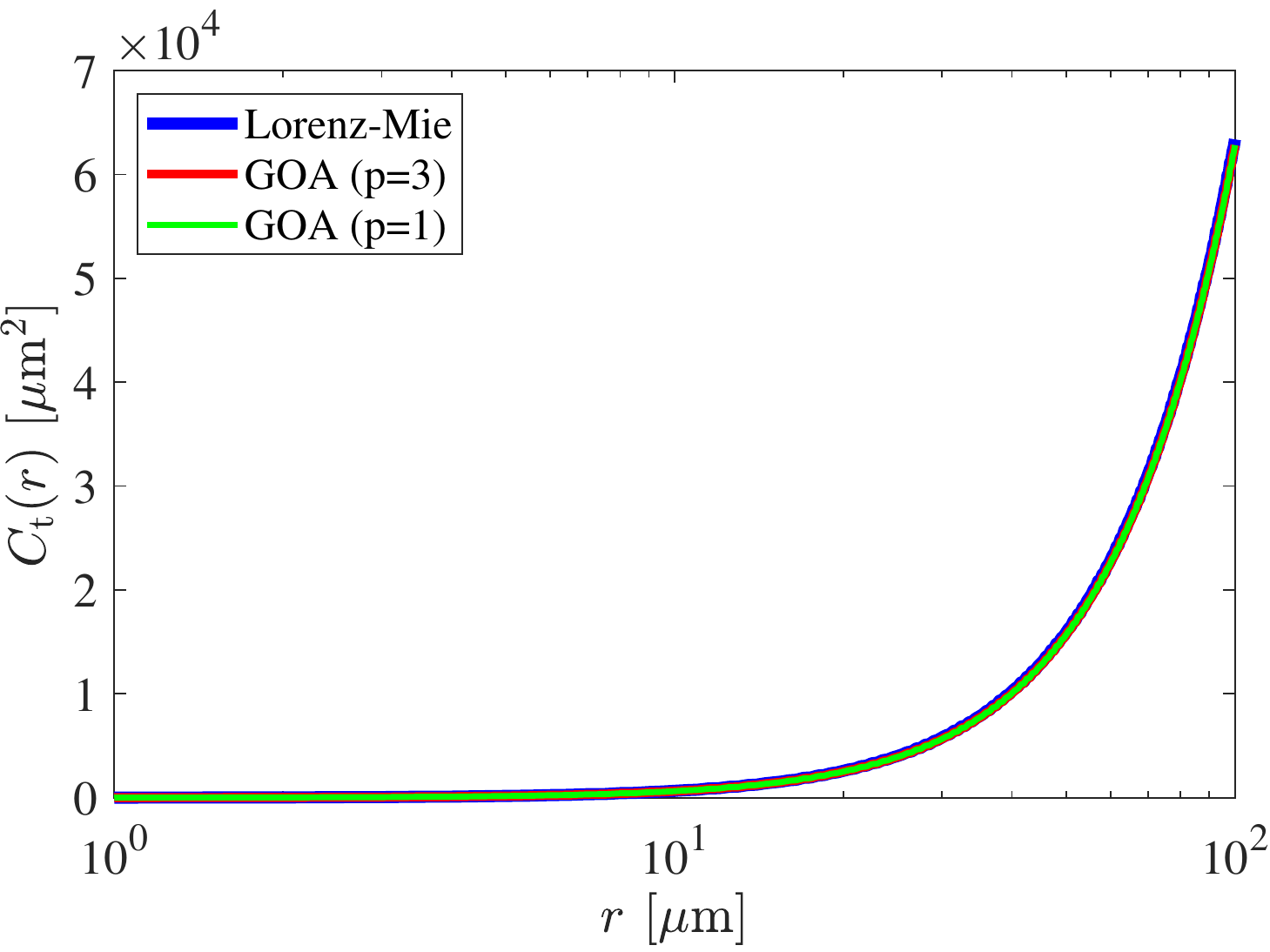}
\end{overpic}
\end{minipage}
}
\caption{\label{fig:Ct_1_100} Variation of the extinction cross section $C_\mathrm{t}$ as a function of $r$. Here, we compare our GOA (red: $p=3$, green: $p=1$) calculation against that of Lorenz-Mie theory (blue) with $\eta = 1.33$ and $\lambda = 0.6~\mathrm{{\mu}m}$. The ripple structures \cite{1977ApOpt} that are well captured by GOA are clearly shown in the left diagram.}
\end{figure}

Moreover, we can further simplify the extinction cross section to
\begin{equation}\label{eq:ct_1}
  C_\mathrm{t}(r) = 2\pi r^2 + \frac{4r\lambda\eta^2}{(\eta+1)^2|\eta-1|}\sin\left(\frac{4\pi r}{\lambda}(1-\eta)\right)
\end{equation}
by setting $p=1$, since $C_\mathrm{t}$ only relies on the value of $S_1$ (or $S_2$) evaluated at $\theta = 0$, and the light rays with $p>1$ contribute little to the forward scattering. This is evidenced in Fig. \ref{fig:Ct_1_100} where the $C_\mathrm{t}$ curves of $p=1$ (green) and $p=3$ (red) are virtually indistinguishable for a very wide range of $r$. In Appendix E, we show that the Relative Mean Squared Error (RelMSE) \footnote{The RelMSE for $C_\mathrm{t}$ is calculated by $(C_\mathrm{t}^{p=3}-C_\mathrm{t}^{p=1})^2/(C_\mathrm{t}^{p=3})^2$} is less than $10^{-5}$ for $\eta=1.33$ and $\lambda = 0.6~\mathrm{{\mu}m}$ using $p=1$ and $p=3$, respectively.

\subsection{Comparisons between GOA and Lorenz-Mie Theory}
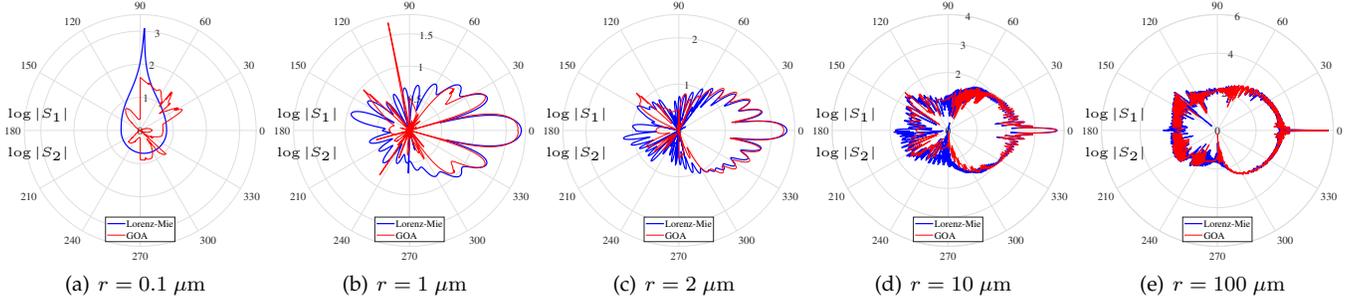
\begin{figure*}[h]
\centering
\subfigure[$r=0.1~\mathrm{{\mu}m}$]{
\begin{minipage}[b]{0.19\linewidth}
\begin{overpic}[width=1.0\linewidth]{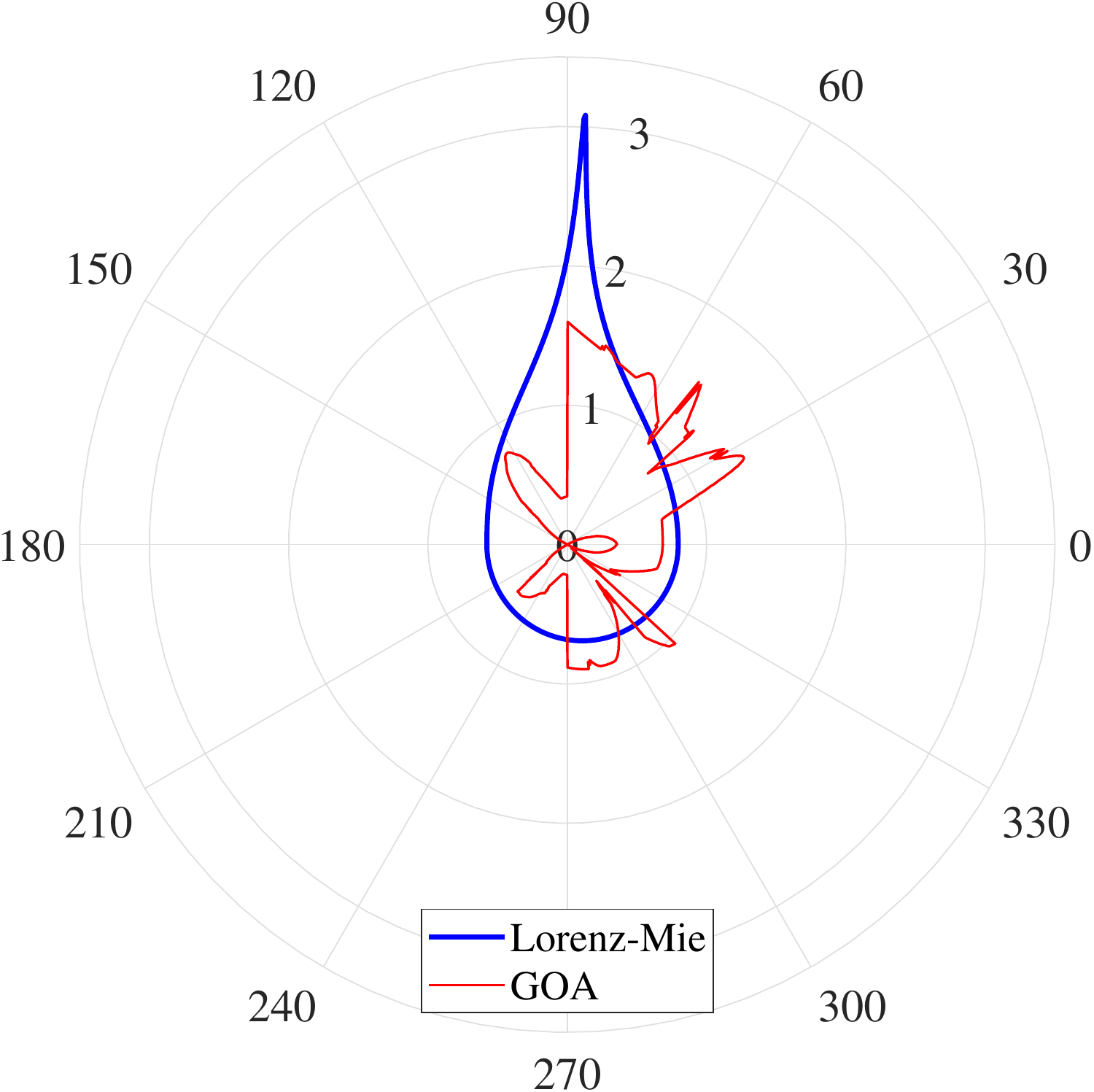}
\put(1,55){\tiny $\log|S_1|$}
\put(1,40){\tiny $\log|S_2|$}
\end{overpic}
\end{minipage}
}
\hspace{-0.1in}
\subfigure[$r=1~\mathrm{{\mu}m}$]{
\begin{minipage}[b]{0.19\linewidth}
\begin{overpic}[width=1.0\linewidth]{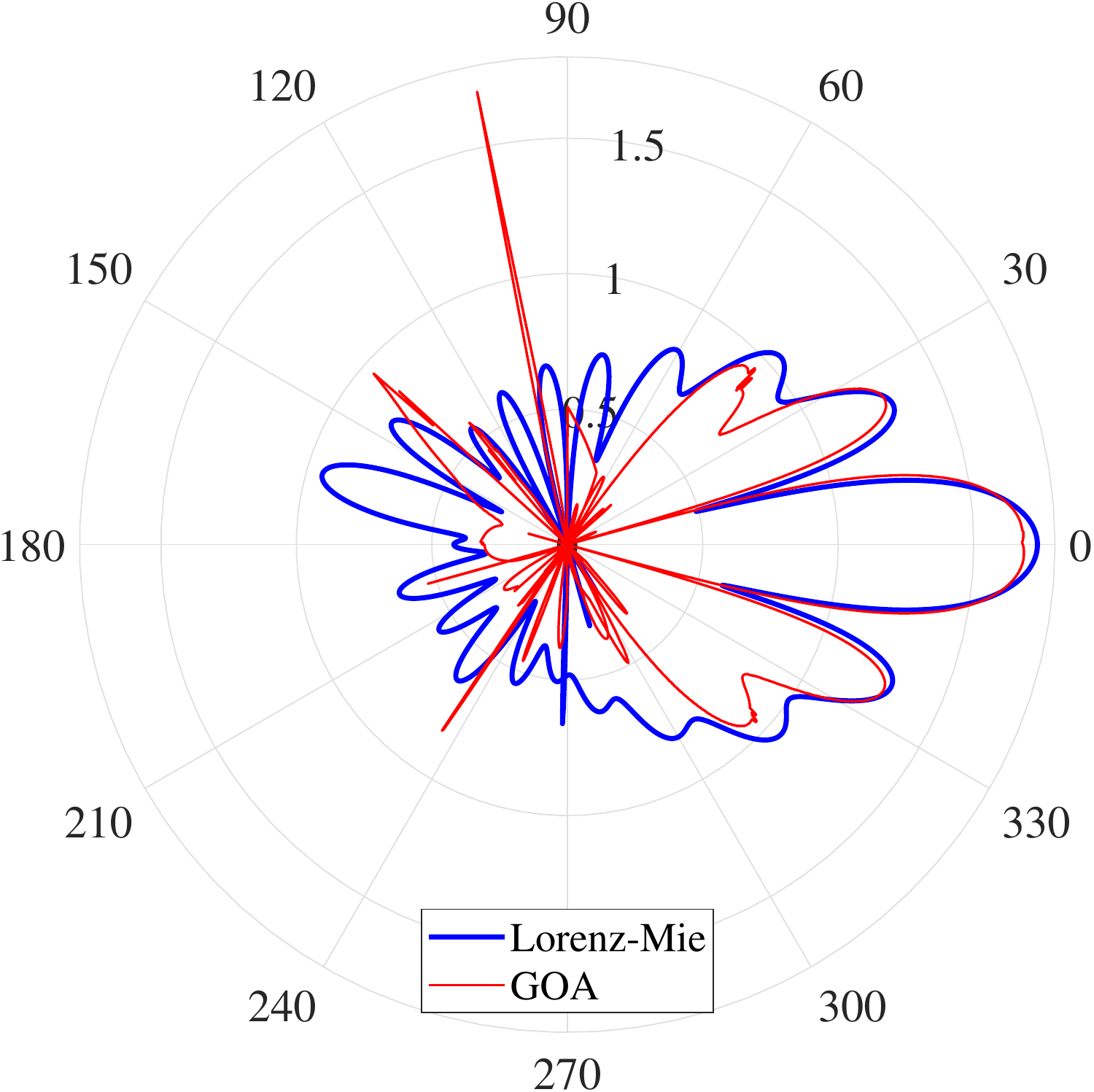}
\put(1,55){\tiny $\log|S_1|$}
\put(1,40){\tiny $\log|S_2|$}
\end{overpic}
\end{minipage}
}
\hspace{-0.1in}
\subfigure[$r=2~\mathrm{{\mu}m}$]{\label{fig:goa_mie2}
\begin{minipage}[b]{0.19\linewidth}
\begin{overpic}[width=1.0\linewidth]{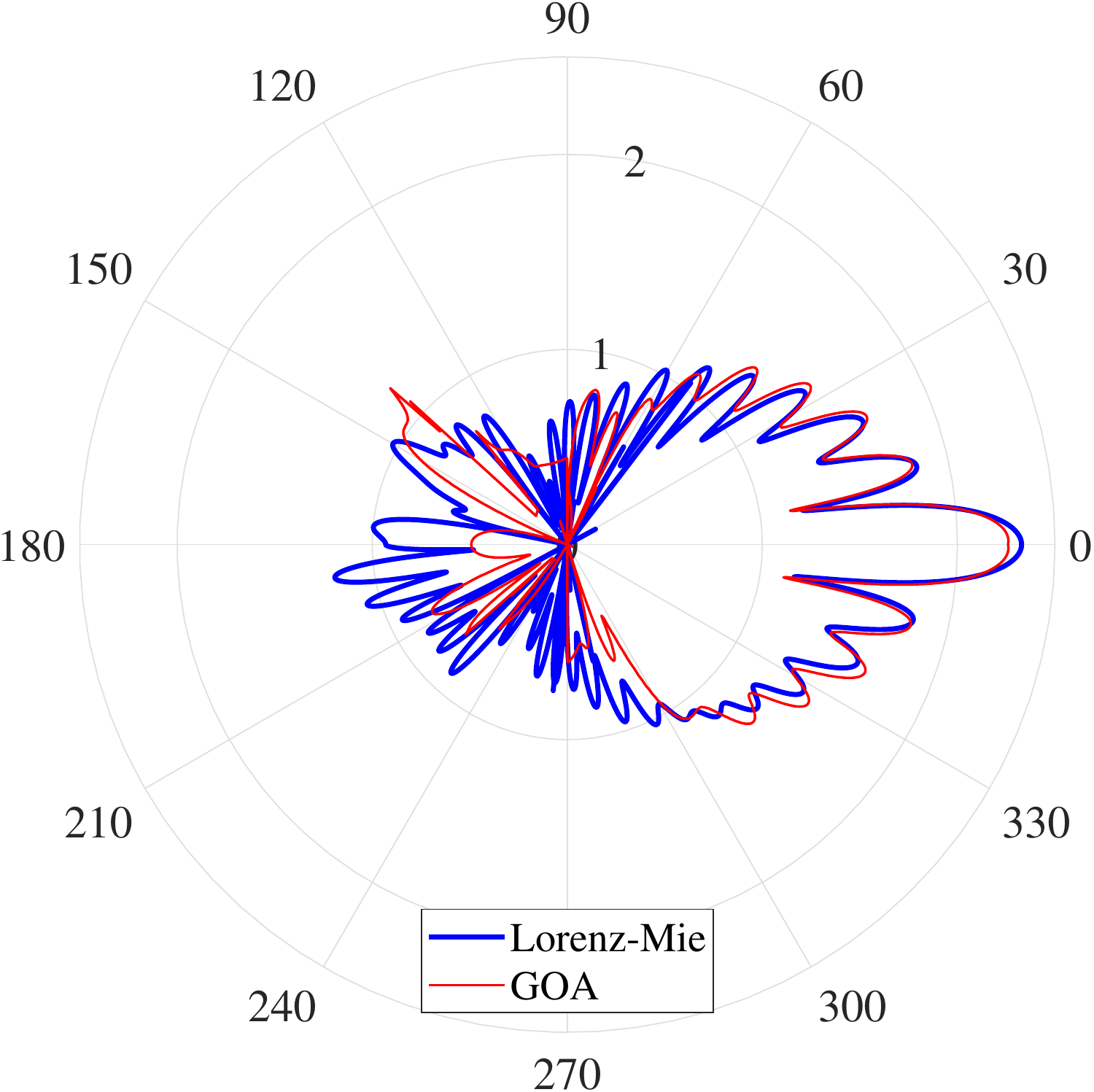}
\put(1,55){\tiny $\log|S_1|$}
\put(1,40){\tiny $\log|S_2|$}
\end{overpic}
\end{minipage}
}
\hspace{-0.1in}
\subfigure[$r=10~\mathrm{{\mu}m}$]{
\begin{minipage}[b]{0.19\linewidth}
\begin{overpic}[width=1.0\linewidth]{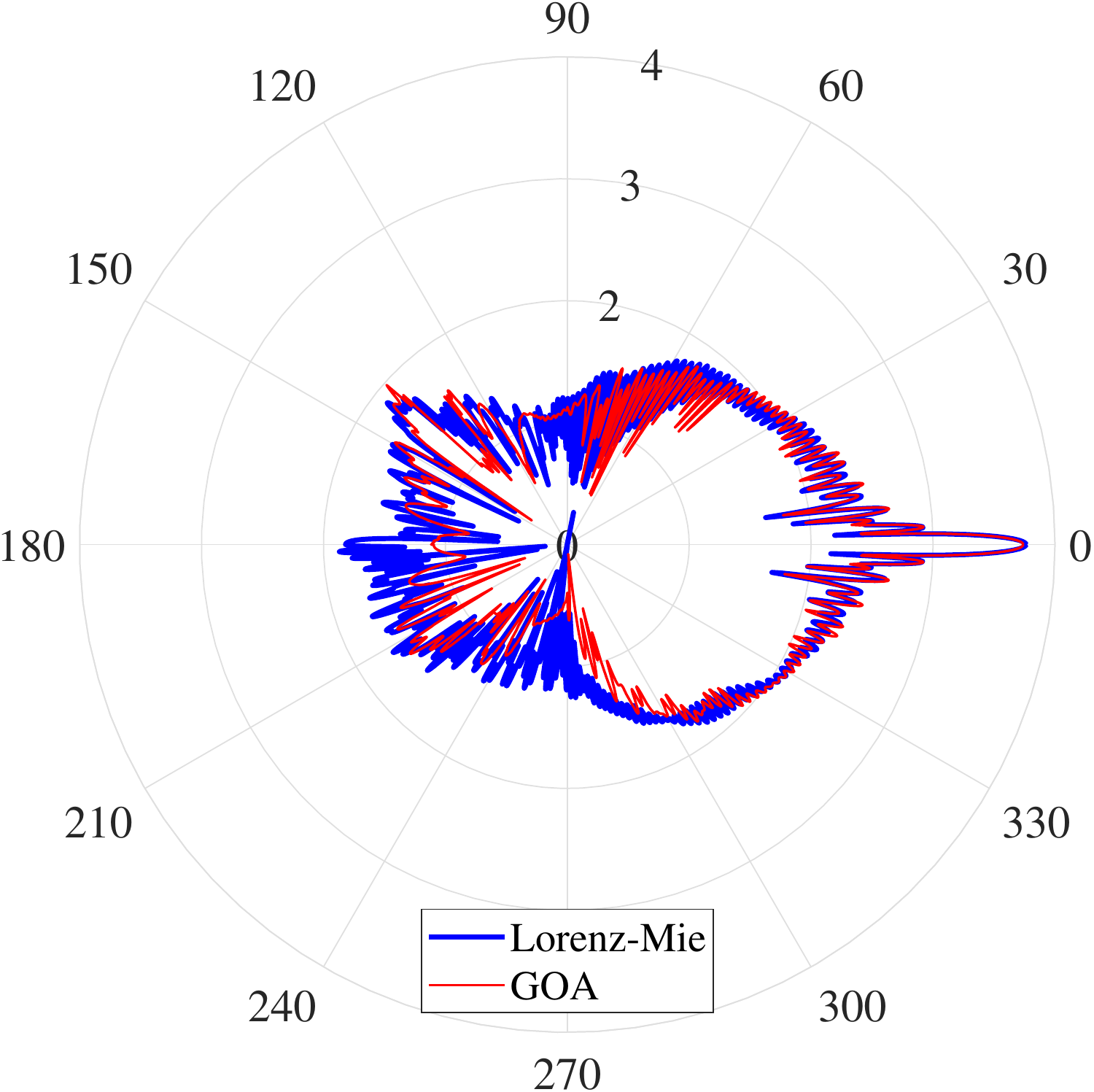}
\put(1,55){\tiny $\log|S_1|$}
\put(1,40){\tiny $\log|S_2|$}
\end{overpic}
\end{minipage}
}
\hspace{-0.1in}
\subfigure[$r=100~\mathrm{{\mu}m}$]{
\begin{minipage}[b]{0.19\linewidth}
\begin{overpic}[width=1.0\linewidth]{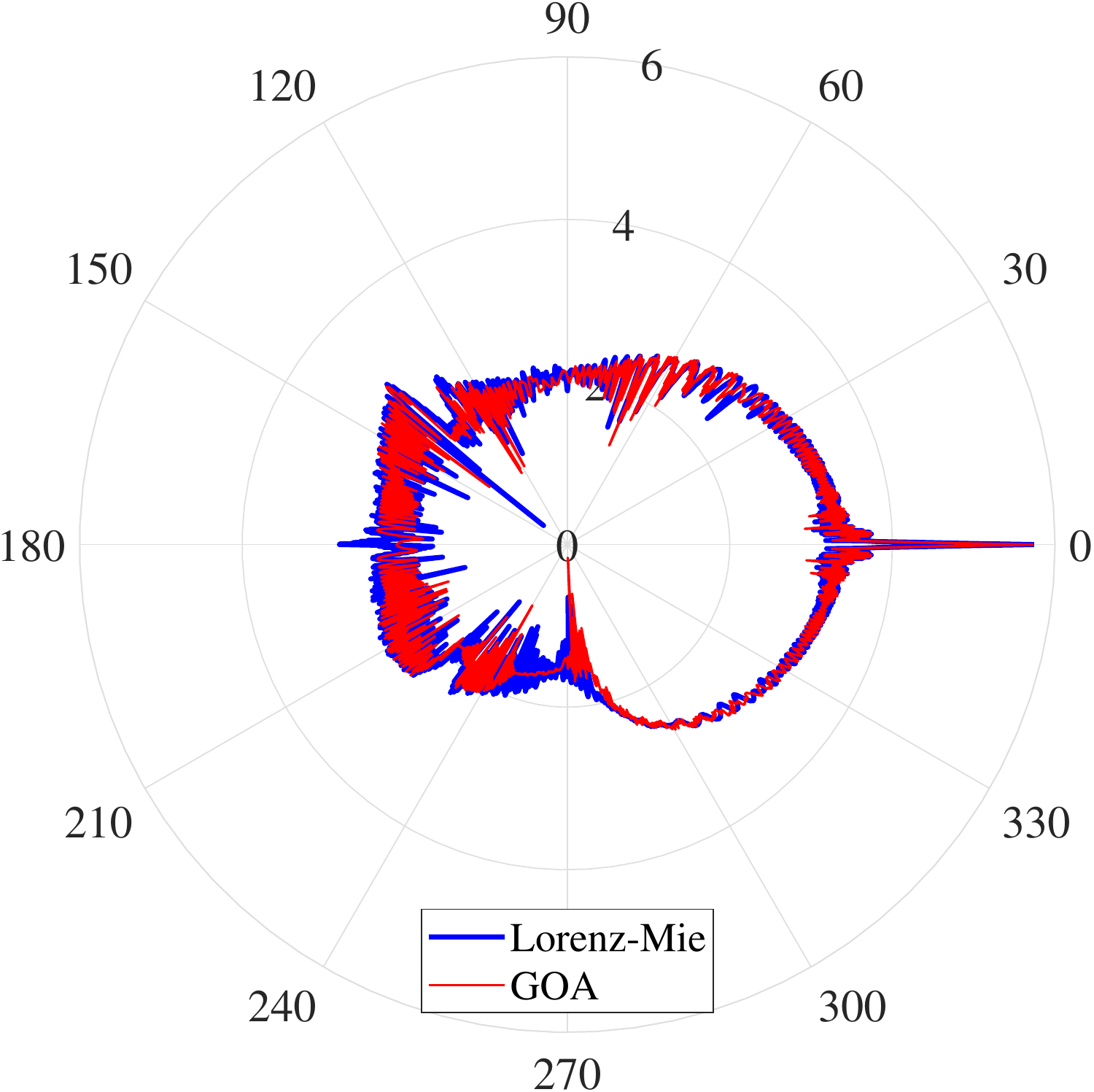}
\put(1,55){\tiny $\log|S_1|$}
\put(1,40){\tiny $\log|S_2|$}
\end{overpic}
\end{minipage}
}
\caption{\label{fig:goa_mie} Visual comparisons of $\log |S_1|$ (upper half) and $\log |S_2|$ (lower half) by Lorenz-Mie calculations (blue curves) with those by GOA (red curves) for $\eta = 1.33$ and $\lambda = 0.6~\mathrm{{\mu}m}$. The particle radius is set to $r=0.1, 1, 2, 10$ and $100~\mathrm{{\mu}m}$, respectively.}
\end{figure*}

\begin{figure}[h]
\centering
\rotatebox[origin=lt]{90}{\scriptsize{ \quad\quad\quad \bf GOA \quad\quad\quad\quad\quad\quad \bf Lorenz-Mie}}
\subfigure[$r=1~\mathrm{{\mu}m}$]{
\begin{minipage}[b]{0.46\linewidth}
\begin{overpic}[width=1.0\linewidth, trim={20px, 60px, 20px, 50px}, clip]{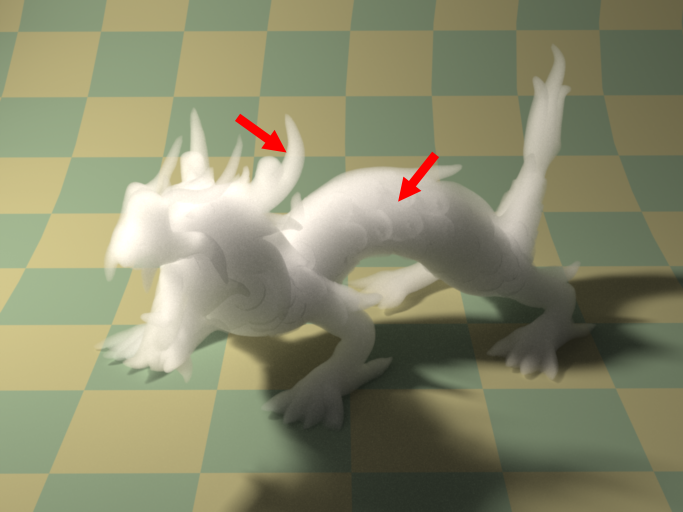}
\end{overpic}
\begin{overpic}[width=1.0\linewidth, trim={20px, 60px, 20px, 50px}, clip]{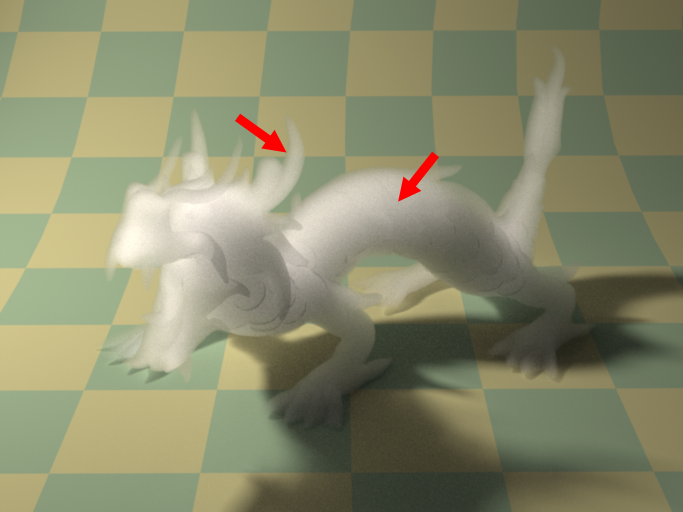}
\end{overpic}
\end{minipage}
}
\hspace{-0.1in}
\subfigure[$r=2~\mathrm{{\mu}m}$]{
\begin{minipage}[b]{0.46\linewidth}
\begin{overpic}[width=1.0\linewidth, trim={20px, 60px, 20px, 50px}, clip]{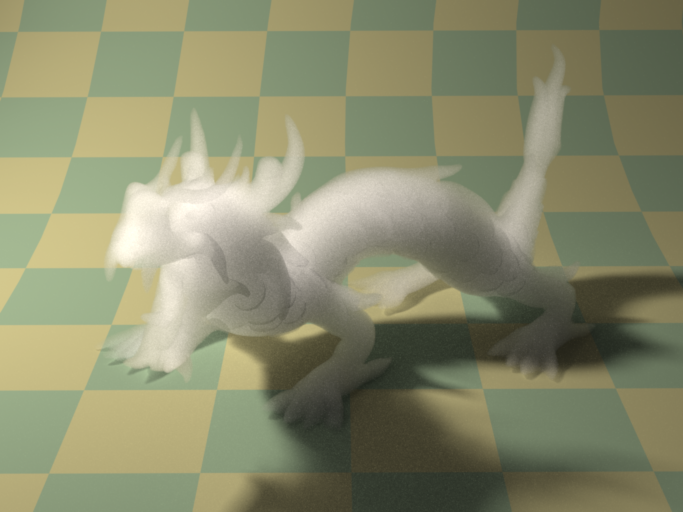}
\end{overpic}
\begin{overpic}[width=1.0\linewidth, trim={20px, 60px, 20px, 50px}, clip]{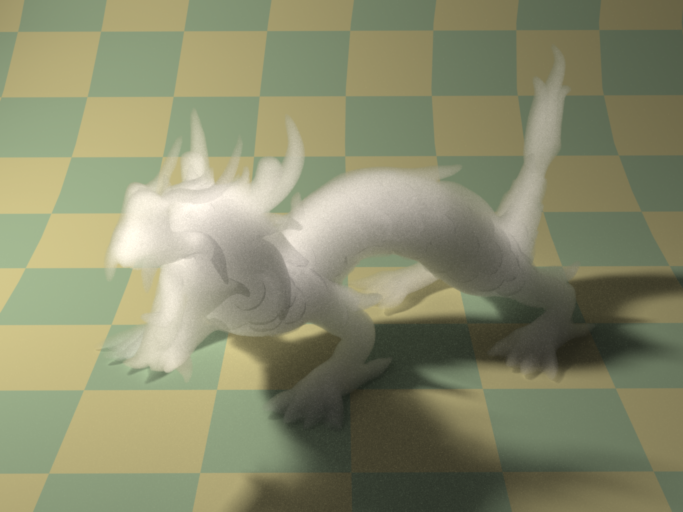}
\end{overpic}
\end{minipage}
}
\caption{\label{fig:goa_mie_dragon} Rendering a smooth medium with phase functions derived from Lorenz-Mie theory (top row) and GOA (bottom row), respectively. Here, $\eta = 1.33$. The red arrows highlight the differences.}
\end{figure}

\begin{figure}[h]
\centering
\subfigure{
\begin{minipage}{0.48\linewidth}
\begin{overpic}[width=1.0\linewidth]{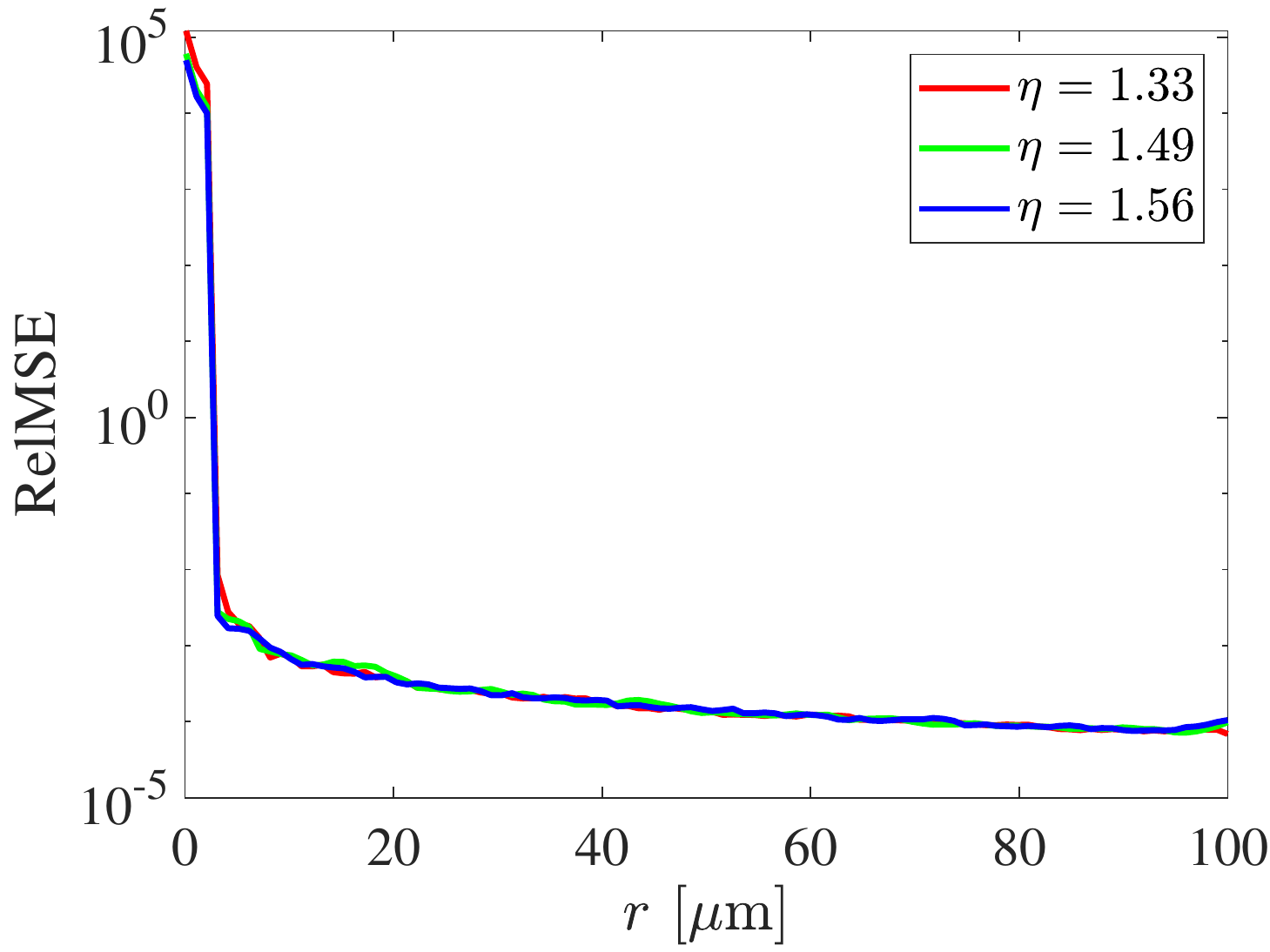}
\put(20, 30){\includegraphics[width=0.5\linewidth]{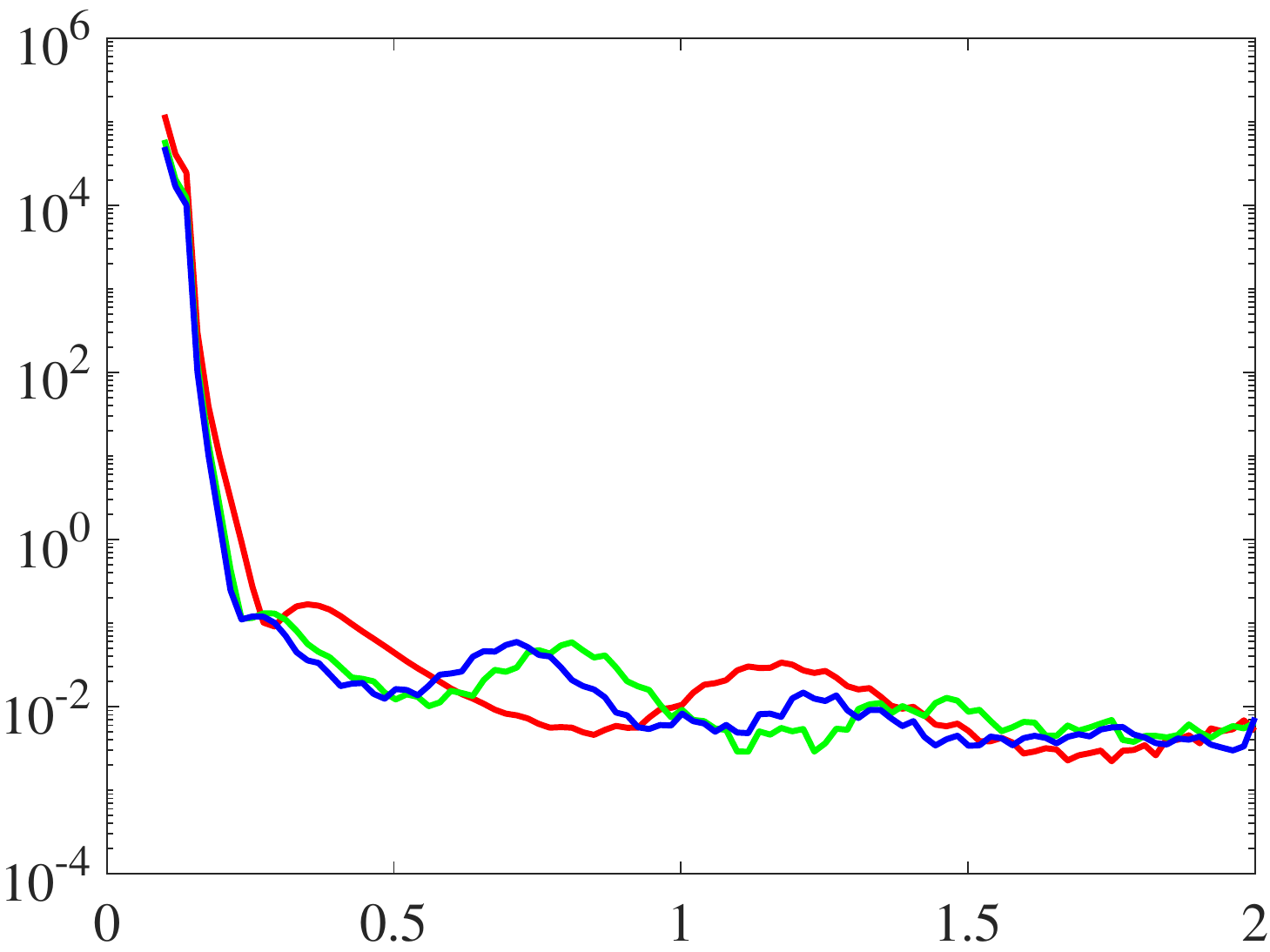}}
\end{overpic}
\end{minipage}
}
\hspace{-0.08in}
\subfigure{
\begin{minipage}{0.48\linewidth}
\begin{overpic}[width=1.0\linewidth]{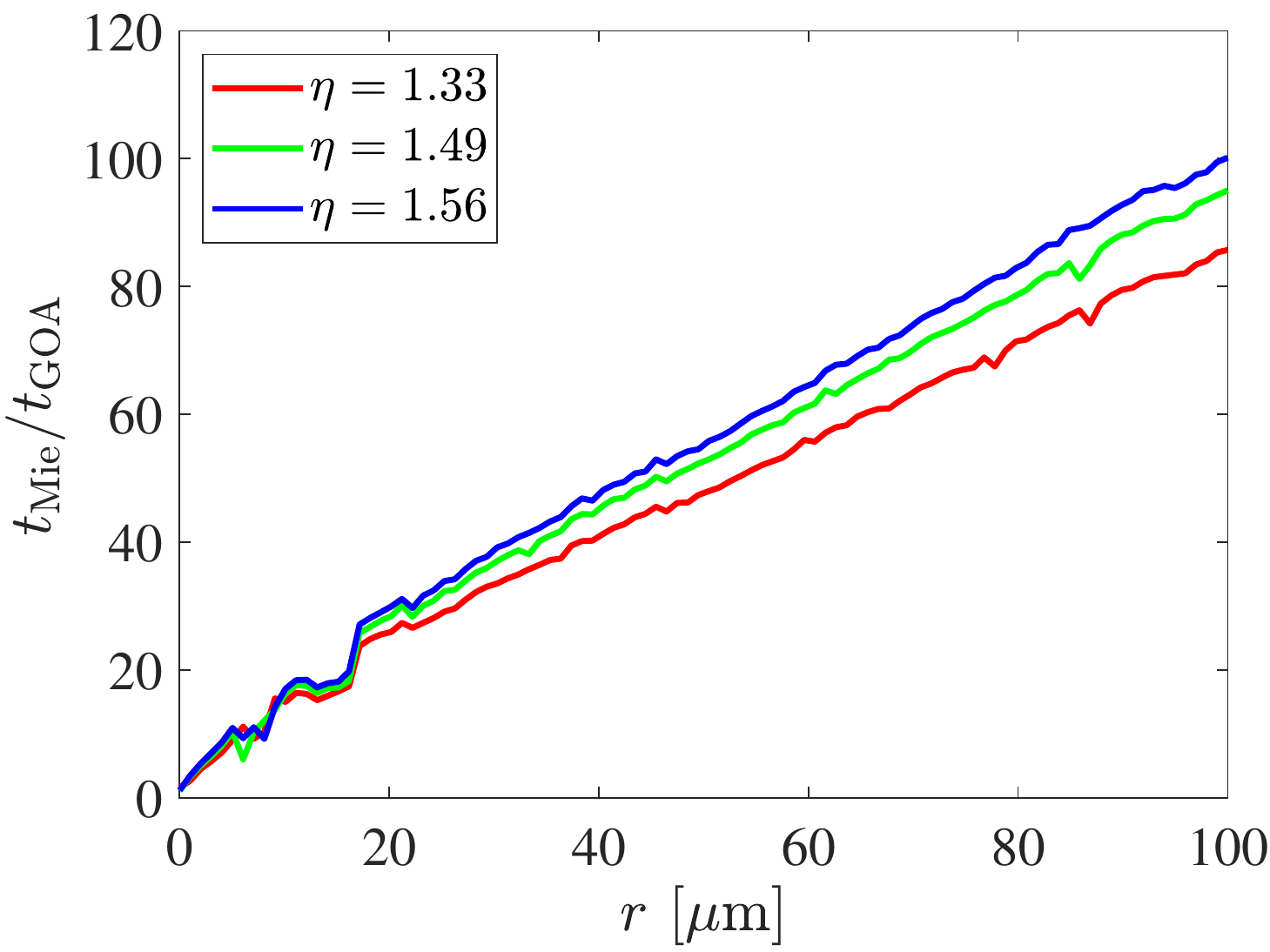}
\end{overpic}
\end{minipage}
}
\caption{\label{fig:goa_mie_mse} Comparisons between GOA and Lorenz-Mie theory in RelMSE and runtime complexity. Left: Variation of RelMSE as a function of $r$. Right: Variation of runtime ratio $t_\mathrm{Mie}/t_\mathrm{GOA}$ as a function of $r$.}
\end{figure}

To investigate the range of validity of GOA in simulating the scattering patterns of spherical particles, we compare the results with the rigorous Lorenz-Mie results on a wide range of particle radii in Fig. \ref{fig:goa_mie}. The wavelength of the incident rays is set to $0.6~\mathrm{{\mu}m}$ and the relative index of refraction is $\eta=1.33$ in all the calculations. Fig. \ref{fig:goa_mie} reveals that the scattering amplitude distributions by GOA align well with those obtained with Lorenz-Mie theory for large particles with $r>1~\mathrm{{\mu}m}$. The agreement of these two methods is especially good in almost all directions when the radius is large enough (e.g., $r=100~\mathrm{{\mu}m}$). However, when $r\approx1~\mathrm{{\mu}m}$, some discrepancies between the two methods appear. These discrepancies become large as the radius of particle decreases further.

To show the influence of these discrepancies on the perception of the translucent appearance, we render a smooth medium comprising monodisperse particles of radius $r$. We determine the phase function using either Lorenz-Mie theory or GOA according to $r$ \footnote{These phase functions are precomputed and stored in tables.}. The extinction coefficient is set to a constant $0.6$ for a fair comparison. The synthesized images are presented in Fig. \ref{fig:goa_mie_dragon} with $r$ setting to $1~\mathrm{{\mu}m}$ and $2~\mathrm{{\mu}m}$, respectively. Clearly, the differences of phase functions between Lorenz-Mie and GOA in the case of $r=1~\mathrm{{\mu}m}$ result in inconsistence of appearance. However, this inconsistence almost disappears completely when $r$ goes up to $2~\mathrm{{\mu}m}$, although there are some mismatches on the backward peaks of $S_1(\theta)$ (see Fig. \ref{fig:goa_mie2}, same for $S_2(\theta)$). Fig. \ref{fig:goa_mie} and Fig. \ref{fig:goa_mie_dragon} together have verified the accuracy of choosing GOA to compute $S_1(\theta)$ and $S_2(\theta)$ when $r\ge2~\mathrm{{\mu}m}$. Please see more comparisons and discussions in Appendix E.


To further show the similarity between Lorenz-Mie theory and GOA in computing $S_1(\theta)$ and $S_2(\theta)$, we report their Relative Mean Squared Error (RelMSE) in Fig. \ref{fig:goa_mie_mse} left. The RelMSE is computed on $(|S_1(\theta)|+|S_2(\theta)|)/2$ and different relative refractive indexes are tested. The results confirm that subtle errors exist when $r$ is large: despite some fluctuations, these calculations of GOA exhibit errors of less than $0.01$ as compared with exact Lorenz-Mie calculations when $r\ge2~\mathrm{{\mu}m}$.

In the calculations of $(|S_1(\theta)|+|S_2(\theta)|)/2$, Lorenz-Mie theory generally consumes much more time than GOA as evidenced in Fig. \ref{fig:goa_mie_mse} right. As $r$ increases, the runtime ratio of Lorenz-Mie theory and GOA $t_\mathrm{Mie}/t_\mathrm{GOA}$ grows linearly with respect to $r$, and can easily reach two orders of magnitude difference in performance. This is explained by the fact that the number of terms in Lorenz-Mie theory is linearly proportional to the size parameter $\alpha$, as we mentioned previously. In comparison, the runtime for GOA is independent of the radius $r$.

Considering the trade-off between accuracy and time complexity, we choose GOA when $r\ge2~\mathrm{{\mu}m}$ and switch to Lorenz-Mie theory otherwise. This makes the runtime of computing $S_1(\theta)$ (or $S_2(\theta)$) almost constant with respect to $r$ while retaining the accuracy as much as possible.





\section{Bulk Optical Properties with Graininess} \label{sec:VRE}
Now, we consider light scattering by a cloud of spherical particles of the same composition but of different sizes. The particles are assumed to be in each other's far-field regimes and their sizes are likely to range from wavelength-scale to the scale much larger than the wavelength. In this section, we first discuss the particle size distribution that may vary spatially and then study the bulk optical properties of the discrete participating medium considering graininess. Thanks to the high efficiency of GOA, we are able to evaluate the bulk optical properties on-the-fly.


\subsection{Particle Size Distribution}
We use the particle size distribution (PSD) $N(r)$ to describe the population of particles in a discrete participating medium. Thus $N(r)\ud r$ is the total concentration (particle number per unit volume) of particles with sizes in the domain $[r, r+\ud r]$. The total particle number concentration within some limited interval $[r_\mathrm{min},r_\mathrm{max}]$ of sizes is obtained by $N_0 = \int_{r_\mathrm{min}}^{r_\mathrm{max}} N(r)\ud r$. To use $N(r)$ as a probability density distribution (PDF), we have to normalize $N(r)$ via $N(r)/N_0$.

\begin{figure}[t]
  \centering
  \includegraphics[width=1.0\linewidth]{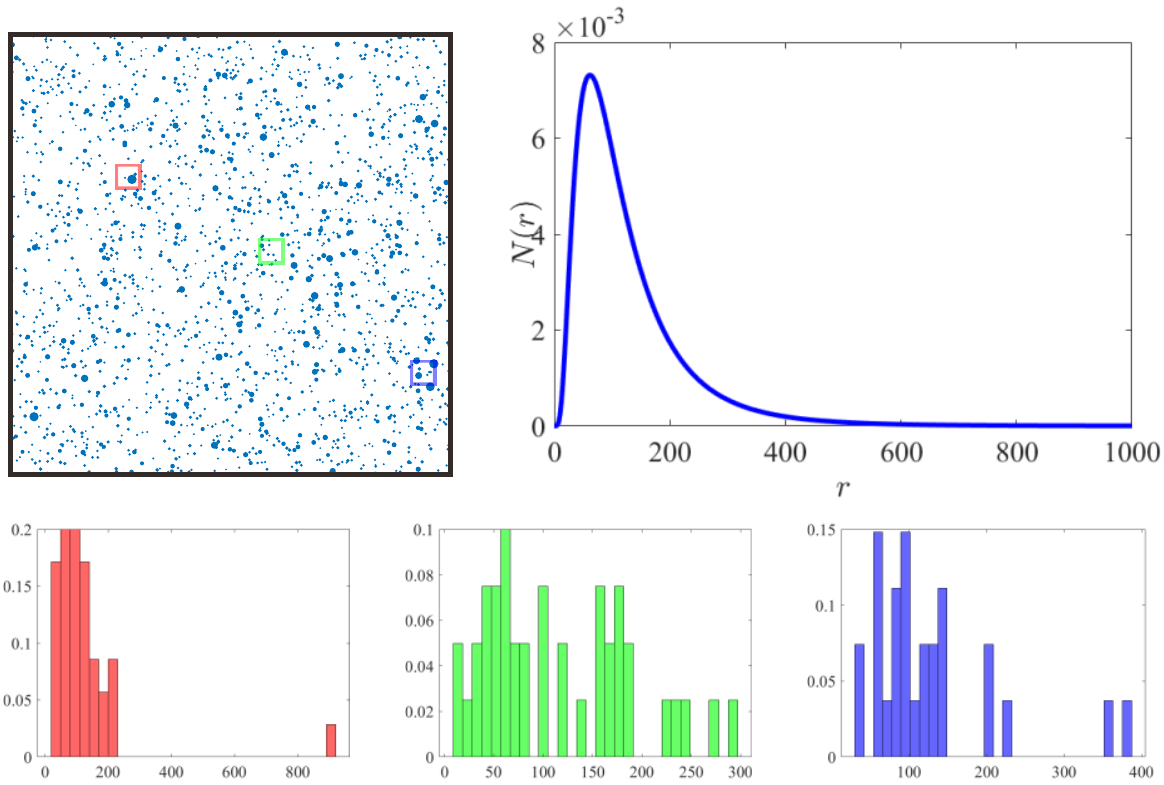}\\
  \caption{Illustration of spatially-varying PSD. First row: A realization of $2000$ polydisperse particles and its global PSD. Second row: Actual PSDs for particles in three small patches.}\label{fig:svpsd}
\end{figure}

It is generally reported that particle sizes follow close to a log-normal distribution \cite{Frisvad:2007:CSP:1275808.1276452,sadeghi11physically}:
\begin{equation}
N(r) = \frac{N_0}{\sqrt{2\pi}r \ln \sigma_\mathrm{g}} \exp\left[-\frac{1}{2}\left(\frac{\ln r - \ln \bar{r}_\mathrm{g}}{\ln \sigma_\mathrm{g}}\right)^2\right]
\end{equation}
in which $\sigma_\mathrm{g}$ is the geometric standard deviation and $\bar{r}_\mathrm{g}$ is the geometric mean radius. Obviously, this statistical tendency stems from the observation of a large number of particles. A small number of particles will give rise to a size distribution deviating from the log-normal distribution. To demonstrate this in 2D, we generate $2000$ particles in a box with uniformly distributed positions and log-normal distributed radii. The visualization of these particles and its PSD are shown in the first row of Fig. \ref{fig:svpsd}. We then extract three small patches from the box and estimate their actual PSDs by binning. As expected, the PSDs plotted in the second row of Fig. \ref{fig:svpsd} vary spatially and contain quite different features. In what follows, we use the notation $N(r,\mbx)$ to emphasize that the PSD varies spatially. Nevertheless, the ensemble average over these spatially-varying PSDs converges to the log-normal distribution shown in the first row of Fig. \ref{fig:svpsd}. Rendering with this global PSD yields a smooth appearance similar to that from a traditional continuous medium.

\subsection{Bulk Optical Properties}
With the spatially-varying PSD $N(r,\mbx)$, we are able to obtain the bulk optical properties of a local area in which many independent particles are immersed. Supposing that $\mathcal{V}_\mbx$ is a small volume centered around $\mbx$, the bulk extinction coefficient $\sigma_\mathrm{t}$ \footnote{In rendering literature, the symbol $\sigma$ refers to the cross section sometimes, while using $\mu$ for the coefficient.} of this volume is evaluated by
\begin{equation}\label{eq:bulk_sigma_t}
  \sigma_\mathrm{t}(\mathcal{V}_\mbx) = \frac{1}{\mu(\mathcal{V}_\mbx)}\int_{r_\mathrm{min}(\mathcal{V}_\mbx)}^{r_\mathrm{max}(\mathcal{V}_\mbx)}C_\mathrm{t}(r)\int_{\mbx\in\mathcal{V}_\mbx}N(r,\mbx)\ud \mbx \mathrm{d}r
\end{equation}
in which $\mu(\mathcal{V}_\mbx)$ is the measurement of $\mathcal{V}_\mbx$. $r_\mathrm{min}(\mathcal{V}_\mbx)$ and $r_\mathrm{max}(\mathcal{V}_\mbx)$ return the minimum and maximum particle radii inside $\mathcal{V}_\mbx$, respectively. For brevity, we simplify both $r_\mathrm{min}(\mathcal{V}_\mbx)$ and $r_\mathrm{max}(\mathcal{V}_\mbx)$ by dropping $\mathcal{V}_\mbx$ henceforth. For the scattering coefficient and the absorption coefficient, they can be defined in a similar way by replacing $C_\mathrm{t}(r)$ with $C_\mathrm{s}(r)$ and $C_\mathrm{a}(r)$, respectively \footnote{These properties can be viewed as the properties at position $\mbx$ when $\mathcal{V}_\mbx$ is infinitely small, i.e., $\sigma_\mathrm{t}(\mbx)=\sigma_\mathrm{t}(\mathcal{V}_\mbx)$, $\sigma_\mathrm{s}(\mbx)=\sigma_\mathrm{s}(\mathcal{V}_\mbx)$ and $\sigma_\mathrm{a}(\mbx)=\sigma_\mathrm{a}(\mathcal{V}_\mbx)$.}. Generally, these properties exhibit multi-scale effects with respect to the size of $\mathcal{V}_\mbx$.

\begin{figure}[t]
\centering
\subfigure{
\begin{minipage}[b]{0.32\linewidth}
\begin{overpic}[width=1.0\linewidth]{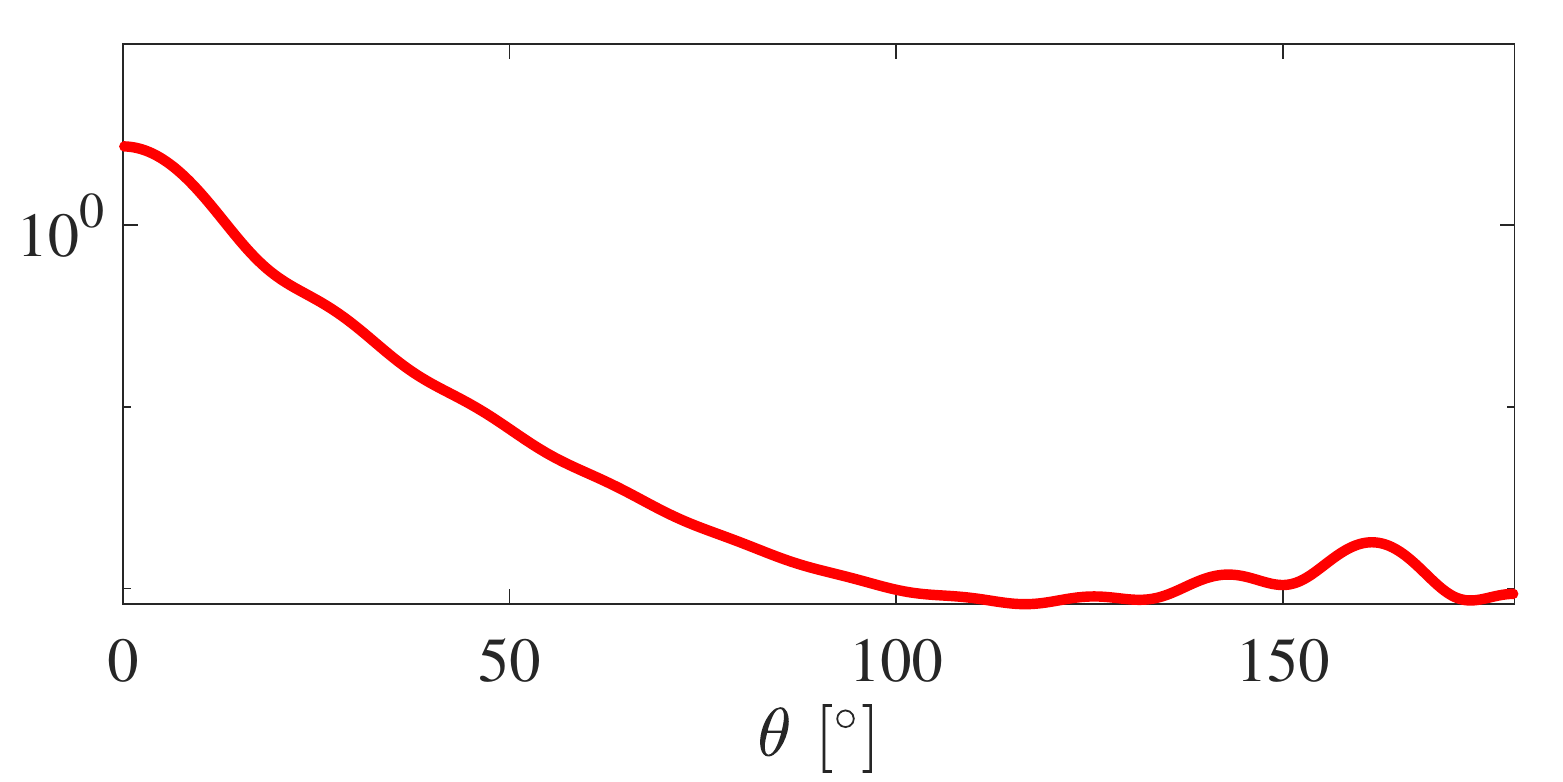}
\end{overpic}\\
\begin{overpic}[width=1.0\linewidth]{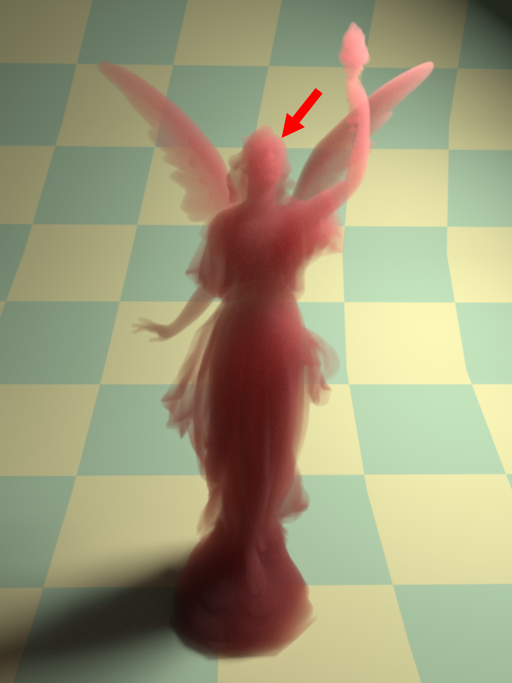}
\put(2,2){\tiny \textcolor[rgb]{1,1,1}{$r_\mathrm{min}=0.1$, $r_\mathrm{max}=1$}}
\end{overpic}
\end{minipage}
}
\hspace{-0.12in}
\subfigure{
\begin{minipage}[b]{0.32\linewidth}
\begin{overpic}[width=1.0\linewidth]{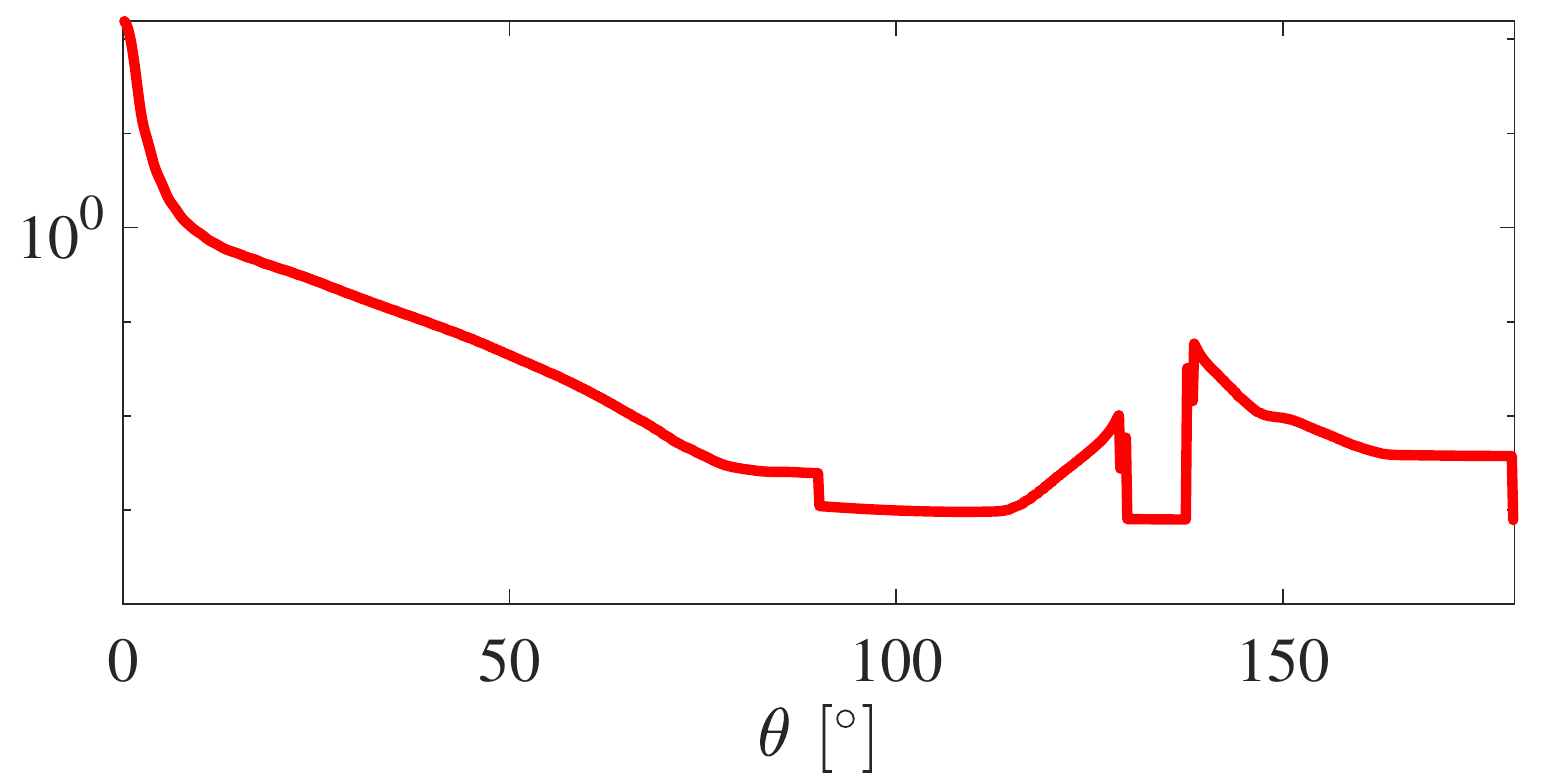}
\end{overpic}\\
\begin{overpic}[width=1.0\linewidth]{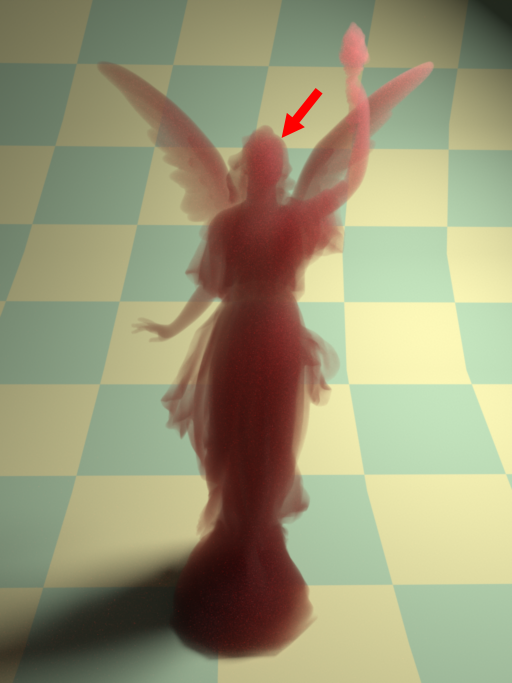}
\put(2,2){\tiny \textcolor[rgb]{1,1,1}{$r_\mathrm{min}=1$, $r_\mathrm{max}=10$}}
\end{overpic}
\end{minipage}
}
\hspace{-0.12in}
\subfigure{
\begin{minipage}[b]{0.32\linewidth}
\begin{overpic}[width=1.0\linewidth]{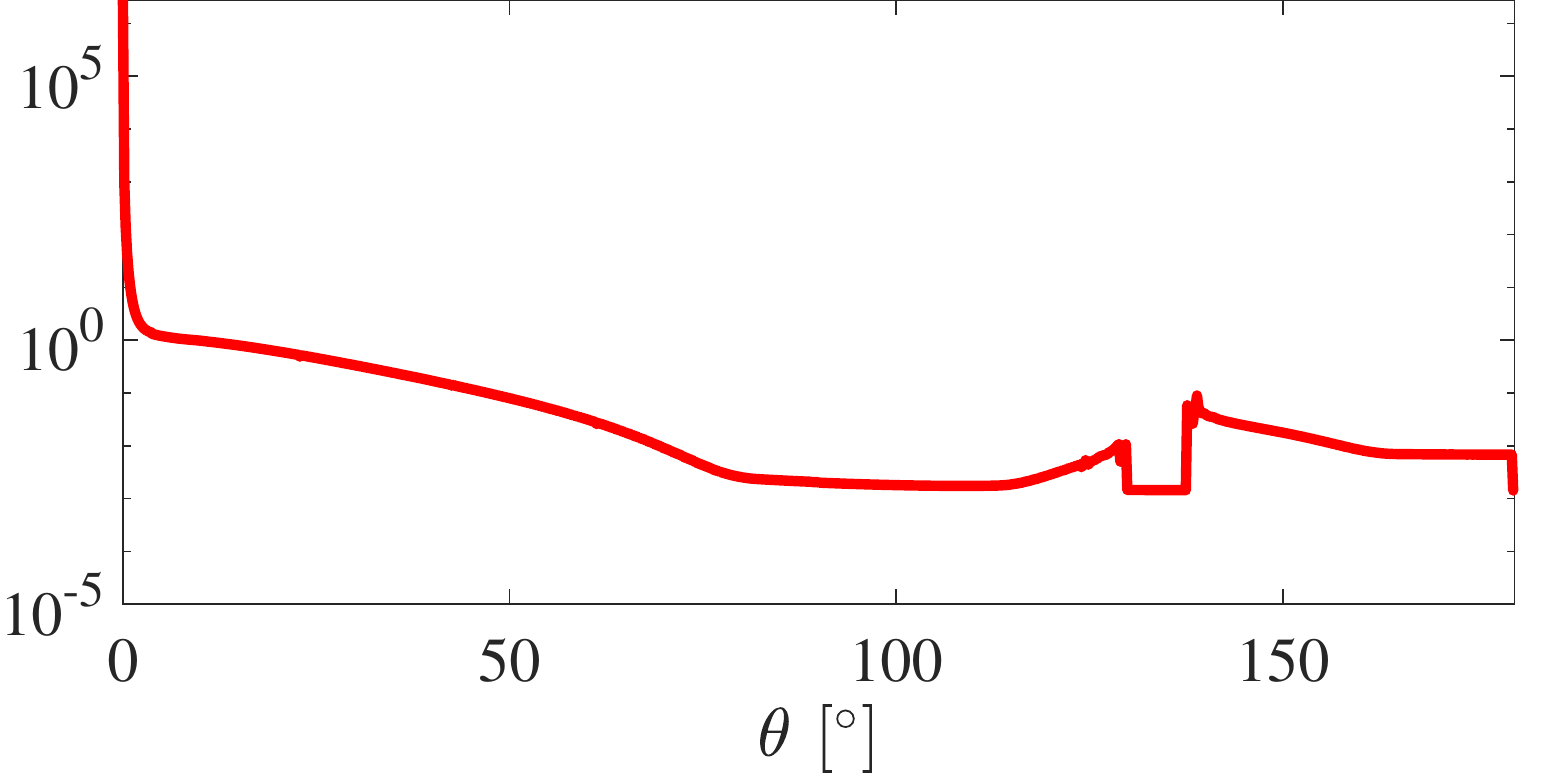}
\end{overpic}\\
\begin{overpic}[width=1.0\linewidth]{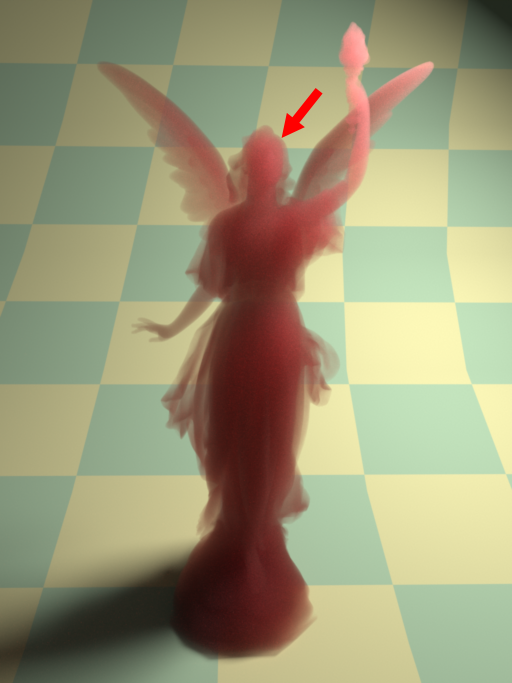}
\put(2,2){\tiny \textcolor[rgb]{1,1,1}{$r_\mathrm{min}=1$, $r_\mathrm{max}=1000$}}
\end{overpic}
\end{minipage}
}
\caption{\label{fig:phase} Top row: Visualizations of the phase functions estimated by our method. Bottom row: The corresponding renderings of a smooth homogeneous medium with these phase functions, while keeping other properties fixed. The red arrows highlight the differences.}
\end{figure}

The ensemble phase function is derived as
\begin{equation}\label{eq:bulk_phase}
\begin{split}
  f(\mathcal{V}_\mbx,\theta) &=\frac{\int_{r_\mathrm{min}}^{r_\mathrm{max}} C_\mathrm{s}(r)f_{\mathrm{p}}(\theta,r)\int_{\mbx\in\mathcal{V}_\mbx}N(r,\mbx)\ud \mbx\mathrm{d}r}{\mu(\mathcal{V}_\mbx)\sigma_\mathrm{s}(\mathcal{V}_\mbx)} \\
  &=\frac{\int_{r_\mathrm{min}}^{r_\mathrm{max}} (|S_1(\theta,r)|^2+|S_2(\theta,r)|^2)\int_{\mbx\in\mathcal{V}_\mbx}N(r,\mbx)\ud \mbx\mathrm{d}r}{2|k|^2\mu(\mathcal{V}_\mbx)\sigma_\mathrm{s}(\mathcal{V}_\mbx)}.
\end{split}
\end{equation}
in which $\sigma_\mathrm{s}(\mathcal{V}_\mbx)$ serves as the normalization factor for $f(\mathcal{V}_\mbx,\theta)$. Fig. \ref{fig:phase} visualizes the phase functions generated by the above formula. Here, we use uniform sampled radii between $r_\mathrm{min}$ and $r_\mathrm{max}$ \footnote{This is similar to the log-normal distribution with a very large $\sigma_\mathrm{g}$.}. By fixing other properties, we show the influence of these phase functions on the final appearance of a smooth homogeneous medium in Fig. \ref{fig:phase}.
Clearly, this formula is only valid for $\sigma_\mathrm{s}(\mathcal{V}_\mbx)\neq 0$.
When $\sigma_\mathrm{s}(\mathcal{V}_\mbx)= 0$, i.e., the volume $\mathcal{V}_\mbx$ is free of particles, $f(\mathcal{V}_\mbx,\theta)$ degenerates into a delta function: $\sigma_\mathrm{s}(\mathcal{V}_\mbx)f(\mathcal{V}_\mbx,\theta)=\delta(\theta)$.

Similarly, we can derive the transmittance along a light beam of length $s$ as
\begin{equation}\label{eq:transmittance_graniness}
  T_\mathcal{A}(s) =  \exp \left\{- \frac{1}{\mu(\mathcal{A})} \int_{r_\mathrm{min}(\mathcal{C})}^{r_\mathrm{max}(\mathcal{C})}C_\mathrm{t}(r)\int_{\mbx\in \mathcal{C}}N(r,\mbx)\ud \mbx \mathrm{d}r\right\}
\end{equation}
in which $\mathcal{C} = \mathcal{A}\times s$ represents a small cylinder around the light beam and $\mathcal{A}$ is its cross section. The derivation is provided in Appendix D.

For particles with a monodisperse distribution, the transmittance is simplified to
\begin{equation}\label{eq:transmittance_mono}
  T_\mathcal{A}(s) =  \exp \left\{- \frac{C_\mathrm{t}}{\mu(\mathcal{A})} \int_{\mbx\in\mathcal{C}}N(\mbx)\ud \mbx \right\}
\end{equation}
in which $C_\mathrm{t}$ is a constant and $N(\mbx)$ is a function of $\mbx$ only. The integral in the above formula simply counts the number of particles located in the query region $\mathcal{A}\times s$.

\section{Rendering Solution} \label{sec:render}
With these bulk optical properties, we are able to derive a multi-scale \emph{volumetric rendering equation} (VRE) describing radiative transfer in discrete random media. Then, we develop a Monte Carlo sampling based solution to solve the VRE. This solution only requires the position and the size of each particle, avoiding the explicit tessellation of its shape. In a preprocessing stage, we generate and store $N_\mathrm{tot}$ particles with random positions and log-normal distributed radii for a discrete participating medium. During rendering, the stored particles are queried to determine the optical properties for each traced ray. A uniform grid is developed for acceleration.

\subsection{Multi-scale Volumetric Rendering Equation}
Conventionally, the VRE describing macroscopic light scattering in participating media is written as
\begin{equation}
\begin{split}
  L(\mbx, \mbomega) &= T(\mbx, \mbz) L(\mbz, \mbomega) \\
  &+ \int_0^z T(\mbx, \mby) \int_{\mathbb{S}^2} \sigma_\mathrm{s}(\mby) f(\mathbf{y}, \mbomega', \mbomega) L_\mathrm{i}(\mathbf{y},\mbomega')\ud \mbomega' \ud y
\end{split}
\end{equation}
in which $L(\mathbf{x},\mbomega)$ is the radiance arriving at $\mbx\in\mathbb{R}^3$ along a direction $\mbomega\in\mathbb{S}^2$, $f$ represents the scattering phase function characterizing the probability of radiation incident from $\mbomega'$ being scattered into direction $\mbomega$, $z$ is the distance through the medium to the nearest boundary at $\mathbf{z} = \mathbf{x}-z\mbomega$ and $\mathbf{y} = \mathbf{x}-y\mbomega$ is a point at distance $y\in(0,z)$. The conventional transmittance between $\mbx$ and $\mby$ is computed as $T(\mbx, \mby) = \exp\{-\int_0^y \sigma_\mathrm{t}(\mbx-s\mbomega) \ud s\}$.

By substituting the multi-scale properties into the above equation, we arrive at a multi-scale version of the VRE:
\begin{equation}\label{eq:multi_VRE}
\begin{split}
  L(\mathbf{x},\mbomega) &= T_\mathcal{A}(\mbx, \mbz) L(\mbz, \mbomega)\\
  &+ \int_0^z T_\mathcal{A}(\mbx, \mby) \int_{\mathbb{S}^2} L_\mathrm{i}(\mathbf{y},\mbomega') Q(\theta)\mathrm{d}\mbomega' \ud y
\end{split}
\end{equation}
with
\begin{equation}\label{eq:phase_Q}
Q(\theta) =
                                        \left\{ \begin{array}{ll}
                                        \sigma_\mathrm{s}(\mathcal{V}_\mby)f(\mathcal{V}_\mby, \theta) &  N(\mathcal{V}_\mby) \neq 0\\
                                        \delta(\theta)  & N(\mathcal{V}_\mby) = 0.
                                        \end{array}
                                        \right.
\end{equation}
Since this multi-scale VRE is a general extension to the conventional one, it naturally supports multiple scattering.

\subsection{Query Cylinder}
\begin{figure}[t]
  \centering
  \includegraphics[width=1.0\linewidth]{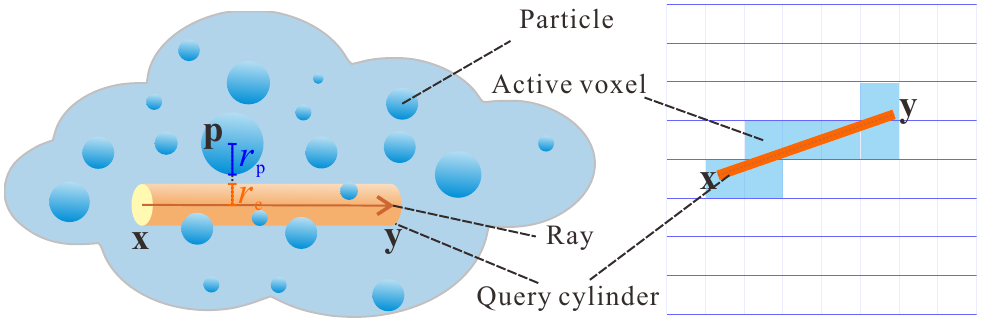}\\
  \caption{Illustration of a query cylinder and the active voxels in our ray traversal algorithm. Only the particles contained in the active voxels will be queried.}\label{fig:cylinder}
\end{figure}

In our multi-scale VRE, every optical property depends on a PSD while the PSD is defined on a differential volume.
To evaluate the transmittance $T_\mathcal{A}(\mbx, \mby)$ between any two positions $\mbx$ and $\mby$ in the medium, we need a differential volume around the ray $\mbx\to\mby$. This volume is used to query particles which contribute to the transmittance $T_\mathcal{A}(\mbx, \mby)$. In our implementation, we design it to be a thin cylinder centered around $\mbx\to\mby$, as illustrated in Fig. \ref{fig:cylinder}. We name such a cylinder as a query cylinder. In this sense, we view each ray as a ``fat ray'' which gathers small particles along its trajectory. This is quite different from the implementation of rendering continuous media in which the optical properties are determined globally, without explicitly querying particles in a local area.

In theory, the cross section $\mathcal{A}$ should be infinitely small. However, a too small one may have numerical issues and cause large variance.
Conversely, bias will be introduced in $T_\mathcal{A}(\mbx, \mby)$ when $\mathcal{A}$ is very large. In practice, we select $\mathcal{A}$ as follows and keep it unchanged as the ray traverses the medium. Supposing that $z_\mathrm{near}$ and $z_\mathrm{med}$ respectively denote the depth of the near plane and the smallest depth of the medium in the view frustum, the size of $\mathcal{A}$ is selected according to
\begin{equation}
  S_\mathcal{A}=k S_\mathrm{pix} \frac{z_\mathrm{med}}{z_\mathrm{near}}.
\end{equation}
Here, $S_\mathrm{pix}$ is the pixel's size and $k$ can be viewed as the percentage of the pixel's footprint at the distance $z_\mathrm{med}$. Typically, satisfactory results are obtained when $k$ is in the range $[0.5,1]$. The influence of $k$ on the visual effect is discussed in the next section.

\subsection{Gathering Particles}
Gathering particles within a query cylinder (central ray: $\mbx\to\mby$, radius: $r_\mathrm{c}$ and cross section: $\mathcal{A}$) requires conducting sphere-cylinder intersection test for every particle in the medium. Given a particle with the position $\mbp$ and radius $r_\mathrm{p}$, it is supposed to be inside the query cylinder if the distance from $\mbp$ to the central ray $\mbx\to\mby$ is smaller than $r_\mathrm{p}+r_\mathrm{c}$, as illustrated in Fig. \ref{fig:cylinder} left.

Testing all particles of the medium is notoriously time-consuming. To boost the performance, we accelerate the process of ray traversing the medium using the 3D digital differential analyzer (3D-DDA) \cite{Amanatides87afast,10.1145/1141911.1141913}. Specifically, we construct a uniform grid for the medium and adopt a ray traversal algorithm, similar to that in \cite{Amanatides87afast}, to find the active voxels intersected by the query cylinder. Fig. \ref{fig:cylinder} right illustrates all the active voxels corresponding to the orange query cylinder. Only those particles inside the active voxels will be tested against the query cylinder. We determine the active voxels simply by the ray $\mbx\to\mby$. This introduces negligible bias, because the radius of the cross section is more than two orders of magnitude smaller than the side length of the voxel. This is significantly different to the thick beams used in beam radiance estimation \cite{Jarosz:2008:Beam}. After collecting all the particles inside the query cylinder, we accumulate their contributions to the transmittance $T_\mathcal{A}(\mbx, \mby)$ according to Eq. (\ref{eq:transmittance_graniness}).

To construct a uniform grid, its resolution should be carefully determined. We have observed by experiments that high performance is achieved when roughly one particle resides in each voxel after space subdivision.


\subsection{Computing $Q(\theta)$}
To solve the multi-scale VRE, we also have to compute $Q(\theta)$ at any sample position $\mby$.
$Q(\theta)$ describes the angular distribution of scattering at $\mby$. To quickly compute it, a small query region around $\mby$ should be defined. We set this query region to a small sphere with radius $r_\mathrm{c}$. If this query region contains $N_q>0$ particles, we evaluate $Q(\theta)$ with the following formula:
\begin{equation}\label{eq:qs}
  Q(\theta) = \frac{1}{2|k|^2 \mu(\mathcal{V}_\mby)}\sum_{i=1}^{N_q}(|S_1(\theta, r_i)|^2+|S_2(\theta, r_i)|^2).
\end{equation}


\subsection{Importance Sampling}
Similar to the simplifications used in rendering surface glints \cite{Jakob:2014:DSM:2601097.2601186,10.1111:cgf.13476}, importance sampling is performed according to the global optical properties of the medium, assuming it to be continuous. Specifically, we use the global extinction coefficient for free-flight sampling and use the tabulated global phase function for angular sampling. These global optical properties only need to be determined once in the preprocessing stage, assuming the entire bounding box of the medium to be $\mathcal{V}_\mbx$ in Eq. (\ref{eq:bulk_sigma_t}) and Eq. (\ref{eq:bulk_phase}).


\section{Results}
We have implemented the rendering solution on top of the Mitsuba renderer \cite{Mitsuba}, with spectral rendering enabled. We use 8 spectral samples in the range of the visible spectrum at equally-spaced locations \cite{10.1145/3197517.3201351}. After rendering, we convert the spectral values to the sRGB color space. All synthesized images are created on a PC with an Intel 16-core i7-6900K CPU and 16G RAM.

To compute the bulk optical properties of a discrete participating medium, we need to specify the complex refractive index ($\eta$) for the particles involved and the global PSD ($\bar{r}_g$ and $\sigma_g$). The physical unit for the particle radius is $\mu$m. We also provide an upper bound to the particle size ($r_\mathrm{max}=2000$) to avoid unreasonably large particles which are unusual and are no longer suitable to be treated as participating media. As mentioned previously, we use GOA in the calculation of the scattering amplitude functions when $r\ge2$ and switch to Lorenz-Mie theory otherwise. Except the \textsc{Staircase} scene, the refractive index is set according to the data of ice selected from \cite{Frisvad:2007:CSP:1275808.1276452}, and the host medium is set to be air with $\eta_\mathrm{m}=1$.

\subsection{Comparisons Against Explicit Path Tracing}
We first compare our method with the traditional path tracing. Previous methods simulating the grainy appearance of discrete participating media mostly rely on explicit path tracing (EPT), with potential approximations to simplify the computation of high-order scattering \cite{Moon:2007:RDR:2383847.2383878,Meng:2015:MMR:2809654.2766949,Muller:2016:ERH:2980179.2982429}. However, EPT and other approximations are restricted to geometric optics. This means that only surface reflection and refraction are properly handled. In principle, the mesh of every particle should be explicitly generated and costly ray-object intersections are required.

\begin{figure}[t]
\centering
\rotatebox[origin=lt]{90}{\scriptsize{ \quad\quad \bf Ours \quad\quad\quad\quad\quad\quad \bf EPT}}
\subfigure[$\bar{r}_g=100$ and $\sigma_g=6$]{
\begin{minipage}[b]{0.46\linewidth}
\begin{overpic}[width=1.0\linewidth, trim={0px, 120px, 0px, 120px}, clip]{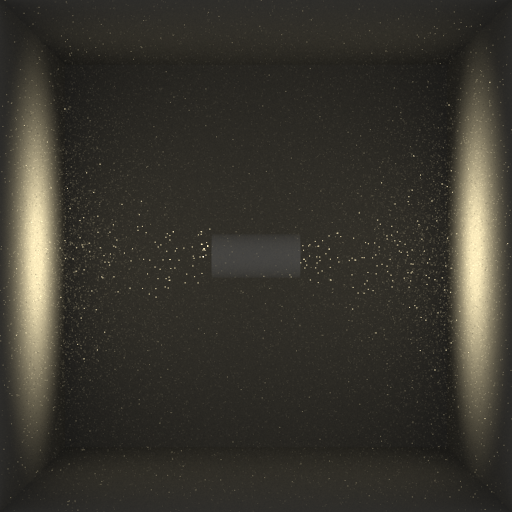}
\end{overpic}
\begin{overpic}[width=1.0\linewidth, trim={0px, 120px, 0px, 120px}, clip]{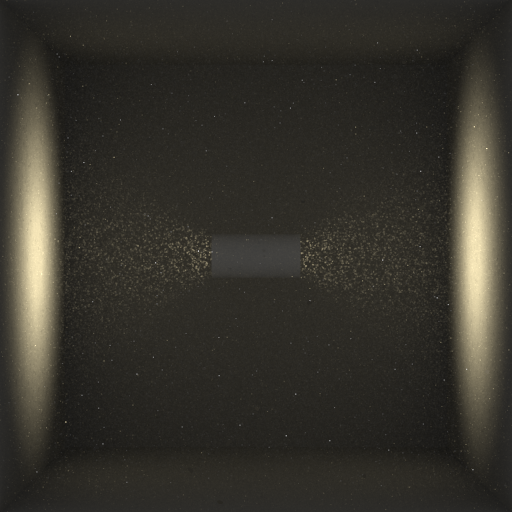}
\end{overpic}
\end{minipage}
}
\hspace{-0.1in}
\subfigure[$\bar{r}_g=600$ and $\sigma_g=3$]{
\begin{minipage}[b]{0.46\linewidth}
\begin{overpic}[width=1.0\linewidth, trim={0px, 120px, 0px, 120px}, clip]{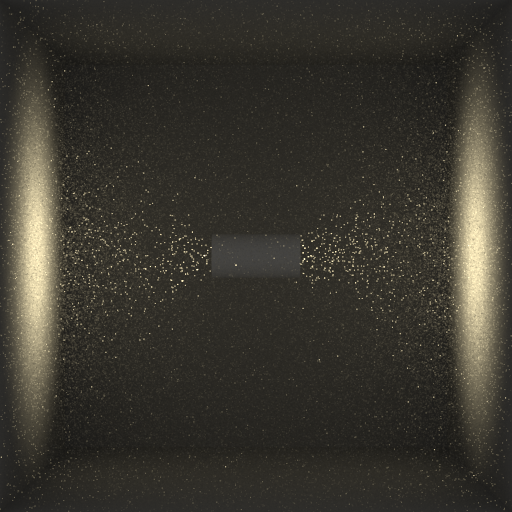}
\end{overpic}
\begin{overpic}[width=1.0\linewidth, trim={0px, 120px, 0px, 120px}, clip]{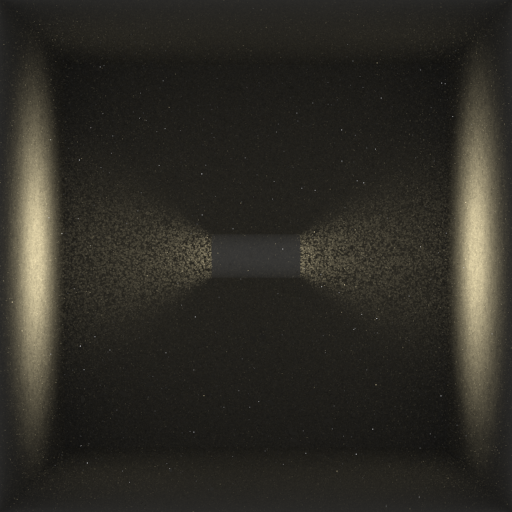}
\end{overpic}
\end{minipage}
}
\caption{\label{fig:cornell_new_ept} Comparisons between EPT and our method in simulating the grainy appearance of discrete participating media with different PSDs ($N_\mathrm{tot}=5\cdot 10^5$).}
\end{figure}

\begin{figure}[t]
\centering
\subfigure[$r=10$ $\mu$m]{
\includegraphics[width=0.32\linewidth]{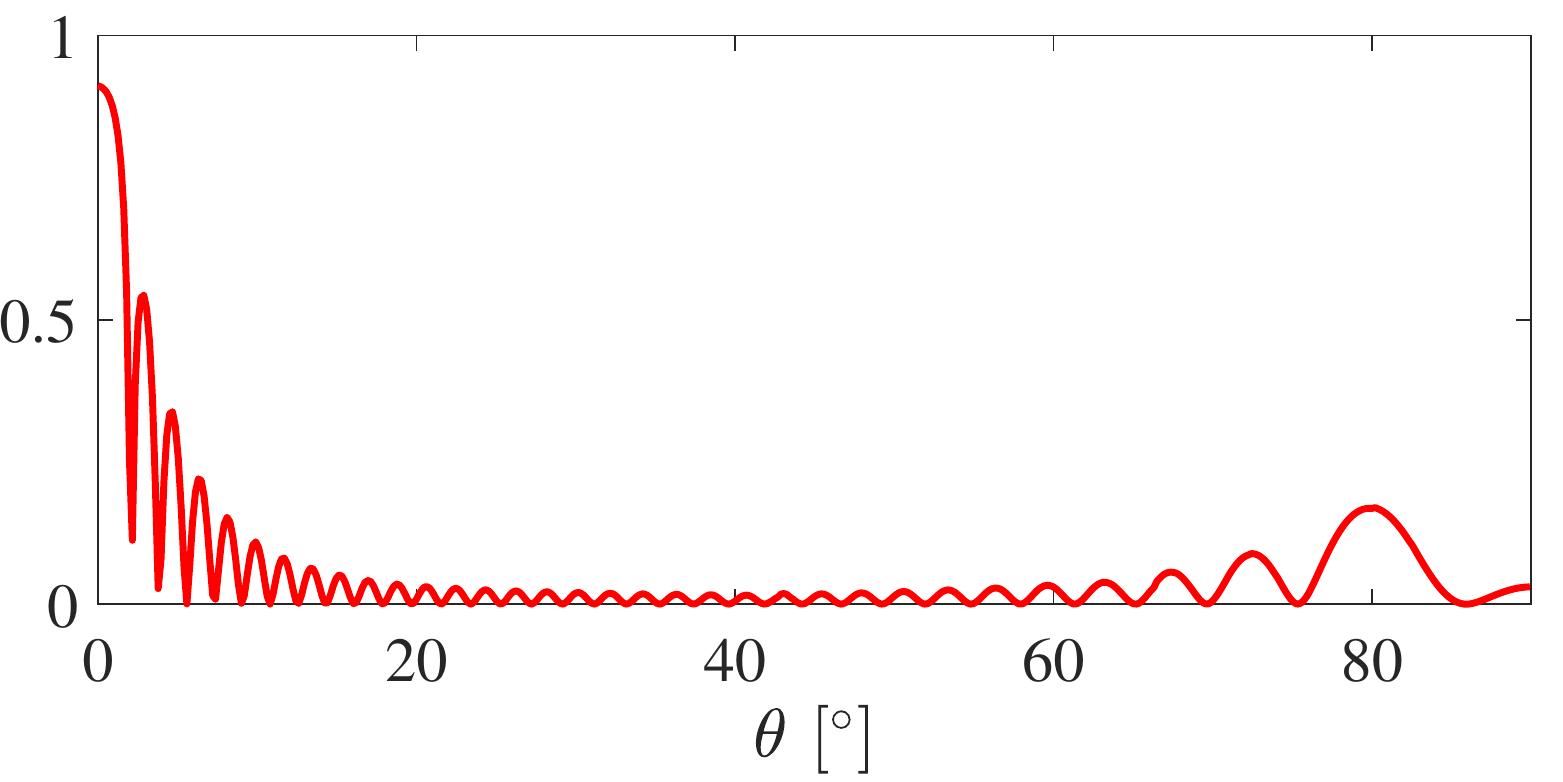}
}
\hspace{-0.12in}
\subfigure[$r=100$ $\mu$m]{
\includegraphics[width=0.32\linewidth]{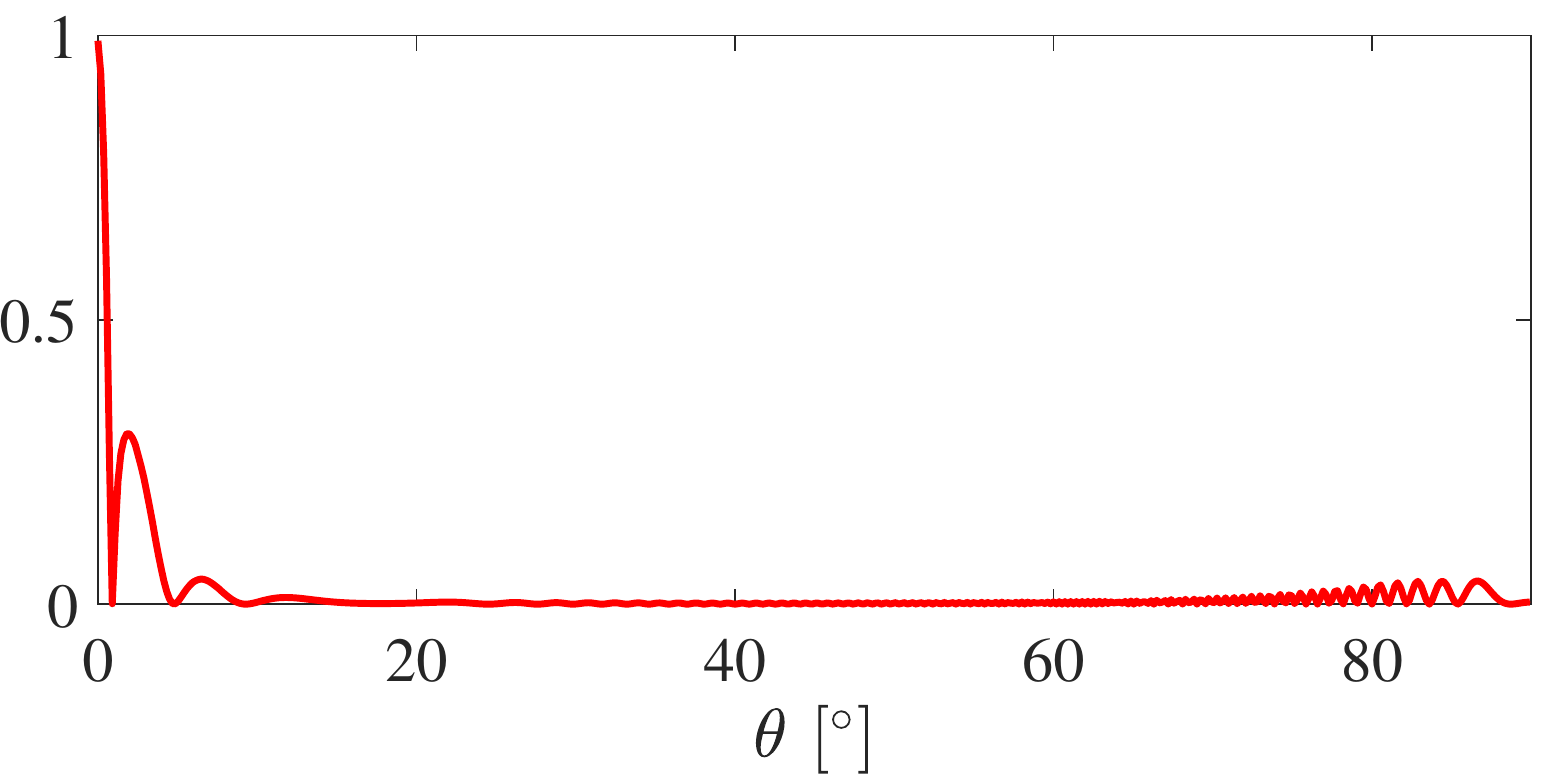}
}
\hspace{-0.12in}
\subfigure[$r=1000$ $\mu$m]{
\includegraphics[width=0.32\linewidth]{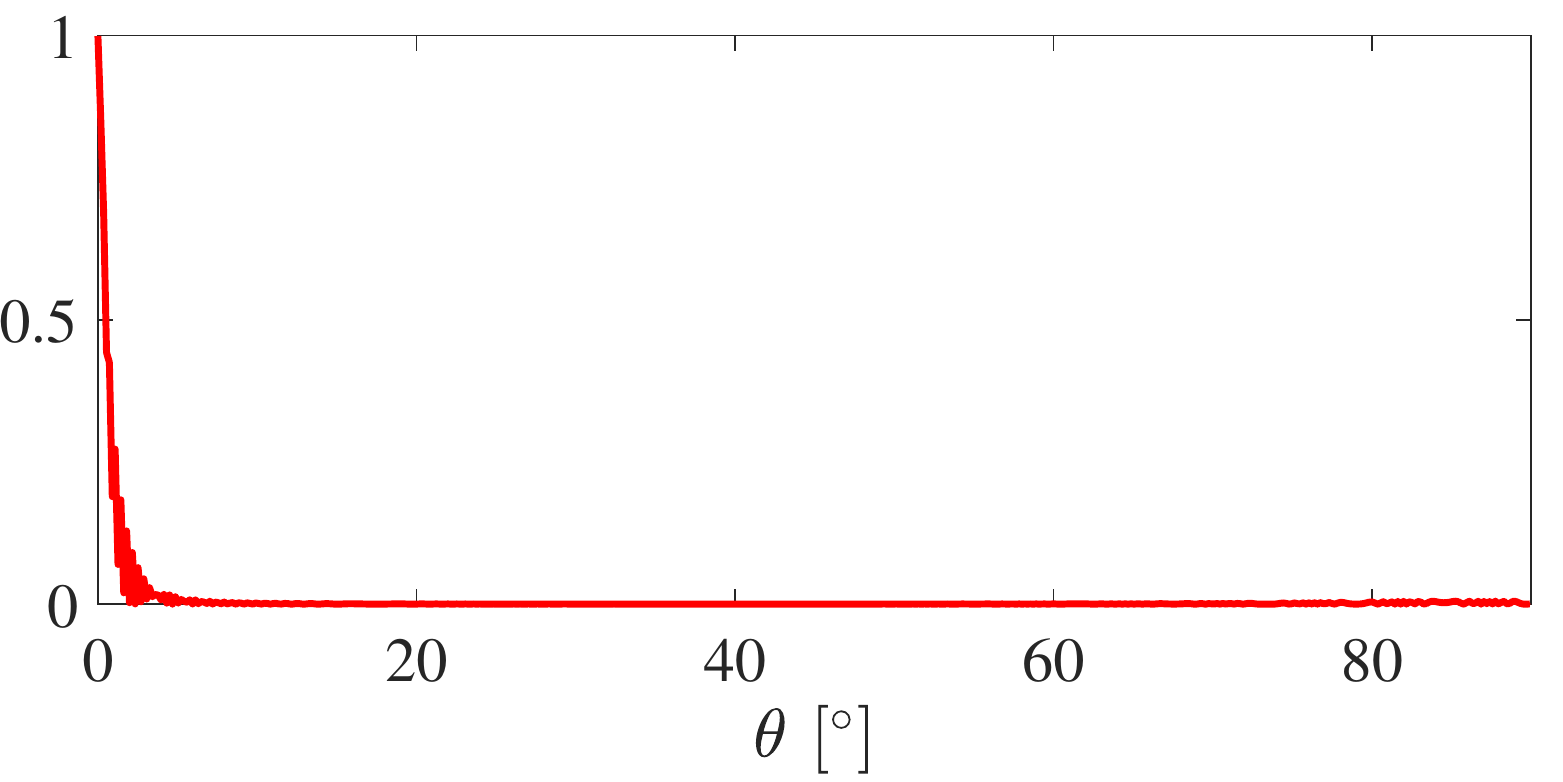}
}
\caption{\label{fig:percentage_sd} Angular percentages of energy contributed by the Fraunhofer diffraction for different sized particles.}
\end{figure}

\begin{figure}[t]
\centering
\setlength{\fboxrule}{1pt}
\setlength{\fboxsep}{0cm}%
\subfigure{
\begin{minipage}[b]{0.32\linewidth}
\begin{overpic}[width=1.0\linewidth]{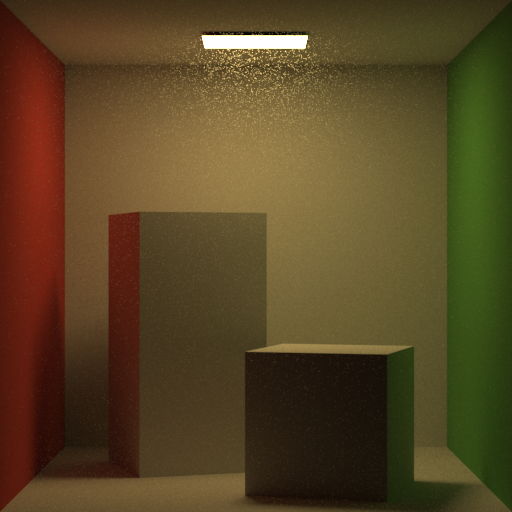}
\put(2,2){\scriptsize {\textcolor[rgb]{1,1,1}{\bf EPT}}}
\end{overpic}
\cfbox{yellow}{\begin{overpic}[width=0.99\linewidth, trim={120px, 374px, 120px, 38px}, clip]{figs/cornell-box/cornell-box-EPT-100000-1000.png}
\end{overpic}}
\end{minipage}
}
\hspace{-0.12in}
\subfigure{
\begin{minipage}[b]{0.32\linewidth}
\begin{overpic}[width=1.0\linewidth]{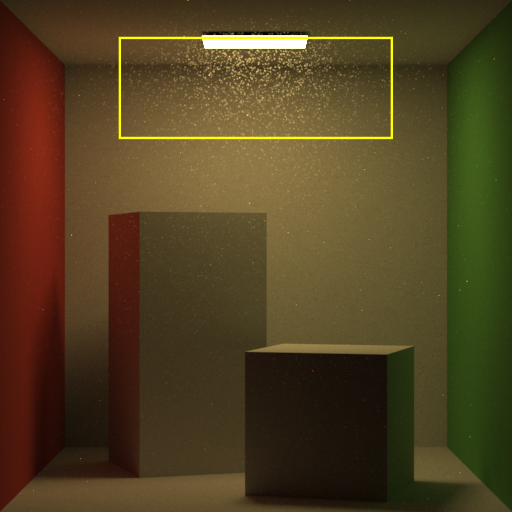}
\put(2,2){\scriptsize {\textcolor[rgb]{1,1,1}{\bf Ours}}}
\end{overpic}
\cfbox{yellow}{\begin{overpic}[width=0.99\linewidth, trim={120px, 374px, 120px, 38px}, clip]{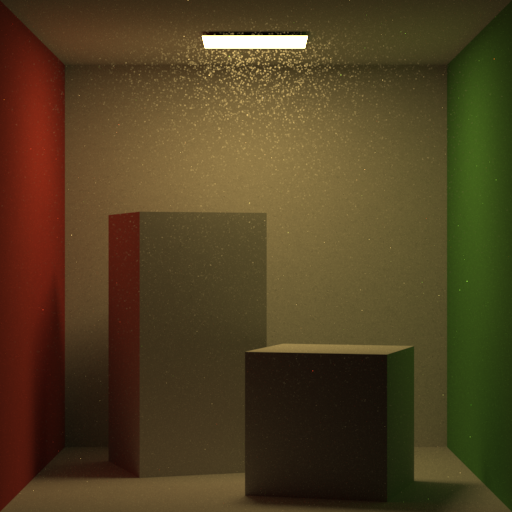}
\end{overpic}}
\end{minipage}
}
\hspace{-0.12in}
\subfigure{
\begin{minipage}[b]{0.32\linewidth}
\begin{overpic}[width=1.0\linewidth]{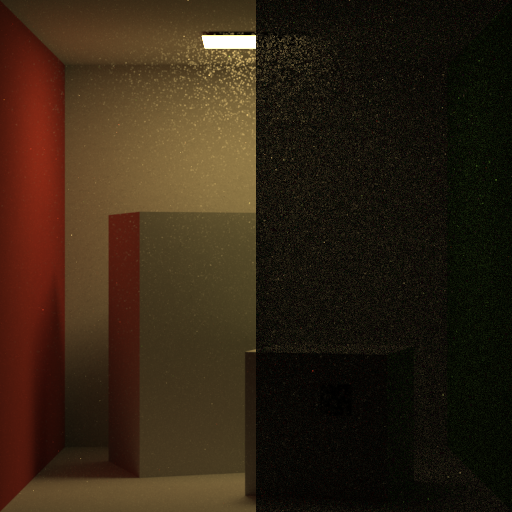}
\put(2,2){\scriptsize {\textcolor[rgb]{1,1,1}{\bf Ours (-FD)\quad\quad~ FD only}}}
\end{overpic}
\cfbox{yellow}{\begin{overpic}[width=0.99\linewidth, trim={120px, 374px, 120px, 38px}, clip]{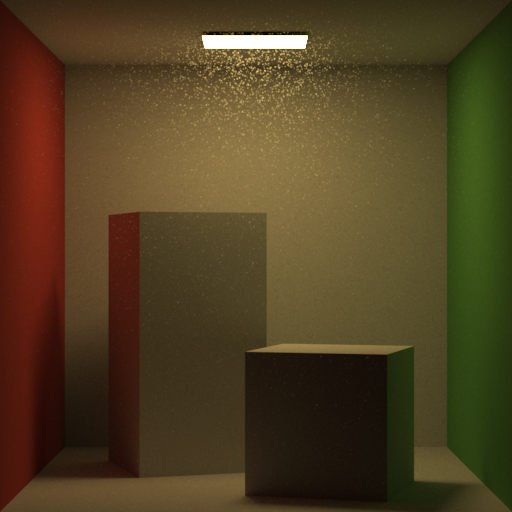}
\end{overpic}}
\end{minipage}
}
\caption{\label{fig:cornell_EPT} Comparisons between EPT and our method with or without the Fraunhofer diffraction (-FD) in rendering $10^5$ large particles of the same size (1000 $\mu$m). The impact of the Fraunhofer diffraction is highlighted in the difference image (FD only).}
\end{figure}

\begin{figure}
\centering
\setlength{\fboxrule}{1pt}
\setlength{\fboxsep}{0cm}%
\begin{minipage}{0.39\linewidth}
\begin{overpic}[width=1.0\linewidth]{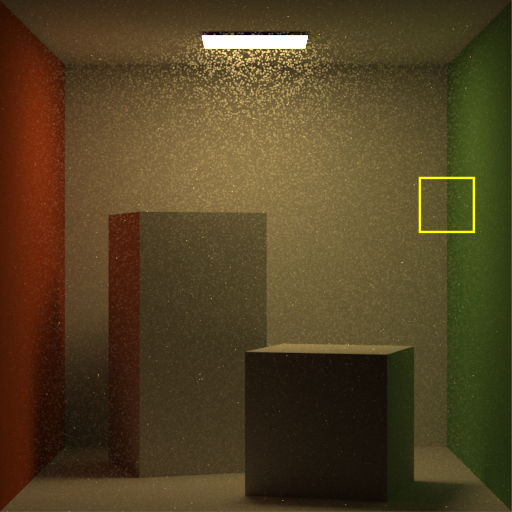}
\put(2,2){\scriptsize {\textcolor[rgb]{1,1,1}{\bf HG}}}
\end{overpic}
\end{minipage}
\hspace{-0.05in}
\begin{minipage}{0.12\linewidth}
\cfbox{yellow}{\begin{overpic}[width=1.0\linewidth, trim={420px, 280px, 38px, 178px}, clip]{figs/cornell-box/cornell-box-EGP-100000-1000.png}
\put(2,2){\scriptsize {\textcolor[rgb]{1,1,1}{\bf Ours}}}
\end{overpic}}
\cfbox{yellow}{\begin{overpic}[width=1.0\linewidth, trim={420px, 280px, 38px, 178px}, clip]{figs/cornell-box/cornell-box-EPT-100000-1000.png}
\put(2,2){\scriptsize {\textcolor[rgb]{1,1,1}{\bf EPT}}}
\end{overpic}}
\cfbox{yellow}{\begin{overpic}[width=1.0\linewidth, trim={420px, 280px, 38px, 178px}, clip]{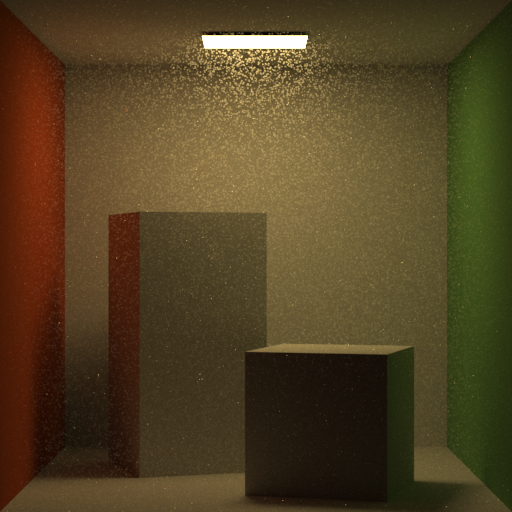}
\put(2,2){\scriptsize {\textcolor[rgb]{1,1,1}{\bf HG}}}
\end{overpic}}
\end{minipage}
\hspace{-0.05in}
\begin{minipage}{0.46\linewidth}
\begin{overpic}[width=1.0\linewidth]{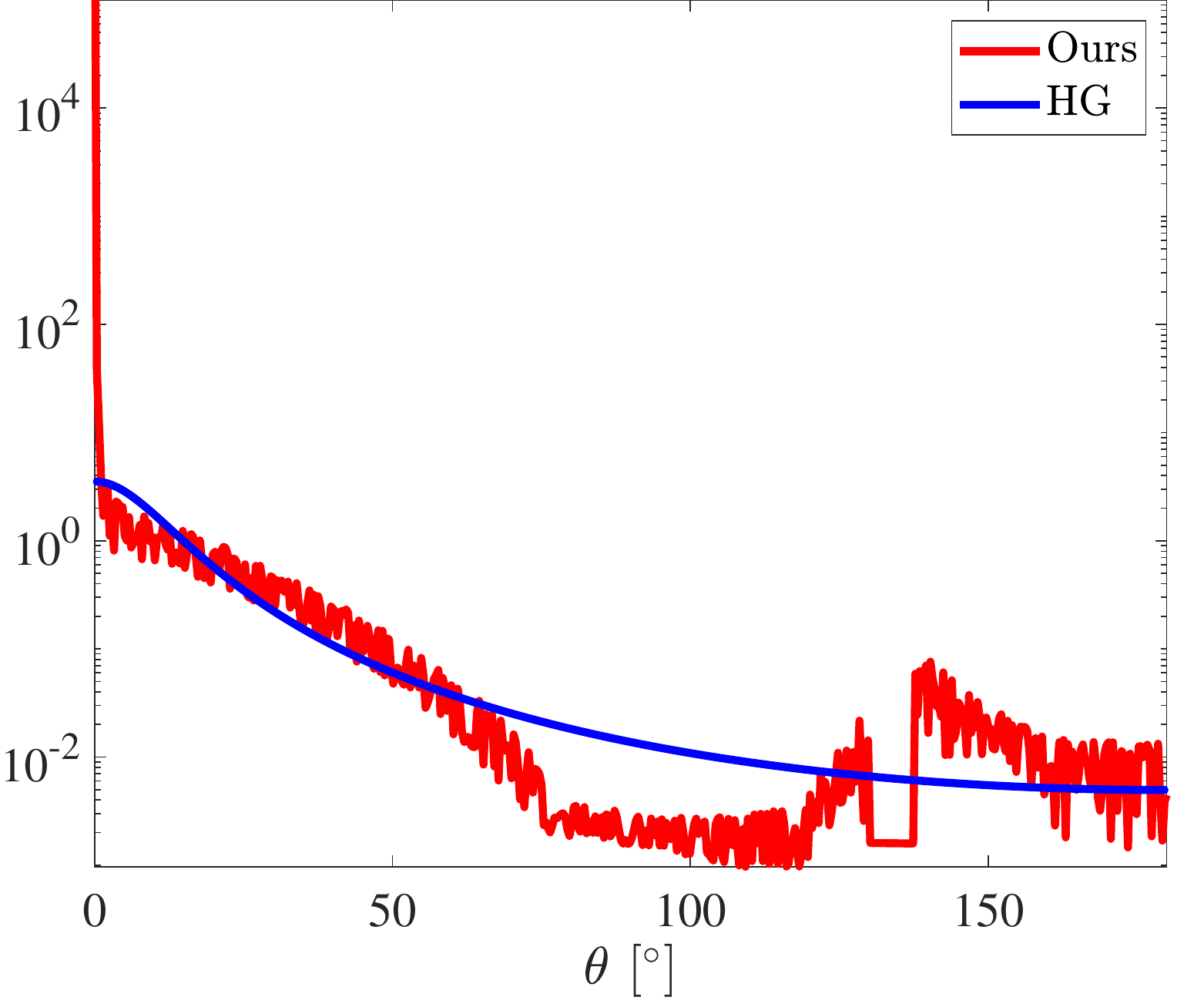}
\end{overpic}
\end{minipage}
\caption{\label{fig:goa_HG} Comparisons between our directional scattering model and the best-fit Henyey and Greenstein (HG) model \cite{HG1941}. Due to obvious discrepancies between our estimated phase function (red curve in the right diagram) and the HG phase function (blue curve in the right diagram), the renderings are slightly differen (see the closeups).}
\end{figure}

\begin{figure*}[t]
\centering
\rotatebox[origin=lt]{90}{\scriptsize{\bf Continuous medium  \quad \bf Discrete medium}}
\subfigure[$r=1000$ $\mu$m, $N_\mathrm{tot}=10^5$]{
\begin{minipage}[b]{0.19\linewidth}
\begin{overpic}[width=1.0\linewidth, trim={0px, 136px, 0px, 0px}, clip]{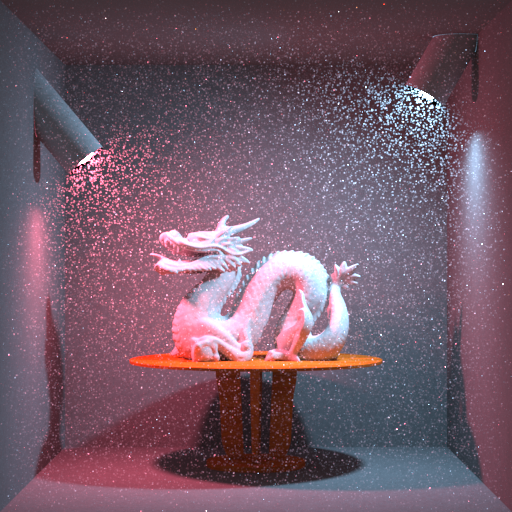}
\end{overpic}
\begin{overpic}[width=1.0\linewidth, trim={0px, 136px, 0px, 0px}, clip]{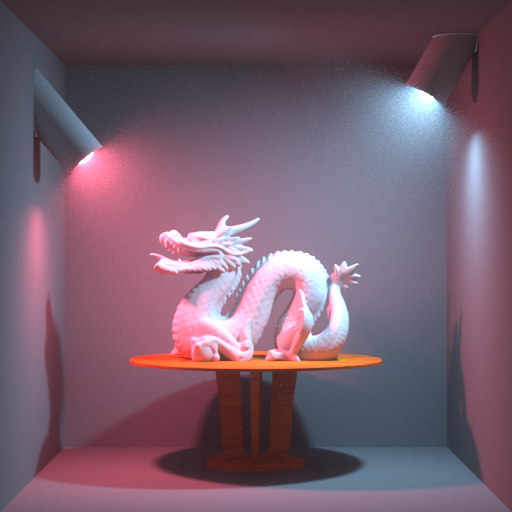}
\end{overpic}
\end{minipage}
}
\hspace{-0.112in}
\subfigure[$r=400$ $\mu$m, $N_\mathrm{tot}=10^6$]{
\begin{minipage}[b]{0.19\linewidth}
\begin{overpic}[width=1.0\linewidth, trim={0px, 136px, 0px, 0px}, clip]{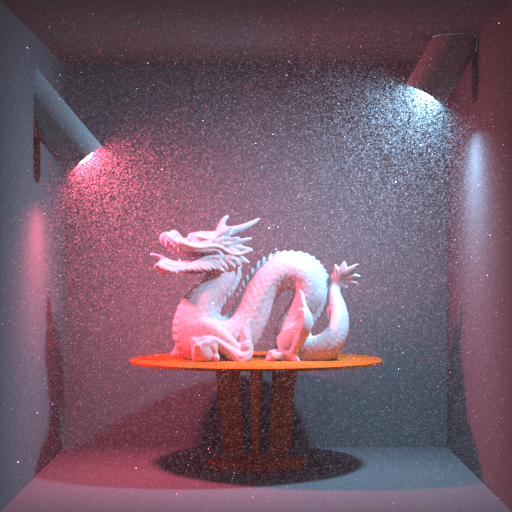}
\end{overpic}
\begin{overpic}[width=1.0\linewidth, trim={0px, 136px, 0px, 0px}, clip]{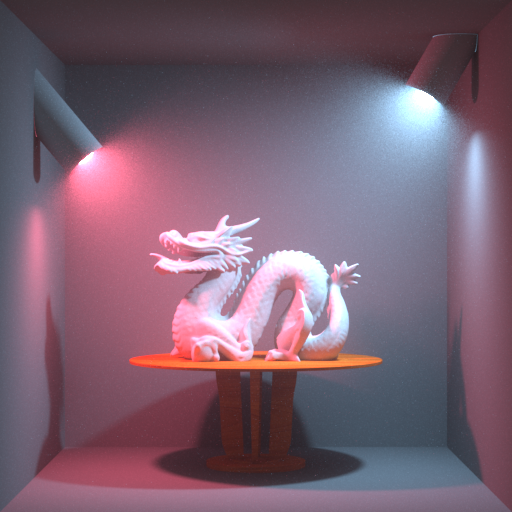}
\end{overpic}
\end{minipage}
}
\hspace{-0.112in}
\subfigure[$r=800$ $\mu$m, $N_\mathrm{tot}=10^6$]{
\begin{minipage}[b]{0.19\linewidth}
\begin{overpic}[width=1.0\linewidth, trim={0px, 136px, 0px, 0px}, clip]{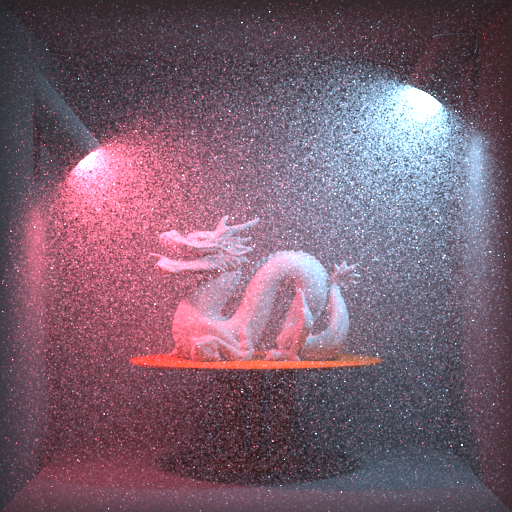}
\end{overpic}
\begin{overpic}[width=1.0\linewidth, trim={0px, 136px, 0px, 0px}, clip]{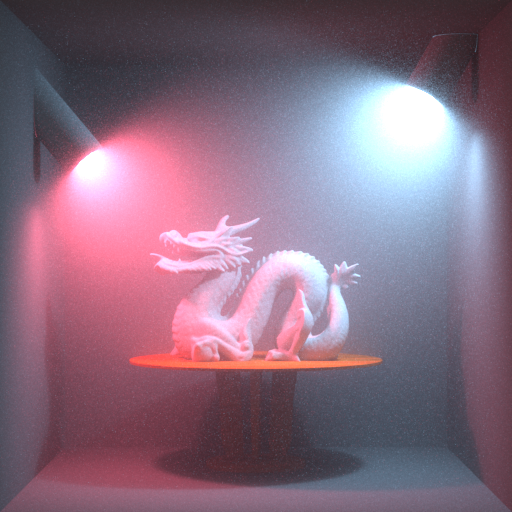}
\end{overpic}
\end{minipage}
}
\hspace{-0.112in}
\subfigure[$r=200$ $\mu$m, $N_\mathrm{tot}=10^7$]{
\begin{minipage}[b]{0.19\linewidth}
\begin{overpic}[width=1.0\linewidth, trim={0px, 136px, 0px, 0px}, clip]{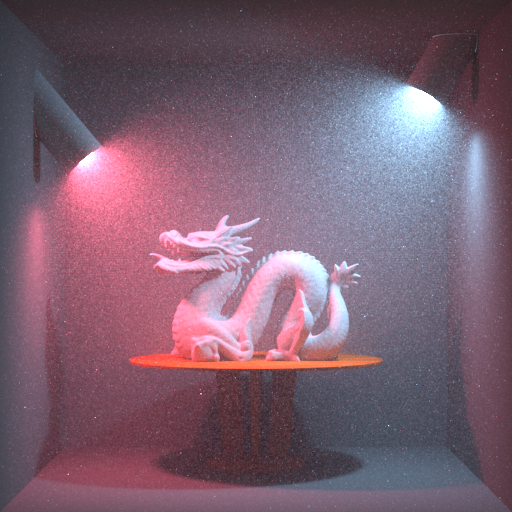}
\end{overpic}
\begin{overpic}[width=1.0\linewidth, trim={0px, 136px, 0px, 0px}, clip]{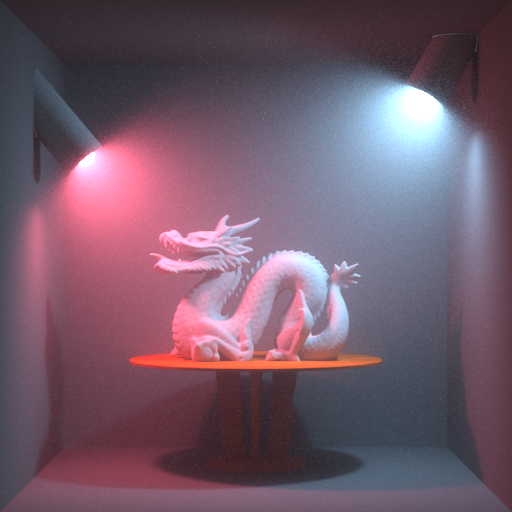}
\end{overpic}
\end{minipage}
}
\hspace{-0.112in}
\subfigure[$r=400$ $\mu$m, $N_\mathrm{tot}=10^7$]{
\begin{minipage}[b]{0.19\linewidth}
\begin{overpic}[width=1.0\linewidth, trim={0px, 136px, 0px, 0px}, clip]{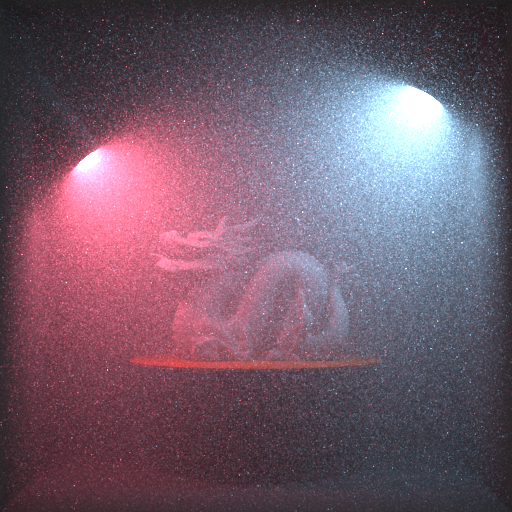}
\end{overpic}
\begin{overpic}[width=1.0\linewidth, trim={0px, 136px, 0px, 0px}, clip]{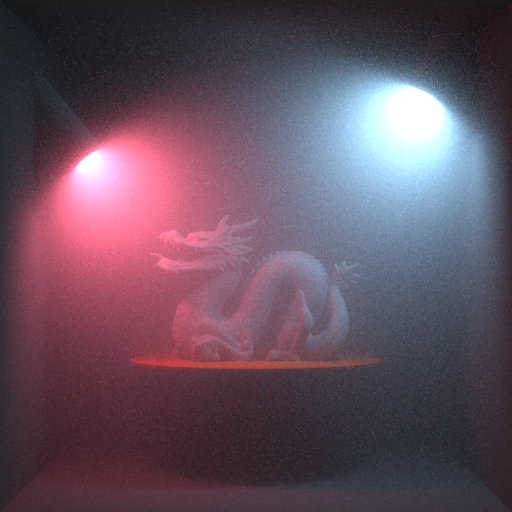}
\end{overpic}
\end{minipage}
}
\caption{\label{fig:lamp} Visual comparisons against continuous media with varying optical properties. As the particle number $N_\mathrm{tot}$ increases, the appearance of the discrete medium exhibits obvious low-frequency haziness and converges to that of a continuous medium.}
\end{figure*}

\begin{figure}
\centering
\subfigure[Discrete medium $N_\mathrm{tot}=10^{8}$]{\label{fig:sponza1}
\begin{minipage}{0.48\linewidth}
\begin{overpic}[width=1.0\linewidth]{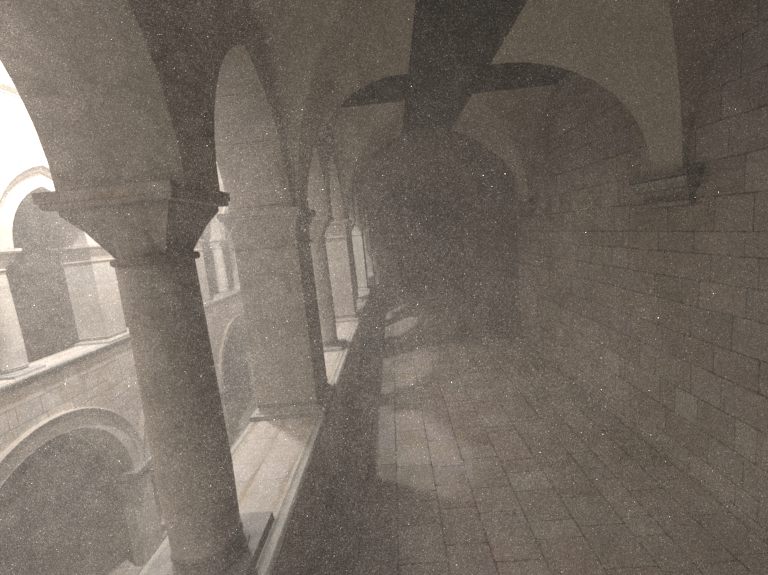}
\end{overpic}
\end{minipage}
}
\hspace{-0.1in}
\subfigure[Continuous medium]{\label{fig:sponza2}
\begin{minipage}{0.48\linewidth}
\begin{overpic}[width=1.0\linewidth]{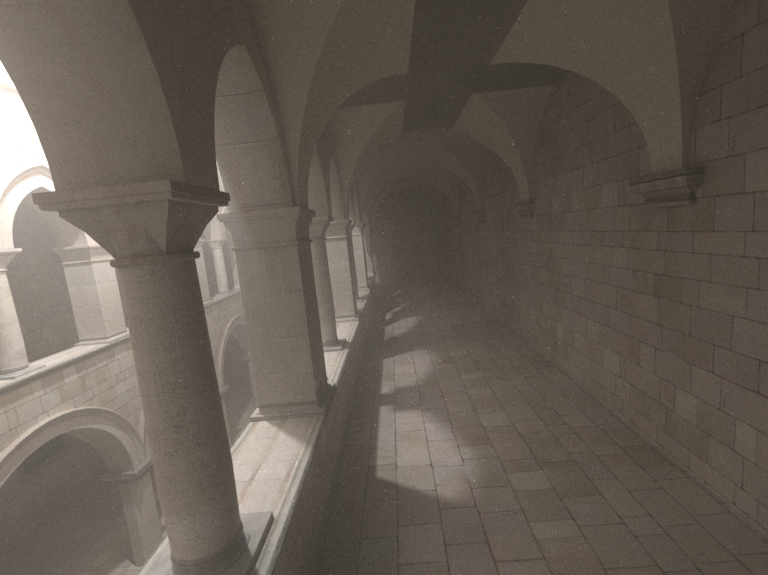}
\end{overpic}
\end{minipage}
}
\caption{\label{fig:sponza} Comparison between our rendering solution and its continuous counterpart on a very dense medium ($r=400$ $\mu$m and $N_\mathrm{tot}=10^{8}$).}
\end{figure}

Compared with EPT, our method offers at least two benefits. First, particle scattering is considered which includes the Fraunhofer diffraction and phase differences, etc. This significantly expands the range of particles that can be handled. Notably, these optical phenomena are quite important in correctly simulating light scattering, especially for very small particles. To verify this, we produce and render $5\cdot 10^5$ spherical particles with different PSDs in Fig. \ref{fig:cornell_new_ept}. These particles have random positions and log-normal sampled radii. For EPT, small transparent balls are instantiated in the scene. Since only surface reflection and refraction are computed for EPT (the top row), the energy inherently belonging to the Fraunhofer diffraction is not correctly captured by EPT, resulting in overly sparse and specular volumetric glints in this \textsc{Lamp} scene. The importance of the Fraunhofer diffraction is visualized in Fig. \ref{fig:percentage_sd}. This figure plots the percentages of energy contributed by the Fraunhofer diffraction at different scattering angles and for different sized particles. Obviously, the Fraunhofer diffraction cannot be ignored especially for small particles. Moreover, EPT easily misses many small particles that are hard to be gathered along an ordinary light path, leading to a slow convergence rate. Also, even substantially increasing the sampling rate or using ``fat ray'' tracing similar to ours, the appearance is still quite sparse. On the contrary, our method (in the bottom row) preserves these energy from the Fraunhofer diffraction and produces smoother appearance that is closer to real scenarios, thanks to the Airy's pattern \footnote{The Fraunhofer diffraction will distort light passing through the particle, making it visually large. The smaller the particle is, the larger the light's distribution is.} caused by the Fraunhofer diffraction and the efficient rendering solution tailored for sparse media.

When only very large particles exist in the medium, our method and EPT tend to produce the similar grainy appearance, as shown in Fig. \ref{fig:cornell_EPT}. Here, we render a discrete medium with $10^5$ particles of the same size (1000 $\mu$m). Since the radius is sufficiently large, pure geometric optics becomes applicable and serves as a valid approximation to the particle scattering. This is further verified in the third column of Fig. \ref{fig:percentage_sd}. We see that the contribution of the Frauhofer diffraction concentrates in a very narrow angle in this case, making it hard to be observed. Consequently, if we remove the Fraunhofer diffraction in our model and leave only reflection and refraction (the third colume of Fig. \ref{fig:cornell_EPT}), we will achieve the grainy appearance similar to that of our full model. However, there are still subtle differences (highlighted in the difference image) contributed by the Fraunhofer diffraction.

The second strength of our rendering solution lies in the runtime performance. For rendering these sparse media in the \textsc{Lamp} scene and the \textsc{Cornell Box} scene in Figs. \ref{fig:cornell_new_ept} and \ref{fig:cornell_EPT}, our method achieves a $1.2\times$ speedup over EPT in the rendering time.

\subsection{Comparisons Against the Henyey and Greenstein model}
In our framework, we derive the phase function from Lorenz-Mie theory and GOA. In computer graphics, it is more common to adopt an empirical model, e.g., the Henyey and Greenstein (HG) model \cite{HG1941}, because of its simplicity and well-defined behavior.
However, the HG model is not very accurate, as pointed out by various literature \cite{Toublanc:96,10.1145/1073204.1073266,10.1145/2508363.2508377}. In Fig. \ref{fig:goa_HG} we compare our estimated phase function with the best-fit HG phase function in rendering the same scene as in Fig. \ref{fig:cornell_EPT}. Since the HG phase function cannot faithfully encode the scattering pattern from these relatively large particles ($r=1000$ $\mu$m), the rendering with the best-fit HG phase function is slightly different from ours, as compared in the insets. Recall that our rendering is close to that generated using EPT in this specific scene with large particles.

\subsection{Comparisons Against Continuous Media}
We also compare the grainy appearance of discrete media simulated by our method with the smooth appearance of continuous media. In Fig. \ref{fig:lamp}, we assume that the particles in the \textsc{Dragon} scene possess the same radius $r$ and are randomly distributed in a cube of the volume $V$. Under this configuration, it is easy to derive the extinction coefficient and the scattering coefficient of the continuous media as $C_\mathrm{t}(r)N_\mathrm{tot}/V$ and $C_\mathrm{s}(r)N_\mathrm{tot}/V$, respectively. The phase function can be computed in a similar way and stored in a table. With these global properties, traditional volumetric path tracing is applicable to render these continuous media. Generally, each discrete medium and its paired continuous medium have the same overall brightness. For the discrete media, when the number of particles $N_\mathrm{tot}$ is small, we can clearly observe individual particles lit by the lamps. As $N_\mathrm{tot}$ increases, the appearance tends to become hazy, and the rendering result gets closer to the corresponding continuous medium. When $N_\mathrm{tot}$ is sufficiently large (e.g., $N_\mathrm{tot}=10^7$), the discrete medium and its continuous counterpart will achieve quite similar appearance. In Fig. \ref{fig:sponza}, a much denser medium ($N_\mathrm{tot}=10^{8}$) is rendered by our method, which again produces smooth appearance matching that from a continuous medium. These pair-wise comparisons demonstrate that our rendering solution converges to the traditional volumetric rendering of continuous media in the limit of particle concentration.

\begin{figure}[t]
\centering
\rotatebox[origin=lt]{90}{\scriptsize{ $N_\mathrm{tot}=10^5$  \quad\quad\quad $N_\mathrm{tot}=10^4$}}
\subfigure[$k=0.2$]{
\begin{minipage}[b]{0.22\linewidth}
\begin{overpic}[width=1.0\linewidth]{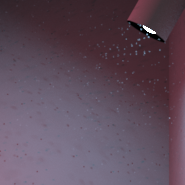}
\end{overpic}
\begin{overpic}[width=1.0\linewidth]{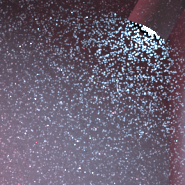}
\end{overpic}
\end{minipage}
}
\hspace{-0.112in}
\subfigure[$k=0.5$]{
\begin{minipage}[b]{0.22\linewidth}
\begin{overpic}[width=1.0\linewidth]{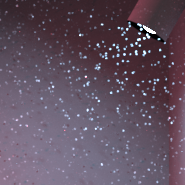}
\end{overpic}
\begin{overpic}[width=1.0\linewidth]{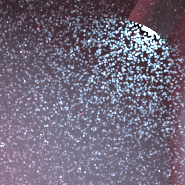}
\end{overpic}
\end{minipage}
}
\hspace{-0.112in}
\subfigure[$k=2.0$]{
\begin{minipage}[b]{0.22\linewidth}
\begin{overpic}[width=1.0\linewidth]{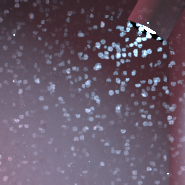}
\end{overpic}
\begin{overpic}[width=1.0\linewidth]{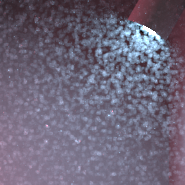}
\end{overpic}
\end{minipage}
}
\hspace{-0.112in}
\subfigure[$k=4.0$]{
\begin{minipage}[b]{0.22\linewidth}
\begin{overpic}[width=1.0\linewidth]{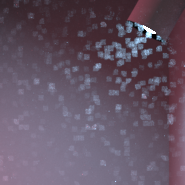}
\end{overpic}
\begin{overpic}[width=1.0\linewidth]{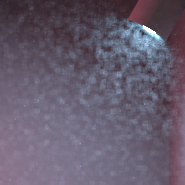}
\end{overpic}
\end{minipage}
}
\caption{\label{fig:lamp2_size} Impact of the query cylinder's cross section on the simulated grainy appearance. Here, $k$ is used in Eq. (\ref{eq:qs}) which determines the percentage of the pixel's footprint at the nearest boundary of the medium.}
\end{figure}

\subsection{Choice of Query Cylinder's Cross Section}
We determine the query cylinder's radius $r_\mathrm{c}$ according to Eq. (\ref{eq:qs}) in which the parameter $k$ plays an important role. We suggest to choose its value in the range $[0.5,1]$ which yields reasonable grainy appearance as show in the second row of Fig. \ref{fig:lamp2_size}. Values in this range allow us to faithfully capture almost all grains in the scenes with little bias.
Generally, a too large $k$ will produce uncomfortable aliasing as shown in the last two rows of Fig. \ref{fig:lamp2_size}. In these two cases, although the overall brightness is similar to that of $k=0.5$, the volumetric glints are overly blurred. On the other hand, a too small $k$ will miss many particles during query and is therefore inefficient, especially when the medium is very sparse, e.g., $N_\mathrm{tot}=10^4$. However, $k$ with a value smaller than 0.5 is also acceptable sometimes. For instance, the first image in the bottom row of Eq. \ref{fig:lamp2_size} is rendered with $k=0.2$ which achieves the similar effect with that of $k=0.5$ for this relatively dense medium ($N_\mathrm{tot}=10^5$).

\begin{figure}[t]
\centering
\subfigure{
\begin{overpic}[width=0.52\linewidth]{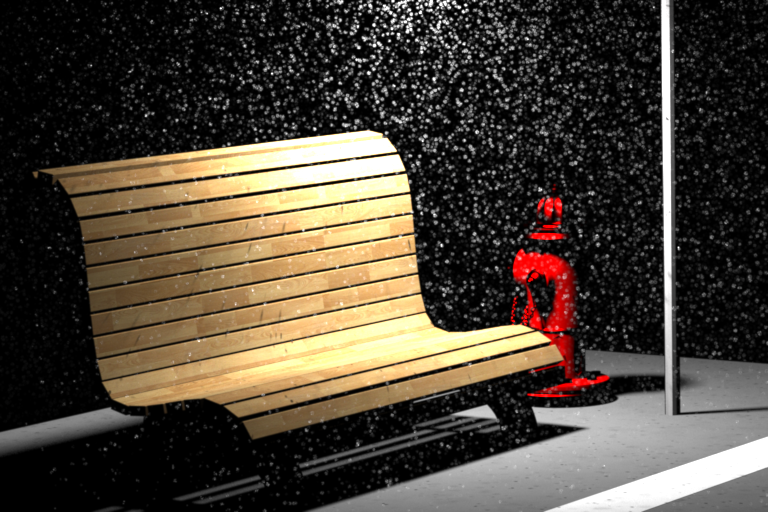}
\end{overpic}
}
\hspace{-0.1in}
\subfigure{
\begin{overpic}[width=0.43\linewidth]{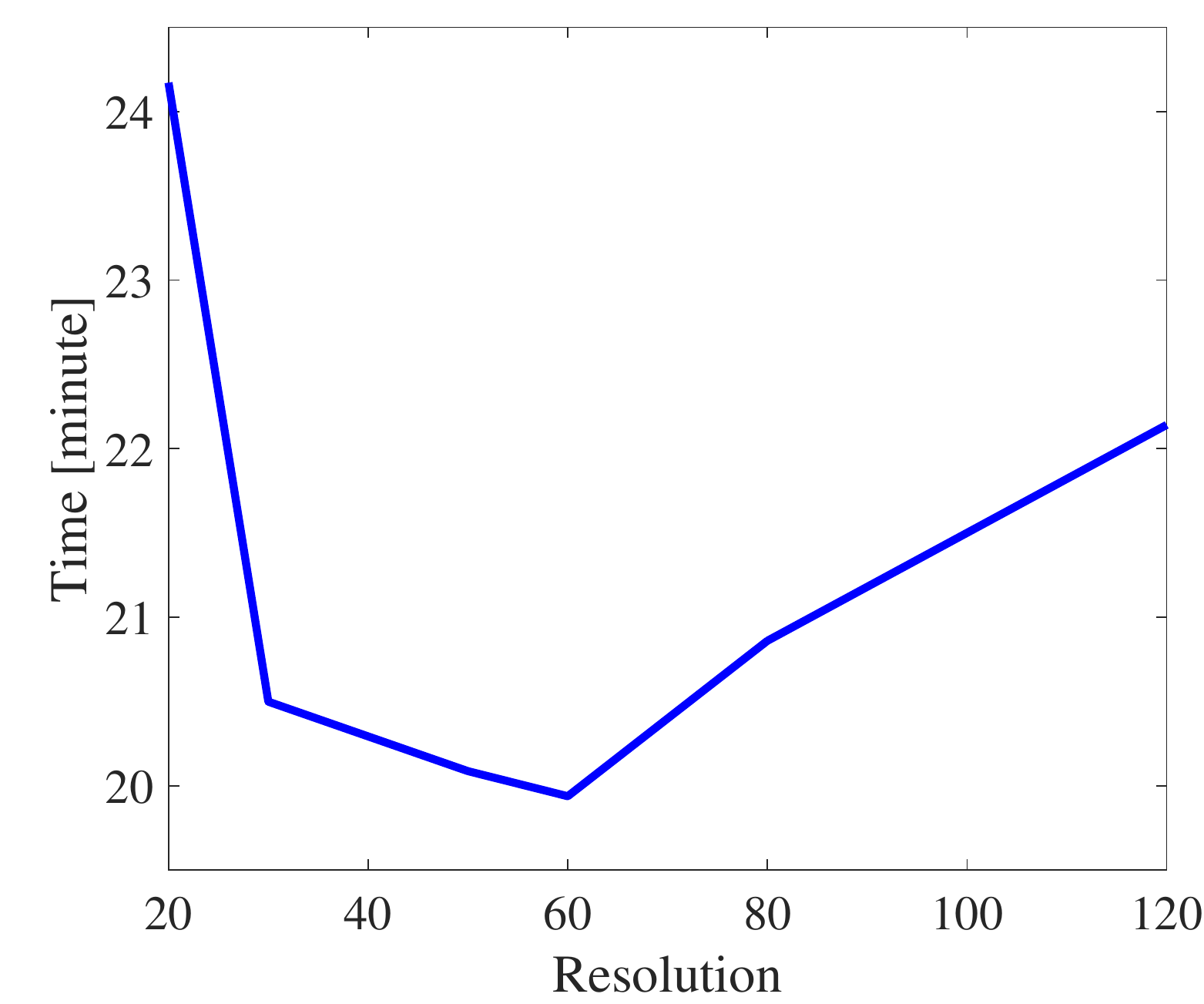}
\end{overpic}
}
\caption{\label{fig:road_res} Evolution of the rendering time (at 2048 spp) over the grid resolution. This \textsc{Bench} scene contains $N_\mathrm{tot}=2\cdot 10^5$ particles ($\bar{r}_g=800$ and $\sigma_g=4$) and achieves the lowest rendering time at a resolution of $60\times60\times60$ which is roughly one particle per voxel.}
\end{figure}

\subsection{Choice of Grid Resolution}
We employ a uniform grid to accelerate the ray traversal process, considering that particles are uniformly distributed in the scene. As shown in the right diagram of Fig. \ref{fig:road_res}, the grid resolution ($Res$) will influence the performance. Although no analytical analysis can be referred to select the best resolution, we empirically observe that a grid resolution yielding roughly one particle per voxel achieves the optimum solution for most scenes. The \textsc{Bench} scene in the left panel of Fig. \ref{fig:road_res} contains $2\cdot 10^5$ particles and achieves the best performance at a resolution of $60\times60\times60$. As the resolution increases, the runtime grows steadily due to the additional cost introduced by the grid. However, a resolution much lower than $60\times60\times60$ will have a very poor performance since too many particles reside in each voxel. For other scenes, a similar conclusion can be drawn.

\begin{figure*}[t]
\centering
\setlength{\fboxrule}{1pt}
\setlength{\fboxsep}{0cm}%
\subfigure{
\begin{minipage}[b]{0.19\linewidth}
\begin{overpic}[width=1.0\linewidth, trim={150px, 0px, 50px, 0px}, clip]{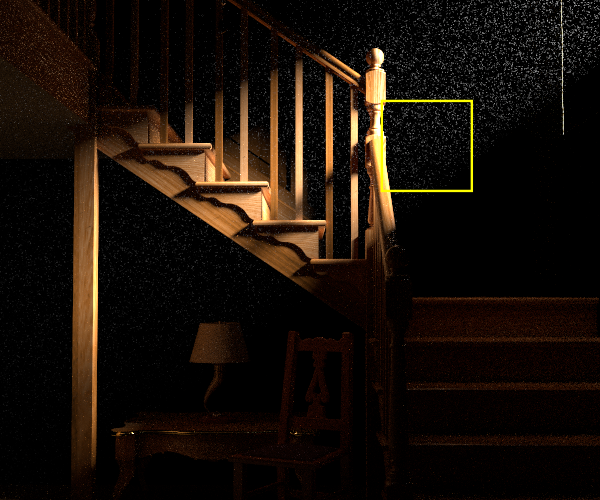}
\put(2,2){\scriptsize {\textcolor[rgb]{1,1,1}{\bf $\eta_\mathrm{i}=0$}}}
\put(31,0){\cfbox{yellow}{\includegraphics[width=0.6\linewidth, trim={382px, 308px, 128px, 102px}, clip]{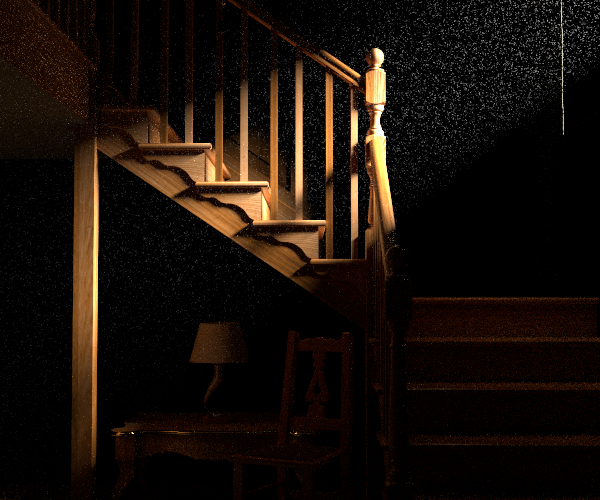}}}
\end{overpic}
\end{minipage}
}
\hspace{-0.1in}
\subfigure{
\begin{minipage}[b]{0.19\linewidth}
\begin{overpic}[width=1.0\linewidth, trim={150px, 0px, 50px, 0px}, clip]{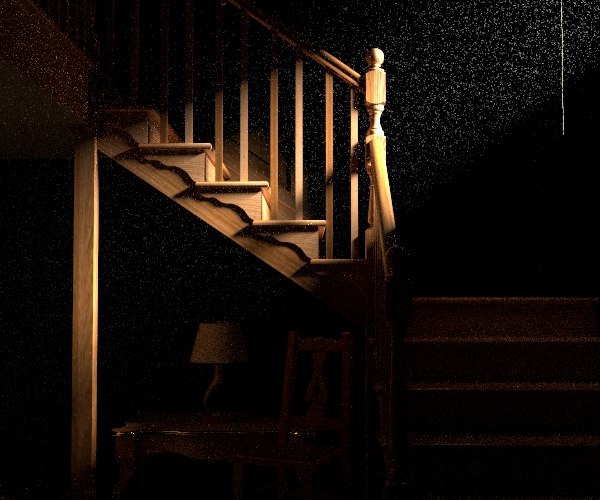}
\put(2,2){\scriptsize {\textcolor[rgb]{1,1,1}{\bf $\eta_\mathrm{i}=2\cdot10^{-5}$}}}
\put(31,0){\cfbox{yellow}{\includegraphics[width=0.6\linewidth, trim={382px, 308px, 128px, 102px}, clip]{figs/stair/staircase_eta1.png}}}
\end{overpic}
\end{minipage}
}
\hspace{-0.1in}
\subfigure{
\begin{minipage}[b]{0.19\linewidth}
\begin{overpic}[width=1.0\linewidth, trim={150px, 0px, 50px, 0px}, clip]{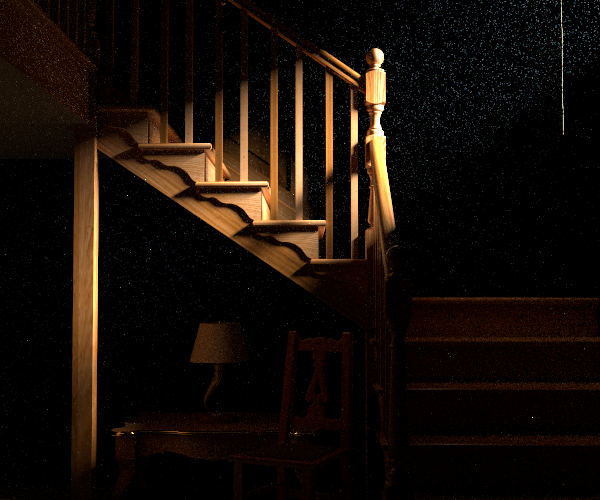}
\put(2,2){\scriptsize {\textcolor[rgb]{1,1,1}{\bf $\eta_\mathrm{i}=4\cdot10^{-5}$}}}
\put(31,0){\cfbox{yellow}{\includegraphics[width=0.6\linewidth, trim={382px, 308px, 128px, 102px}, clip]{figs/stair/staircase_eta2.png}}}
\end{overpic}
\end{minipage}
}
\hspace{-0.1in}
\subfigure{
\begin{minipage}[b]{0.19\linewidth}
\begin{overpic}[width=1.0\linewidth, trim={150px, 0px, 50px, 0px}, clip]{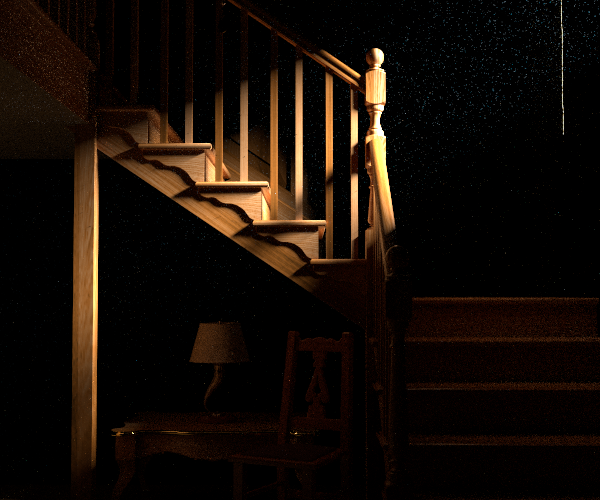}
\put(2,2){\scriptsize {\textcolor[rgb]{1,1,1}{\bf $\eta_\mathrm{i}=6\cdot10^{-5}$}}}
\put(31,0){\cfbox{yellow}{\includegraphics[width=0.6\linewidth, trim={382px, 308px, 128px, 102px}, clip]{figs/stair/staircase_eta3.png}}}
\end{overpic}
\end{minipage}
}
\hspace{-0.1in}
\subfigure{
\begin{minipage}[b]{0.19\linewidth}
\begin{overpic}[width=1.0\linewidth, trim={150px, 0px, 50px, 0px}, clip]{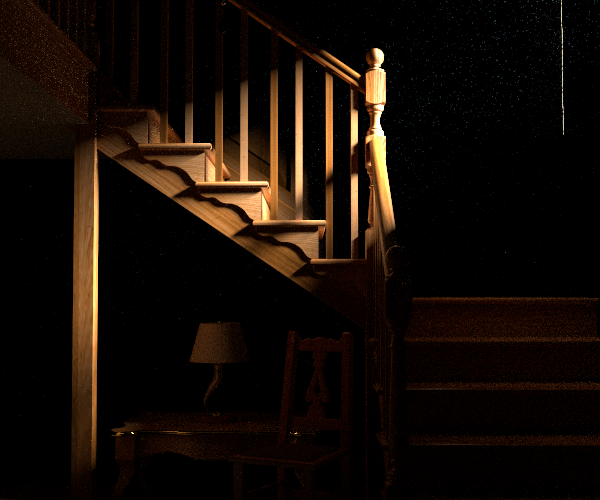}
\put(2,2){\scriptsize {\textcolor[rgb]{1,1,1}{\bf $\eta_\mathrm{i}=8\cdot10^{-5}$}}}
\put(31,0){\cfbox{yellow}{\includegraphics[width=0.6\linewidth, trim={382px, 308px, 128px, 102px}, clip]{figs/stair/staircase_eta4.png}}}
\end{overpic}
\end{minipage}
}
\caption{\label{fig:stair_eta} Impact of the imaginary part of the refractive index ($\eta_\mathrm{i}$) on the \textsc{Staircase} scene. The real part is set to 1.5. Other scene parameters are $N_\mathrm{tot}=3\cdot10^5$, $\bar{r}_g=800$ and $\sigma_g=4$.}
\end{figure*}

\subsection{Impact of the Refractive Index}
The proposed rendering solution can be easily generalized to support absorbing particles. The absorption cross section of each particle is computed by Eq. (\ref{eq:ca}) which varies linearly with the imaginary part of the complex refractive index, i.e., $\eta_\mathrm{i}$. The extinction cross section is slightly modified according to the formulas in Appendix C. The results of changing $\eta_\mathrm{i}$ are shown in Fig. \ref{fig:stair_eta}. This \textsc{Staircase} scene aims to simulate flying dusts in a dirty room lit by a local area light through the window. As expected, an increase of $\eta_\mathrm{i}$ will increasingly dim the intensity of scattering.


\subsection{Impact of the Global PSD}
\begin{figure}[t]
\centering
\subfigure{
\begin{minipage}[b]{0.485\linewidth}
\begin{overpic}[width=1.0\linewidth, trim={0px, 0px, 0px, 0px}, clip]{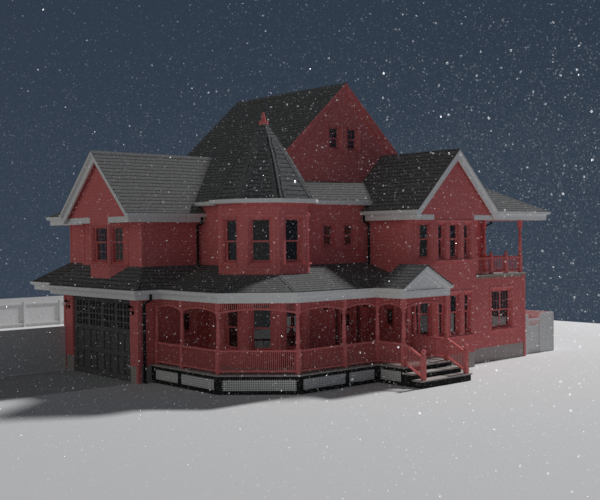}
\put(2,4){\scriptsize {\textcolor[rgb]{0,0,0}{\bf $\bar{r}_g=50$, $\sigma_g=2$}}}
\end{overpic}
\begin{overpic}[width=1.0\linewidth, trim={0px, 0px, 0px, 0px}, clip]{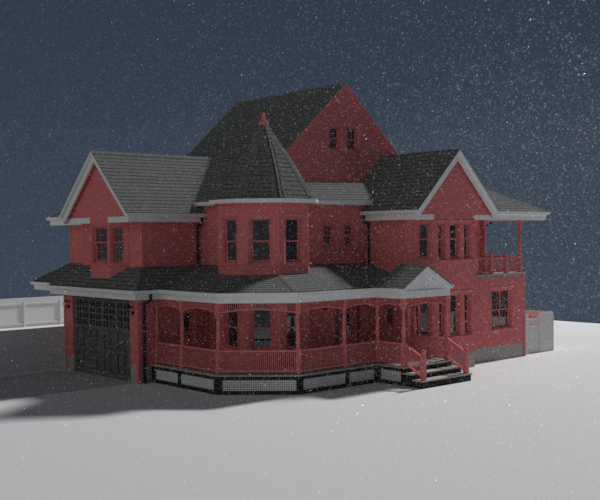}
\put(2,4){\scriptsize {\textcolor[rgb]{0,0,0}{\bf $\bar{r}_g=100$, $\sigma_g=1$}}}
\end{overpic}
\begin{overpic}[width=1.0\linewidth, trim={0px, 0px, 0px, 0px}, clip]{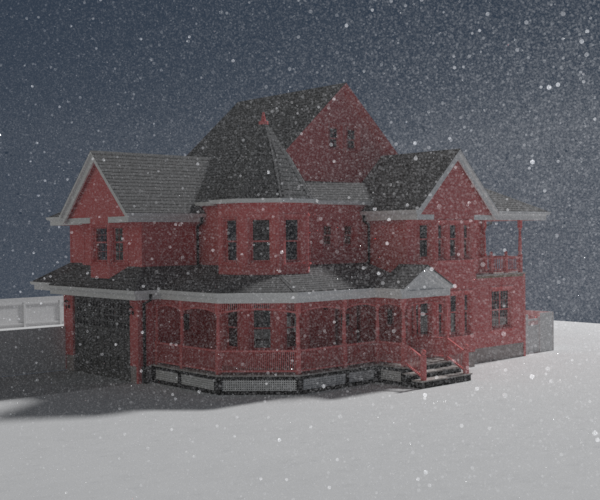}
\put(2,10){\scriptsize {\textcolor[rgb]{0,0,0}{\bf $\bar{r}_{g,1}=100$, $\sigma_{g,1}=1$}}}
\put(2,4){\scriptsize {\textcolor[rgb]{0,0,0}{\bf $\bar{r}_{g,2}=1000$, $\sigma_{g,2}=1.2$}}}
\end{overpic}
\end{minipage}
}
\hspace{-0.12in}
\subfigure{
\begin{minipage}[b]{0.485\linewidth}
\begin{overpic}[width=1.0\linewidth, trim={0px, 0px, 0px, 0px}, clip]{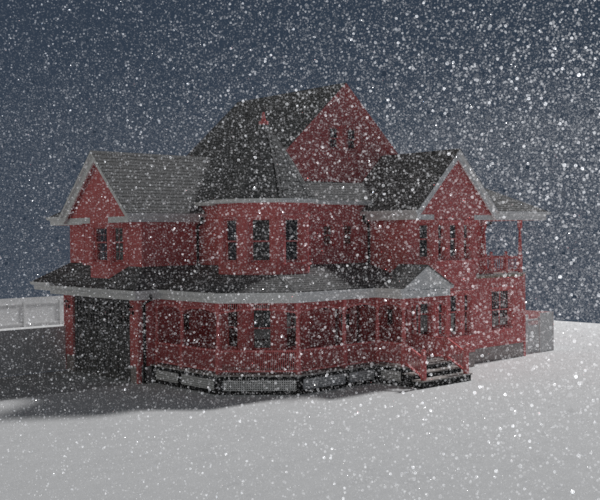}
\put(2,4){\scriptsize {\textcolor[rgb]{0,0,0}{\bf $\bar{r}_g=200$, $\sigma_g=2$}}}
\end{overpic}
\begin{overpic}[width=1.0\linewidth, trim={0px, 0px, 0px, 0px}, clip]{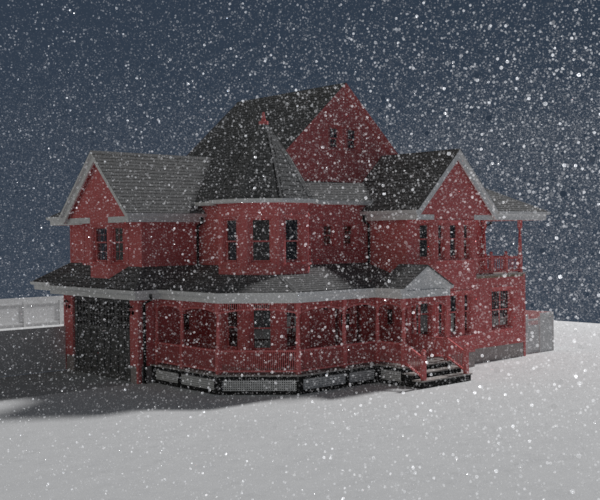}
\put(2,4){\scriptsize {\textcolor[rgb]{0,0,0}{\bf $\bar{r}_g=100$, $\sigma_g=3$}}}
\end{overpic}
\begin{overpic}[width=1.0\linewidth, trim={0px, 0px, 0px, 0px}, clip]{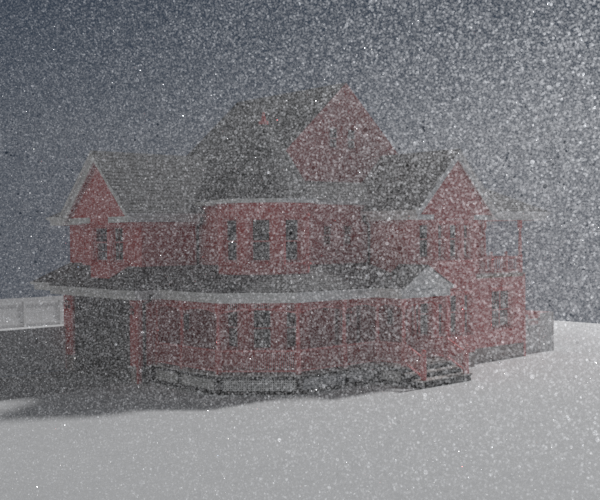}
\put(2,10){\scriptsize {\textcolor[rgb]{0,0,0}{\bf $\bar{r}_{g,1}=200$, $\sigma_{g,1}=1$}}}
\put(2,4){\scriptsize {\textcolor[rgb]{0,0,0}{\bf $\bar{r}_{g,2}=1000$, $\sigma_{g,2}=1.2$}}}
\end{overpic}
\end{minipage}
}
\caption{\label{fig:house_psd} Impact of the global PSD. First row: Changing $\bar{r}_g$ while fixing $\sigma_g$ to 2 and $N_\mathrm{tot}$ to $4\cdot10^6$. Second row: Changing $\sigma_g$ while fixing $\bar{r}_g$ to 100 and $N_\mathrm{tot}$ to $4\cdot10^6$. Third row: Changing $\bar{r}_{g,1}$ in the first mode of the bimodal log-normal distribution while fixing others ($N_\mathrm{tot}=2\cdot10^7$).}
\end{figure}

Fig. \ref{fig:house_psd} analyzes the impact of the global PSD on the appearance of discrete participating media. Here, we generate $N_\mathrm{tot}=4\cdot10^6$ particles with different global PSDs in the \textsc{House} scene. This scene is designed to simulate the appearance of blowing snows in the sky. The first row shows the impact of the geometric mean radius $\bar{r}_g$. The general trend is that the scattering effects become increasingly prominent as $\bar{r}_g$ grows, since the scattering coefficient is positively correlated with the particle radius. Concerning the geometric standard deviation $\sigma_g$, it is responsible for the level of graininess as shown in the second row of Fig. \ref{fig:house_psd}. A small $\sigma_g$ tends to generate smoother appearance than a large one. This is to be expected since a large $\sigma_g$ means local PSDs changing widely, leading to stronger graininess.

In fact, the PSD is not limited to the log-normal distribution. Other distributions also work. For instance, in the third row of Fig. \ref{fig:house_psd}, we show a cloud of blowing snows following a bimodal log-normal distribution. We generate $1.99\cdot10^7$ small particle with $\bar{r}_{g,1}=100$ (or $\bar{r}_{g,1}=200$) and $\sigma_{g,1}=1$, and also generate $10^5$ large particles with $\bar{r}_{g,2}=1000$, $\sigma_{g,2}=1.2$, leading to $2\cdot10^7$ particles in total. Following this complex distribution, we can observe both haziness from massive small particles and graininess from a handful of large particles.

\subsection{Performance Analysis}
\begin{figure}[t]
  \centering
  \includegraphics[width=0.7\linewidth]{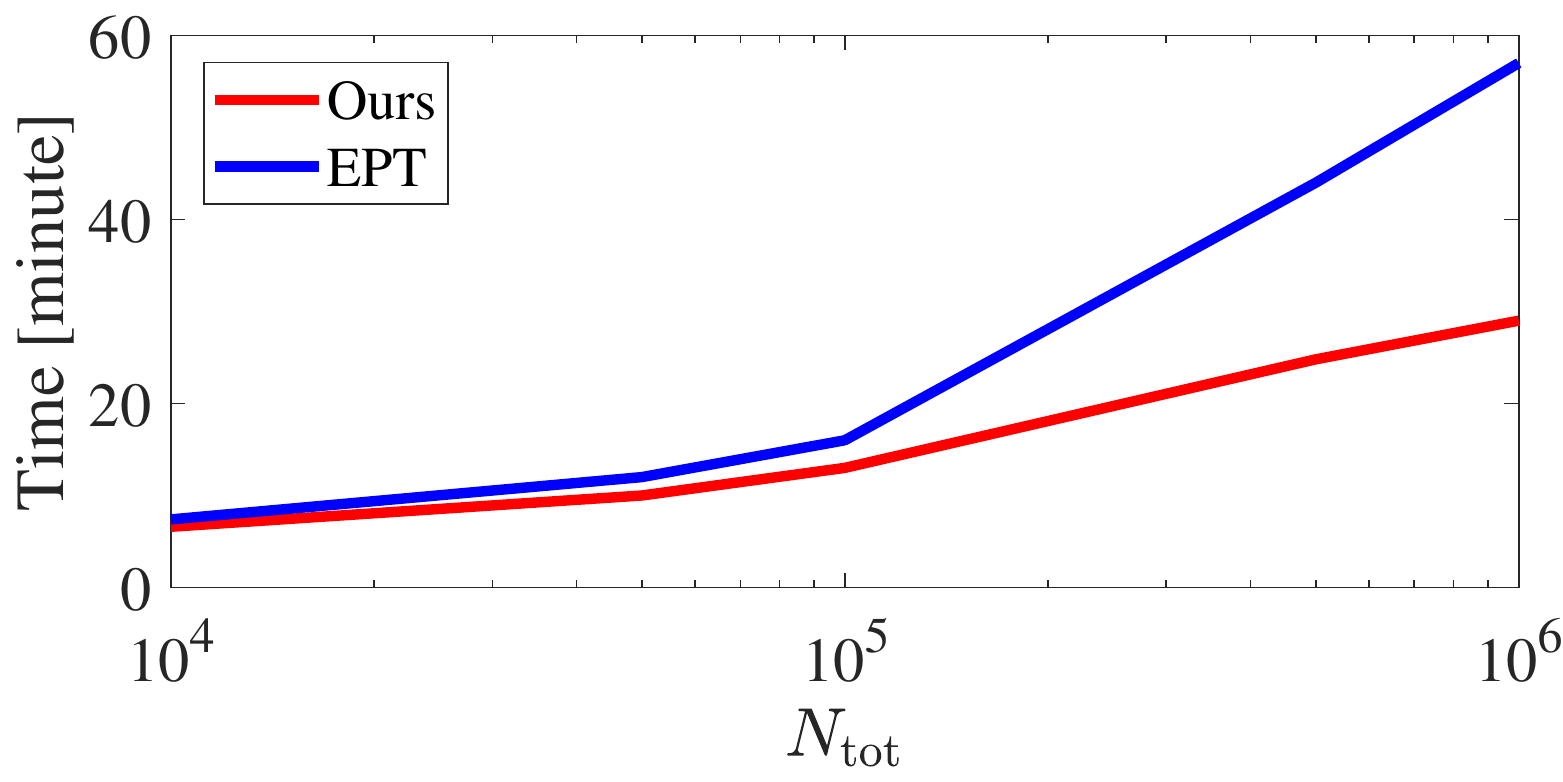}\\
  \caption{Rendering time comparison between our method and EPT on the \textsc{Cornell Box} scene shown in Fig. \ref{fig:cornell_EPT}.}\label{fig:cornell_time}
\end{figure}

The runtime performance of some test scenes can be found in Table \ref{tab:performance}. As we mentioned previously, our rendering solution achieves a roughly $1.2\times$ speed improvement over EPT for the \textsc{Lamp} scene and the \textsc{Cornell Box} scene. As the particle number $N_\mathrm{tot}$ increases, the improvement will be more evident. As shown in Fig. \ref{fig:cornell_time}, $2\times$ speed improvement is achieved when $N_\mathrm{tot}$ increases to $10^6$ in the \textsc{Cornell Box} scene.
Since only the position and the radius of each particle are required, the memory consumption of our rendering solution is affordable even when $N_\mathrm{tot}$ is very large.
The storage scales with the number of particles and the resolution of the grid.
\begin{table}[t]
  \centering
  \caption{Rendering time performance (in minutes) and memory consumption for six typical scenes shown in this paper. Absorption is not included in the computation.}\label{tab:performance}
  \begin{tabular}{lccccc}
    \hline
    \multirow{2}{*}{Scene} & \multirow{2}{*}{$N_\mathrm{tot}$} & \multirow{2}{*}{Spp} & \multirow{2}{*}{Memory} & \multicolumn{2}{c}{Rendering time}\\
    \cline{5-6}
    & & & & EPT & Ours\\
    \hline
    \textsc{Lamp} & $5\cdot10^5$ & 1024 & 49M & 44 & 36 \\
    \textsc{Cornell Box} & $10^5$ & 1024& 60M & 16 & 13\\
    \textsc{Dragon} & $10^5$ & 2048& 397M & --- & 16\\
    \textsc{Bench} & $2\cdot10^5$ & 2048& 52M & --- & 20\\
    \textsc{Staircase} & $3\cdot 10^6$ & 4096& 138M & --- & 53\\
    \textsc{House} & $4\cdot10^6$& 1024& 545M & --- & 38 \\
    \hline
  \end{tabular}
\end{table}

\section{Limitations and Future Work}
Although our framework has successfully simulated the grainy appearance of discrete participating media, it has several limitations deserving further research.


\emph{Non-spherical particles.} Our current framework focuses on discrete participating media composed of spherical particles. However, non-spherical particles are also very common, e.g., in the context of rendering rainbow \cite{sadeghi11physically} or large snow flakes. Extending our framework to non-spherical particles requires to derive new expressions for the scattering amplitude functions $S_1$ and $S_2$. For some special particle shapes, determining $S_1$ and $S_2$ is rather straightforward and analytical expressions exist \cite{Hovenac:91,HE20121467,LU201990}. For more general shapes, precomputation would be required in practice.

\emph{Spatial correlations.} As we assume the particles to be sparsely distributed, spatial correlations between particles are not considered and we directly extend the conventional VRE to support multi-scale graininess. However, for densely packed particles the conventional VRE becomes questionable due to strongly correlated scattering effects \cite{Moon:2007:RDR:2383847.2383878,Meng:2015:MMR:2809654.2766949,Muller:2016:ERH:2980179.2982429}. Recently, new radiative transfer frameworks dedicated for correlated media are available in computer graphics \cite{Jarabo:2018:RTF:3197517.3201282,Bitterli:2018:RTF:3272127.3275103,Guo:2019:FGF:3306346.3323031}. It would be an interesting future work to investigate a more general framework supporting both effects.

\emph{Mixture of different particles.} Another possible direction of future work is to explore an efficient strategy to handle particle mixtures with different compositions. For instance, dusts in real world may be made up of soil particles, textile fibers, human skin cells, etc. A physically-correct participating medium should consider such heterogeneous granular mixtures. For continuous media, this is relatively simple since the concentrations of different grains are fixed \cite{Frisvad:2007:CSP:1275808.1276452}. However, for discrete participating media, the concentrations are dynamic and change spatially in a similar way as PSDs \cite{Muller:2016:ERH:2980179.2982429}. Therefore, determining the bulk optical properties should take spatially-varying concentrations into consideration.

\section{Conclusion}
We have developed a general and physically-based framework for modeling and rendering discrete participating media composed of massive assemblies of independent particles. Notable characteristics of these media include a wide range of PSDs and the appearance of being graininess. To faithfully simulate their appearances, we have derived a novel multi-scale VRE in which a combination of Lorenz-Mie theory and GOA is used to enable high-efficient evaluation of the important optical properties. A Monte Carlo rendering solution is developed to solve the multi-scale VRE with high accuracy and low computational cost. We have extensively evaluated our framework and compared against conventional methods, demonstrating that the proposed framework allows us to reproduce a variety of grainy appearances stemming from different discrete participating media and guarantee temporal coherence in animation. Therefore, for the first time in computer graphics, we have greatly extended the participating media rendering framework to handle a much larger range of particle size statistics.
We believe the proposed framework is a further step in computer graphics to manage the details of participating media and expect a more insightful exploration of this phenomenon in the future research.




%
\bibliographystyle{IEEEtran}

\bibliography{discrete_media_bib}

\appendices

\section{A Brief Introduction of Lorenz-Mie Theory}
In this section, we briefly describe Lorenz-Mie theory \cite{Lorenz1890,Mie1908} which has already been employed in computer graphics \cite{Rushmeier:1995,Callet:96,Jackel1997,Riley:2004:ERA:2383533.2383584,Frisvad:2007:CSP:1275808.1276452}.
For light scattering of an electromagnetic wave from a homogeneous spherical particle, exact solutions of the two scattering amplitude functions $S_1$ and $S_2$ are given by:
\begin{equation}
  S_1(\theta,r) = \sum_{n=1}^{\infty} \frac{2n+1}{n(n+1)}[a_n(r) \pi_n(\cos \theta) + b_n(r) \tau_n(\cos \theta)]
\tag{26}
\end{equation}
\begin{equation}
  S_2(\theta,r) = \sum_{n=1}^{\infty} \frac{2n+1}{n(n+1)}[b_n(r) \pi_n(\cos \theta) + a_n(r) \tau_n(\cos \theta)]
\tag{27}
\end{equation}
which express the scattered fields in terms of an infinite series of spherical multipole partial waves.
Here, $a_n(r)$ and $b_n(r)$ are the Lorenz-Mie coefficients of particle size $r$; $\pi_n(\cos \theta)$ and $\tau_n(\cos \theta)$ are derived from the Legendre functions. Please refer to \cite{Frisvad:2007:CSP:1275808.1276452} for the details and the expressions of $a_n(r)$, $b_n(r)$, $\pi_n(\cos \theta)$, and $\tau_n(\cos \theta)$.

Inserting the expression of $S(0,r)=S_1(0,r)=S_2(0,r)$ into Eq. (3), we can obtain a well-defined form of the extinction cross section as \cite{BOHREN1979215}
\begin{equation}
  C_\mathrm{t}(r) = \frac{\lambda^2}{2\pi} \sum_{n=1}^{\infty} (2n + 1) \mathrm{Re}\left\{\frac{a_n(r)+b_n(r)}{\eta_\mathrm{m}^2}\right\}.
  \tag{28}
\end{equation}
For the scattering cross section, no simple closed-form formula is available. It is generally approximated by \cite{Randrianalisoa:s,Yin:s}
\begin{equation}
  C_\mathrm{s}(r) = \frac{\lambda^2 e^{-4\pi r \mathrm{Im}\{\eta_\mathrm{m}\}/\lambda}}{2\pi\gamma|\eta_\mathrm{m}|^2}\sum_{n=1}^{\infty} (2n + 1) \left(|a_n(r)|^2+|b_n(r)|^2\right)
  \tag{29}
\end{equation}
with $\gamma=2(1+(\beta-1)e^\beta)/\beta^2$ and $\beta=4\pi r \mathrm{Im}\{\eta_\mathrm{m}\}/\lambda$. The notation $\mathrm{Re}$ and $\mathrm{Im}$ take the real and imaginary part of a complex number, respectively.
The phase function is given by \cite{Hulst1981}
\begin{equation}
  f_{\mathrm{p}}(\theta,r) = \frac{|S_1(\theta,r)|^2+|S_2(\theta,r)|^2}{4\pi\sum_{n=1}^{\infty}(2n+1)(|a_n(r)|^2+|b_n(r)|^2)}.
  \tag{30}
\end{equation}

\section{Derivation of $C_\mathrm{t}(r)$ in Eq. (12)}\label{app:C}
Substituting Eq. (11) into Eq. (3), we have
\begin{equation}
\begin{split}
  C_\mathrm{t}(r) &= \frac{4\pi}{|k|^2}\mathrm{Re}\left(S_{\mathrm{D},j}(0,r)+\sum_{p=0}^{\infty} S_j^{(p)}(0, r)\right)\\
  &=\frac{4\pi}{|k|^2}\mathrm{Re}\left(\frac{\alpha^2}{2}+\sum_{p=0}^{\infty}\alpha \epsilon_j(0)\sqrt{\frac{1}{4(p/\eta-2)^2}}e^{\mathrm{i}\phi}\right)\\
  &=\frac{4\pi}{|k|^2}\mathrm{Re}\left(\frac{\alpha^2}{2}+\sum_{p\in \mathcal{P}}\alpha \epsilon_j(0)\sqrt{\frac{1}{4(p/\eta-2)^2}}\cos(\phi)\right)\\
  &=2\pi r^2 + \frac{2\pi r}{|k|} \sum_{p\in \mathcal{P}}  \frac{\epsilon_j(0)}{|p/\eta-1|}\cos\phi
\end{split}
\tag{31}
\end{equation}
in which $\mathcal{P}=\{1,3,5,\cdots\}$.

\section{GOA for Particles with Absorption}\label{app:B}
For particles with absorption, the refractive index is a complex number, which could be written as $\eta_\mathrm{p} = \eta_\mathrm{r}+\eta_\mathrm{i}\mathrm{i}$. Defining the effective refractive index \cite{YU20091178}:
\begin{equation}
\begin{split}
  \eta' &=\bigg\{\frac{1}{2}(\eta_\mathrm{r}^2-\eta_\mathrm{i}^2 +\eta_\mathrm{m}^2\sin^2\theta_\mathrm{i})\\
  &+\frac{1}{2}[4\eta_\mathrm{r}^2\eta_\mathrm{i}^2+(\eta_\mathrm{r}^2-\eta_\mathrm{i}^2-\eta_\mathrm{m}^2\sin^2\theta_\mathrm{i})^2]^\frac{1}{2}\bigg\}^\frac{1}{2}
\end{split}
\tag{32}
\end{equation}
we have
\begin{equation}
  \eta_\mathrm{m}\sin\theta_\mathrm{i} = \eta'\sin \theta'_\mathrm{t}
  \tag{33}
\end{equation}
where $\theta'_\mathrm{t}$ is the effective refractive angle. When particles are absorbing, the refractive angle $\theta_\mathrm{t}$ should be replaced by $\theta'_\mathrm{t}$. The overall phase shift is changed to
\begin{equation}
  \phi = \left\{ \begin{array}{ll}
                                        \phi_\mathrm{p} + \phi_\mathrm{f} + \phi_{\mathrm{r},j} & \quad p=0\\
                                        \phi_\mathrm{p} + \phi_\mathrm{f} + \phi_{\mathrm{t},j}& \quad p>0
                                        \end{array}
                                        \right.  \mbox{\quad} j=1,2.
  \tag{34}
\end{equation}
The analytical expressions of phase shifts due to reflection $\phi_{\mathrm{r},j}$ and refraction $\phi_{\mathrm{t},j}$ are provided in \cite{YU20091178}.

Moreover, the amplitude functions in Eq. (7) should be multiplied with the attenuation factor $\xi_p$ \cite{YU20091178}:
\begin{equation}
  \xi_p = e^{-2\chi p \alpha \cos^2\theta'_\mathrm{t}/\eta_\mathrm{m}}
  \tag{35}
\end{equation}
considering amplitude attenuation in the absorbing particle. Here, $\chi$ is the effective absorption coefficient defined as
\begin{equation}
\begin{split}
  \chi &= \bigg\{\frac{1}{2}(-\eta_\mathrm{r}^2+\eta_\mathrm{i}^2+\eta_\mathrm{m}^2\sin^2\theta_\mathrm{i})\\
  &+\frac{1}{2}[4\eta_\mathrm{r}^2\eta_\mathrm{i}^2+(\eta_\mathrm{r}^2-\eta_\mathrm{i}^2-\eta_\mathrm{m}^2\sin^2\theta_\mathrm{i})^2]^\frac{1}{2}\bigg\}^\frac{1}{2}.
\end{split}
\tag{36}
\end{equation}

\section{Derivation of the Transmittance in Eq. (18)}\label{app:D}
Considering a light ray $\mbx\to\mby$ passing through a discrete participating medium, the transmittance between $\mbx$ and $\mby = \mbx-s \mbomega$ is calculated by
\begin{equation}
\begin{split}
  T(\mbx, \mby) &= \exp\left\{-\int_0^s \sigma_\mathrm{t}(\mathcal{V}_{\mbx-s' \mbomega})\ud s'\right\}\\
  &=\exp\left\{-\int_0^s \frac{\int_{r_\mathrm{min}}^{r_\mathrm{max}}C_\mathrm{t}(r)\int_{\mbx\in\mathcal{V}_{\mbx-s' \mbomega}}N(r,\mbx)\ud \mbx \mathrm{d}r}{\mu(\mathcal{V}_{\mbx-s' \mbomega})}\ud s'\right\}\\
  &=\exp\left\{-\int_0^s \frac{\int_{r_\mathrm{min}}^{r_\mathrm{max}}C_\mathrm{t}(r)\int_{\mbx\in\mathcal{A}\times \ud s'}N(r,\mbx)\ud \mbx \mathrm{d}r}{\mu(\mathcal{A})\times \ud s'}\ud s'\right\}\\
  &=\exp\left\{- \frac{\int_{r_\mathrm{min}}^{r_\mathrm{max}}C_\mathrm{t}(r)\int_{\mbx\in\mathcal{A}\times s}N(r,\mbx)\ud \mbx \mathrm{d}r}{\mu(\mathcal{A})}\right\}.
\end{split}
\tag{37}
\end{equation}
Here we set $\mathcal{V}_{\mbx-s' \mbomega}$ to $\mathcal{A}\times\ud s'$.

\section{More Discussions on $p$}
\begin{figure*}[h]
\centering
\subfigure[$r=1~\mathrm{{\mu}m}$, $\lambda = 0.5~\mathrm{{\mu}m}$]{
\begin{minipage}[b]{0.32\linewidth}
\begin{overpic}[width=1.0\linewidth]{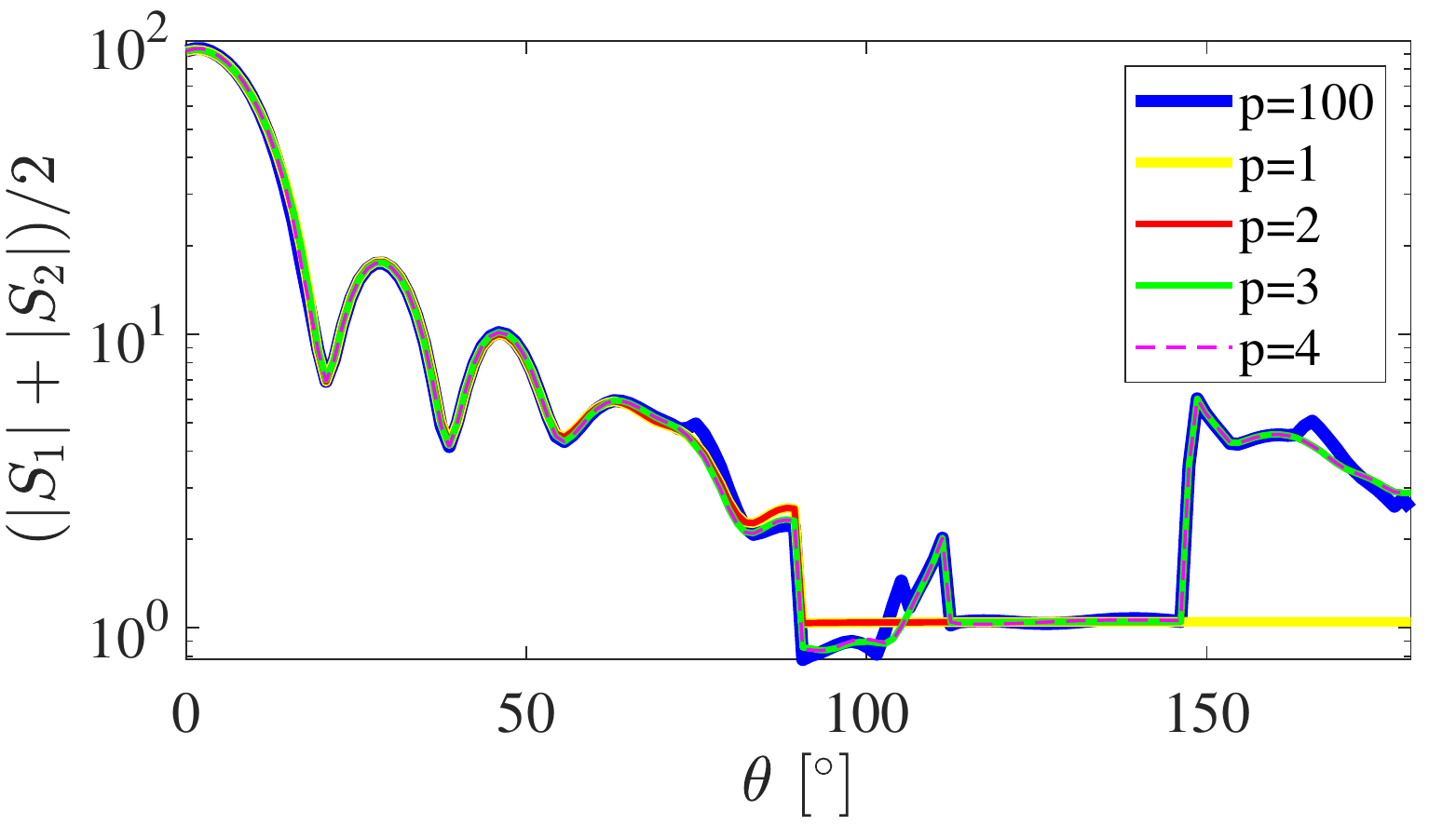}
\end{overpic}
\end{minipage}
}
\hspace{-0.12in}
\subfigure[$r=1~\mathrm{{\mu}m}$, $\lambda = 0.6~\mathrm{{\mu}m}$]{
\begin{minipage}[b]{0.32\linewidth}
\begin{overpic}[width=1.0\linewidth]{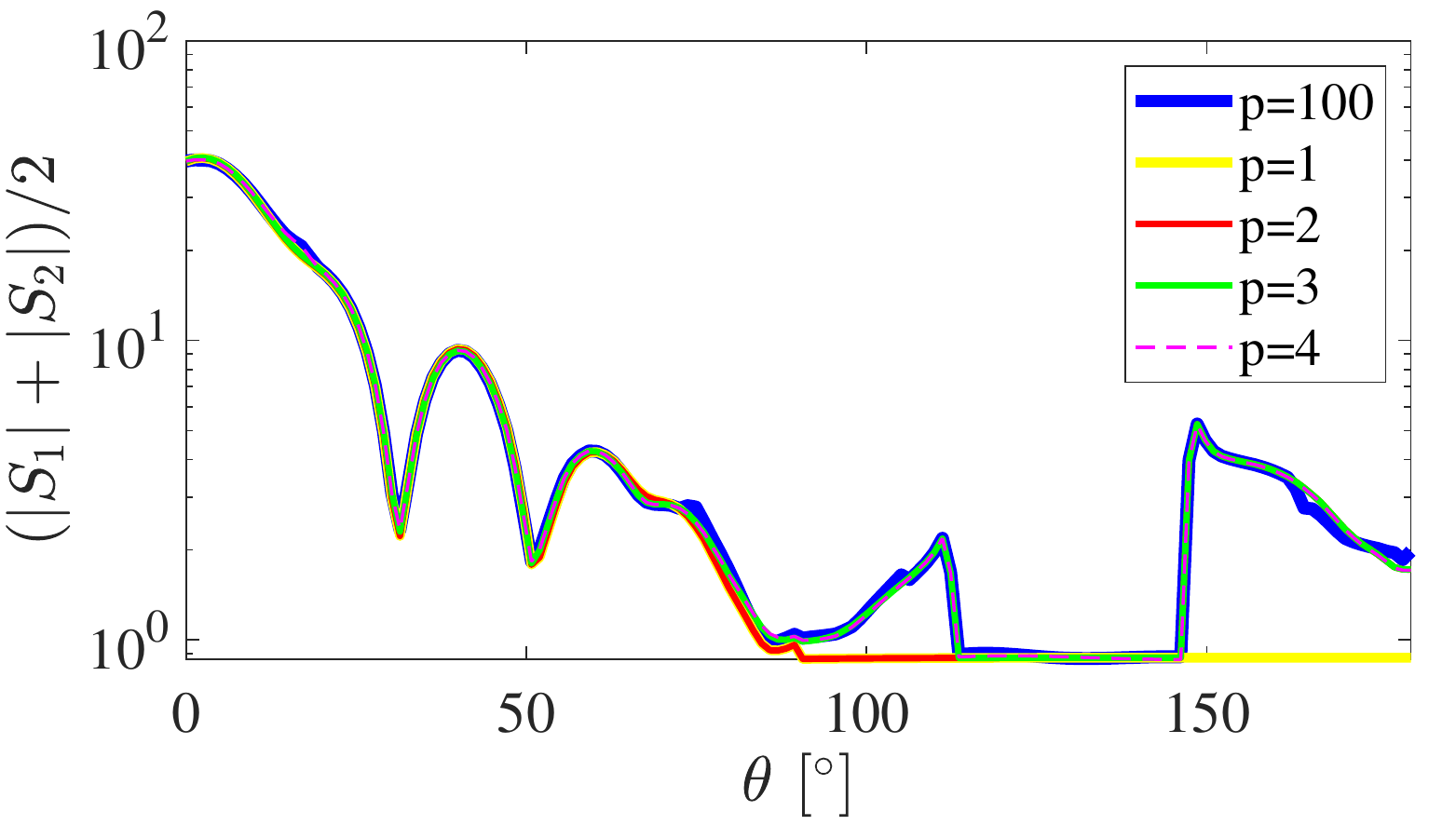}
\end{overpic}
\end{minipage}
}
\hspace{-0.12in}
\subfigure[$r=1~\mathrm{{\mu}m}$, $\lambda = 0.7~\mathrm{{\mu}m}$]{
\begin{minipage}[b]{0.32\linewidth}
\begin{overpic}[width=1.0\linewidth]{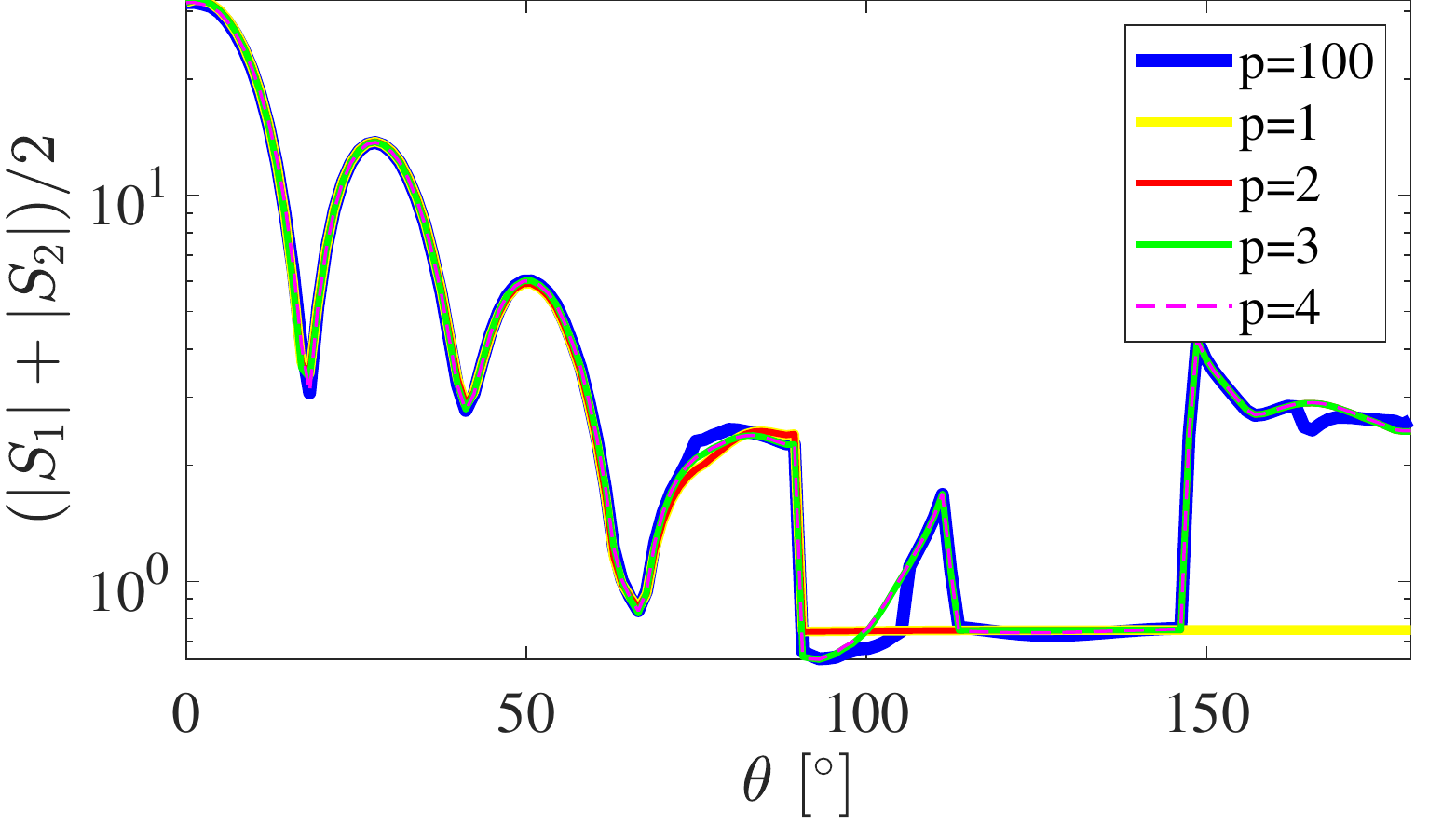}
\end{overpic}
\end{minipage}
}
\subfigure[$r=10~\mathrm{{\mu}m}$, $\lambda = 0.5~\mathrm{{\mu}m}$]{
\begin{minipage}[b]{0.32\linewidth}
\begin{overpic}[width=1.0\linewidth]{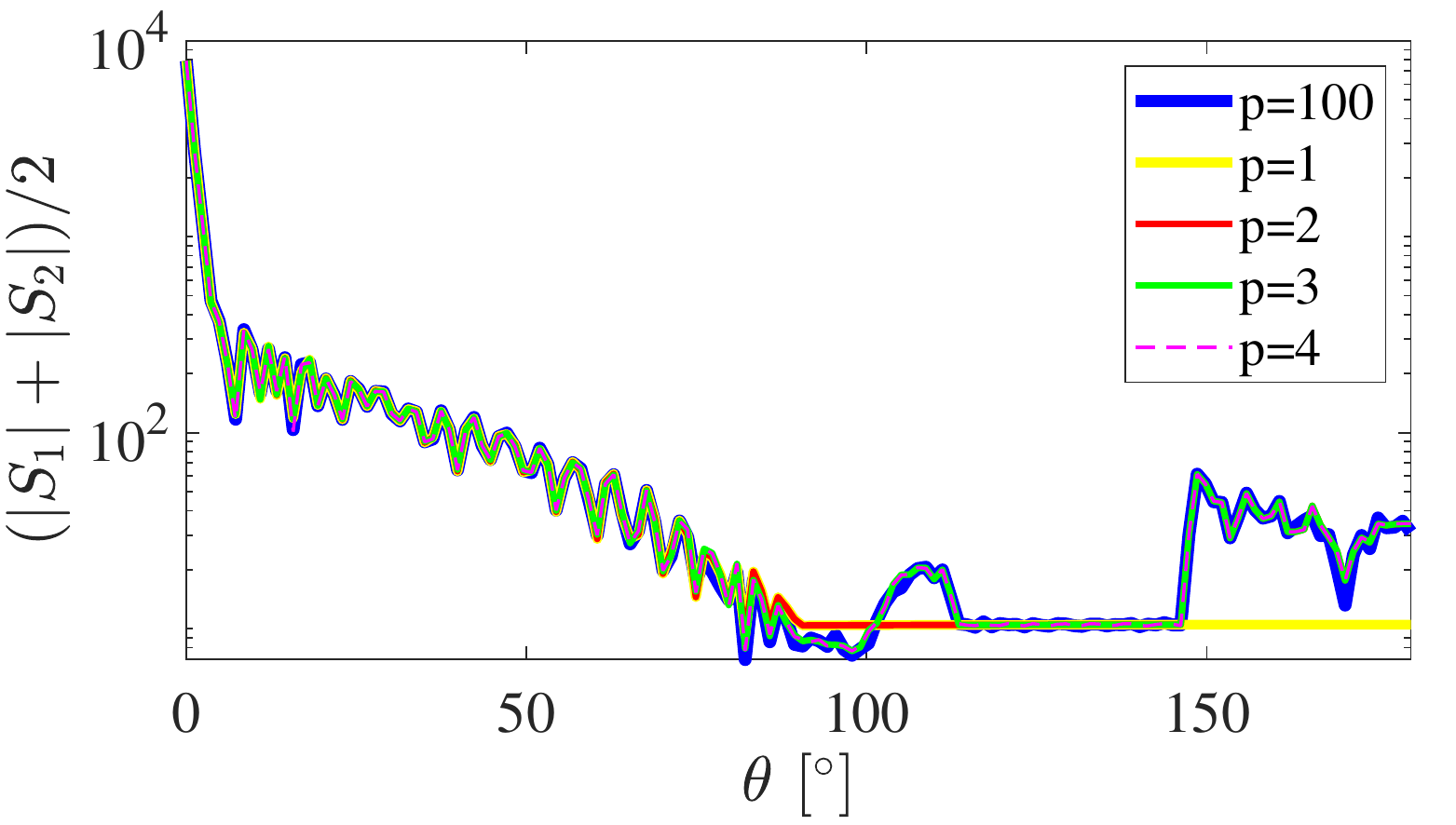}
\end{overpic}
\end{minipage}
}
\hspace{-0.12in}
\subfigure[$r=10~\mathrm{{\mu}m}$, $\lambda = 0.6~\mathrm{{\mu}m}$]{
\begin{minipage}[b]{0.32\linewidth}
\begin{overpic}[width=1.0\linewidth]{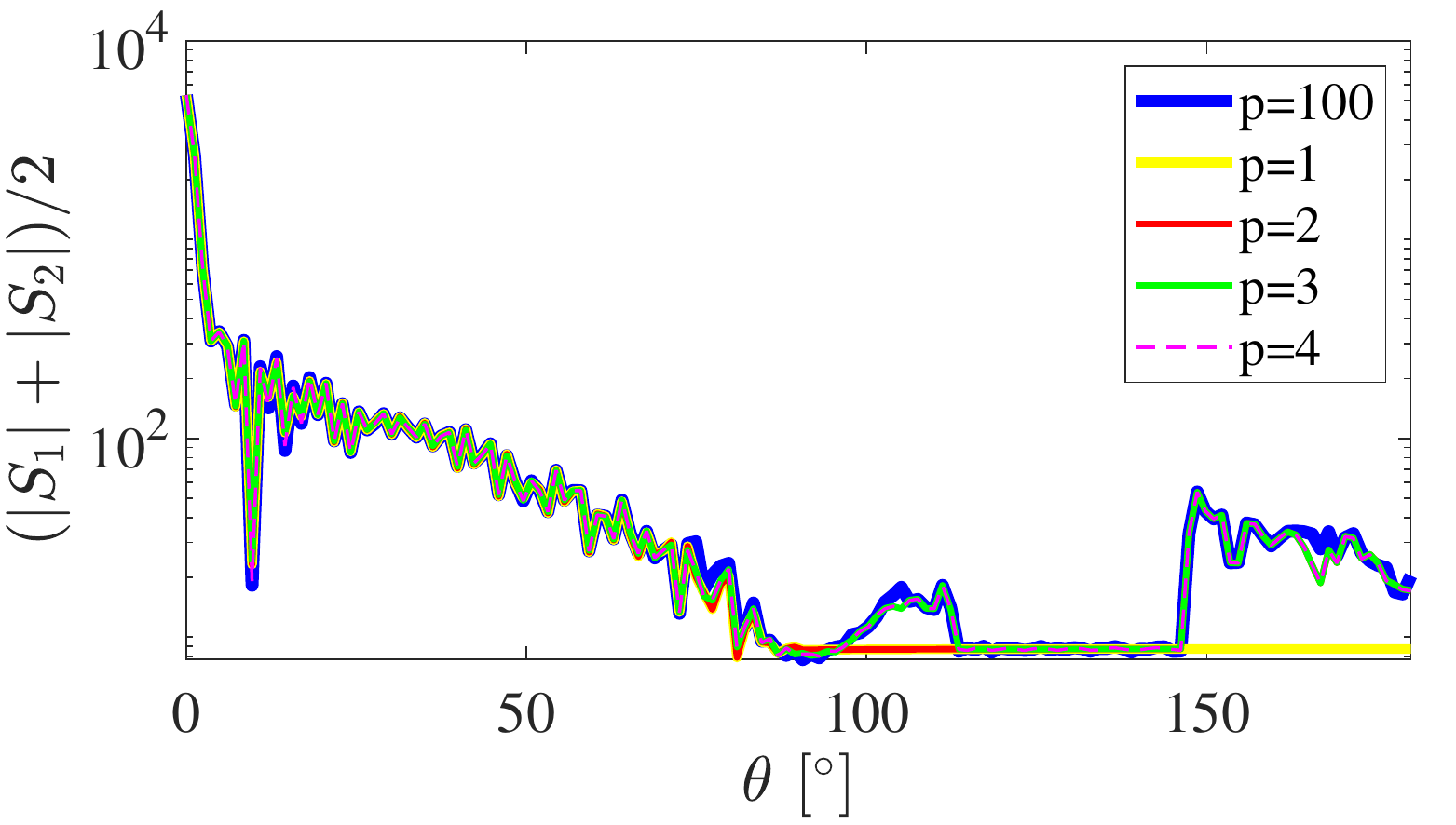}
\end{overpic}
\end{minipage}
}
\hspace{-0.12in}
\subfigure[$r=10~\mathrm{{\mu}m}$, $\lambda = 0.7~\mathrm{{\mu}m}$]{
\begin{minipage}[b]{0.32\linewidth}
\begin{overpic}[width=1.0\linewidth]{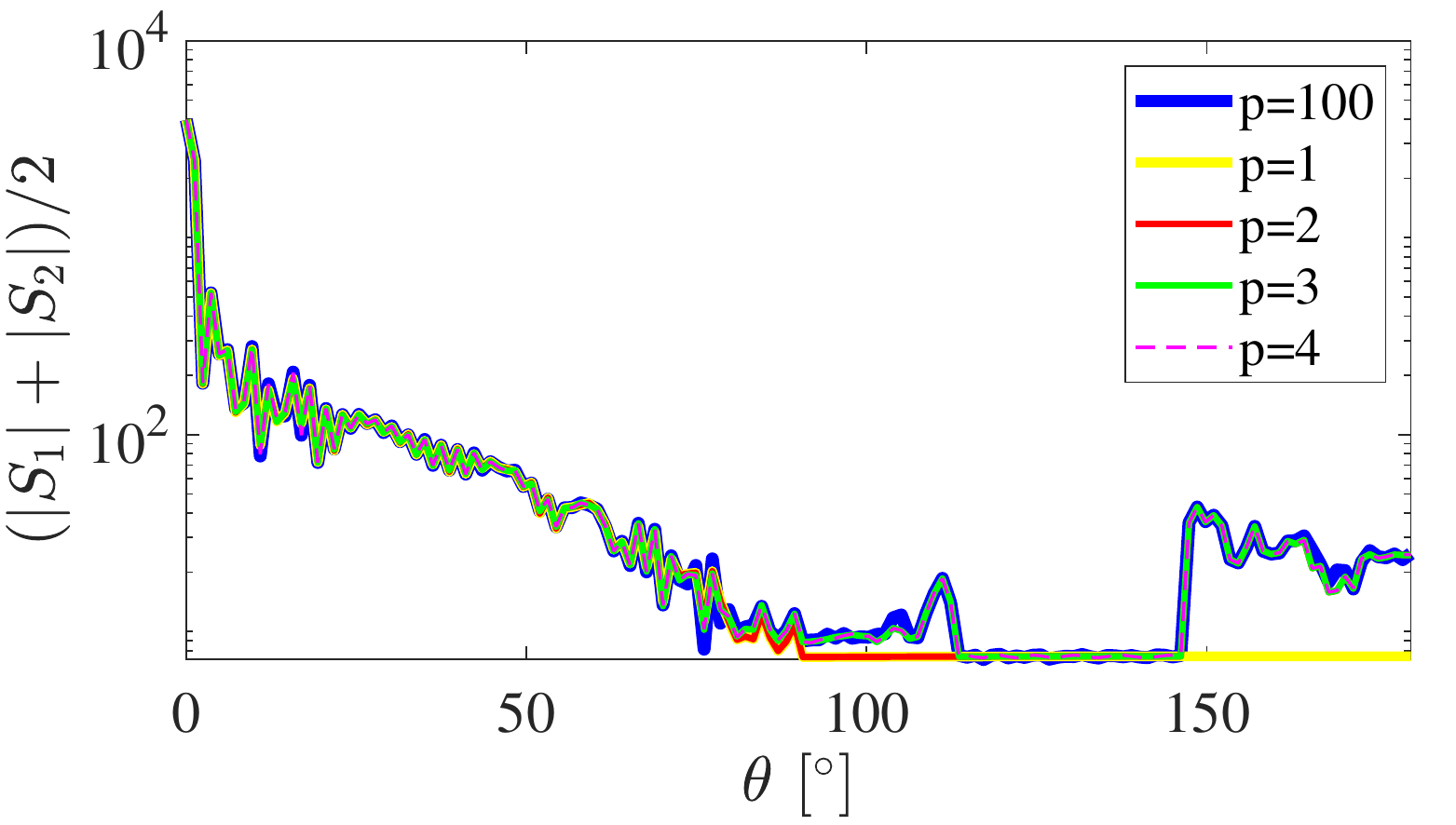}
\end{overpic}
\end{minipage}
}
\subfigure[$r=100~\mathrm{{\mu}m}$, $\lambda = 0.5~\mathrm{{\mu}m}$]{
\begin{minipage}[b]{0.32\linewidth}
\begin{overpic}[width=1.0\linewidth]{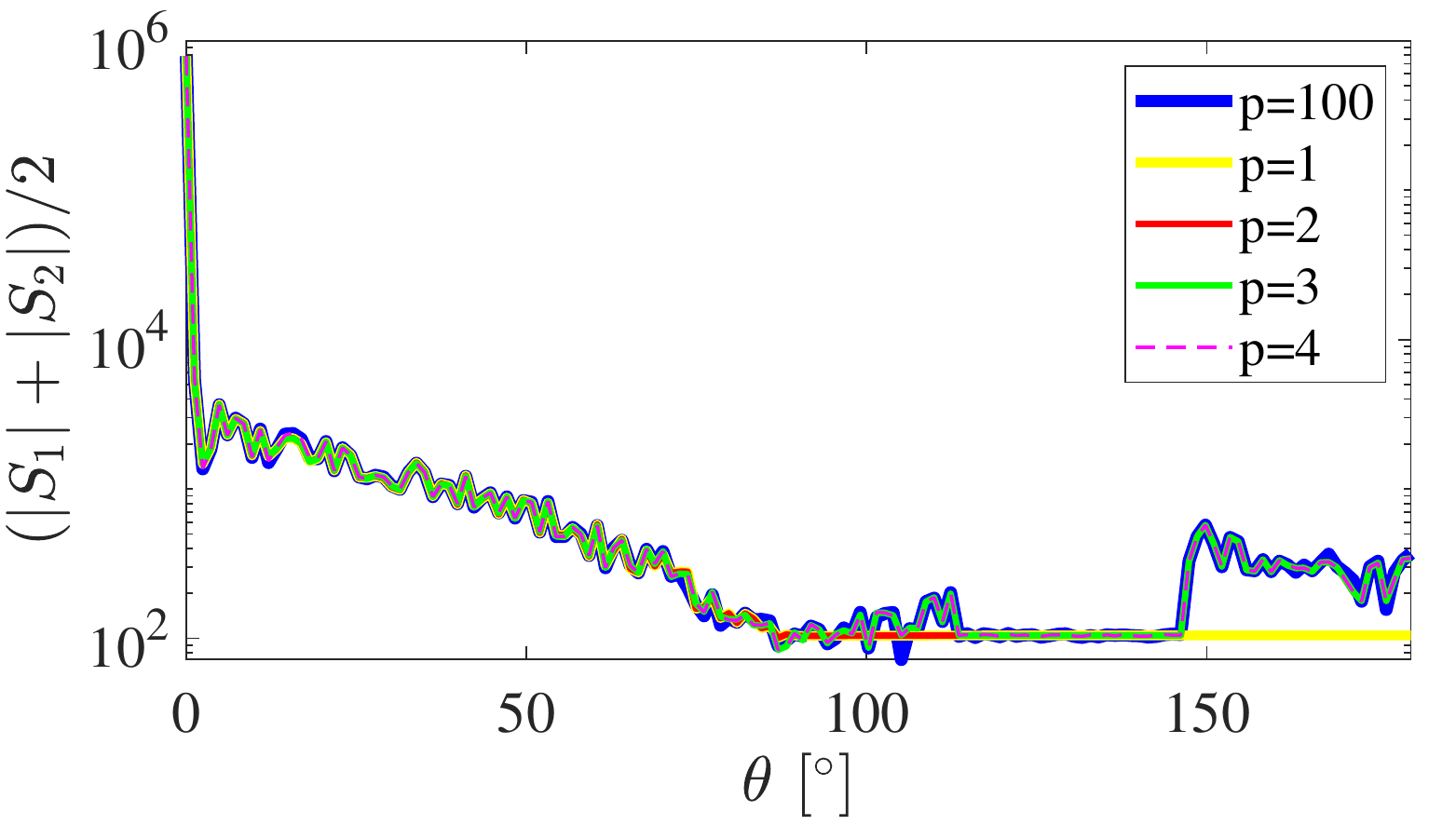}
\end{overpic}
\end{minipage}
}
\hspace{-0.12in}
\subfigure[$r=100~\mathrm{{\mu}m}$, $\lambda = 0.6~\mathrm{{\mu}m}$]{
\begin{minipage}[b]{0.32\linewidth}
\begin{overpic}[width=1.0\linewidth]{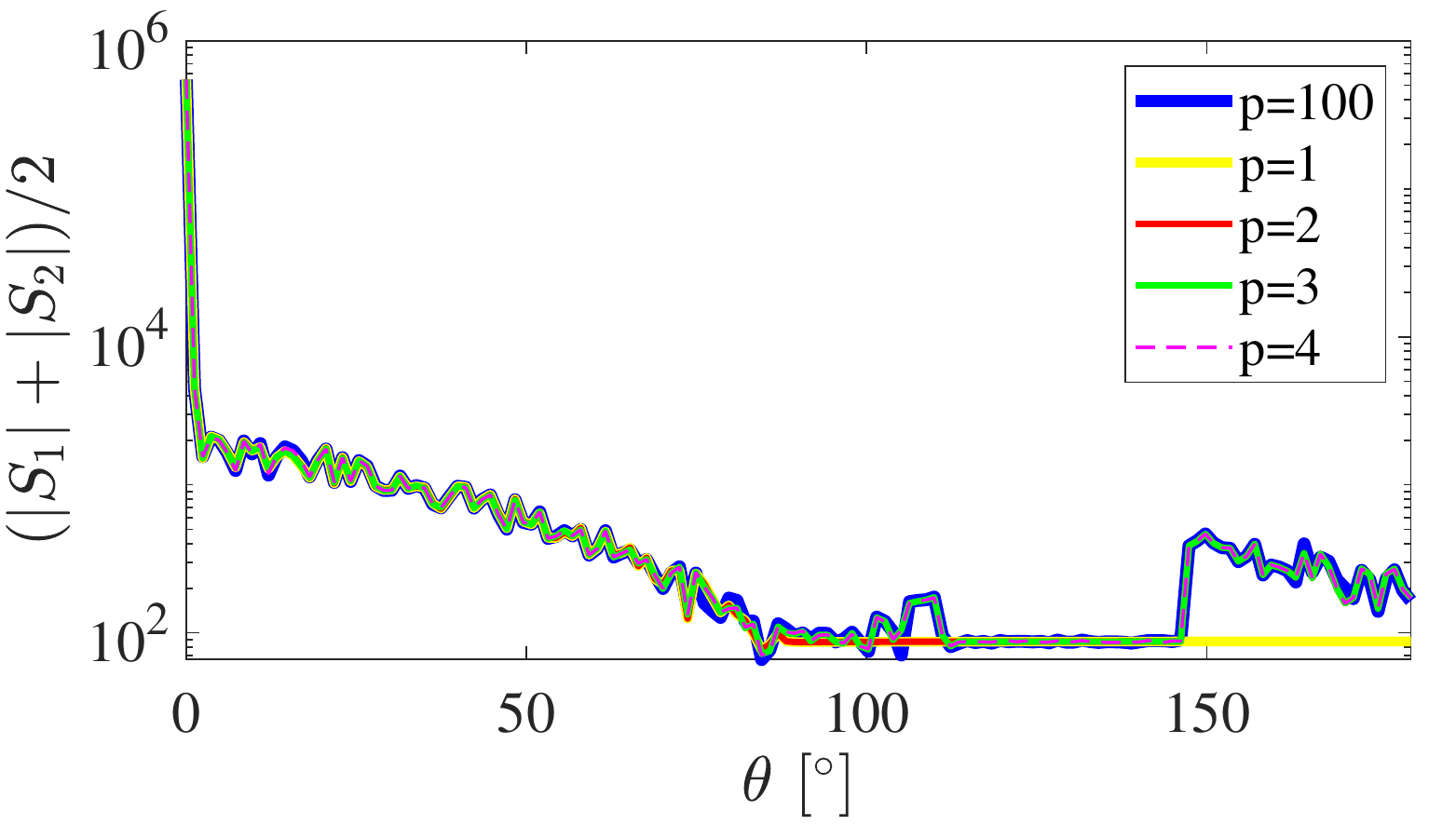}
\end{overpic}
\end{minipage}
}
\hspace{-0.12in}
\subfigure[$r=100~\mathrm{{\mu}m}$, $\lambda = 0.7~\mathrm{{\mu}m}$]{
\begin{minipage}[b]{0.32\linewidth}
\begin{overpic}[width=1.0\linewidth]{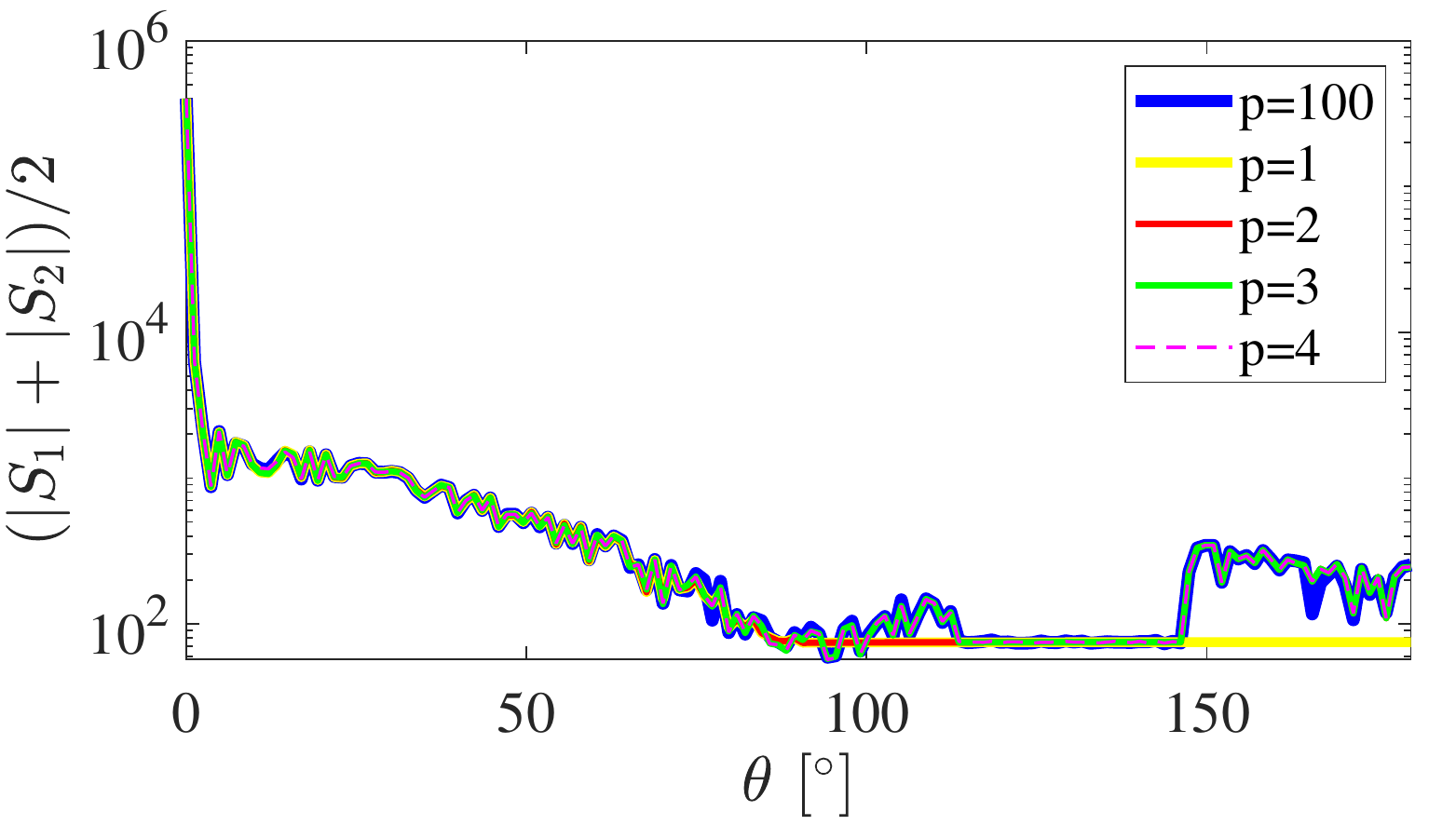}
\end{overpic}
\end{minipage}
}
\subfigure[$r=1000~\mathrm{{\mu}m}$, $\lambda = 0.5~\mathrm{{\mu}m}$]{
\begin{minipage}[b]{0.32\linewidth}
\begin{overpic}[width=1.0\linewidth]{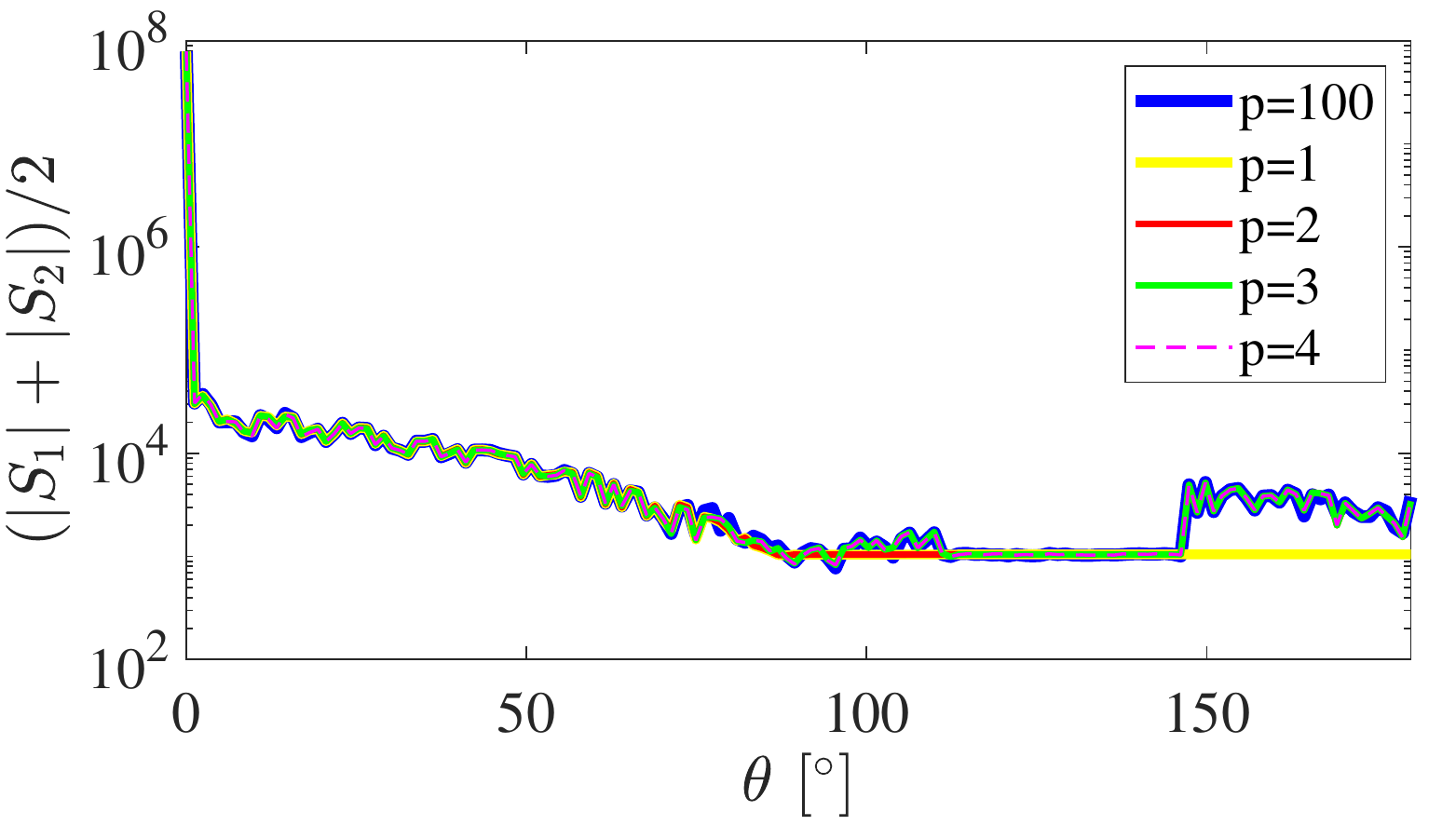}
\end{overpic}
\end{minipage}
}
\hspace{-0.12in}
\subfigure[$r=1000~\mathrm{{\mu}m}$, $\lambda = 0.6~\mathrm{{\mu}m}$]{
\begin{minipage}[b]{0.32\linewidth}
\begin{overpic}[width=1.0\linewidth]{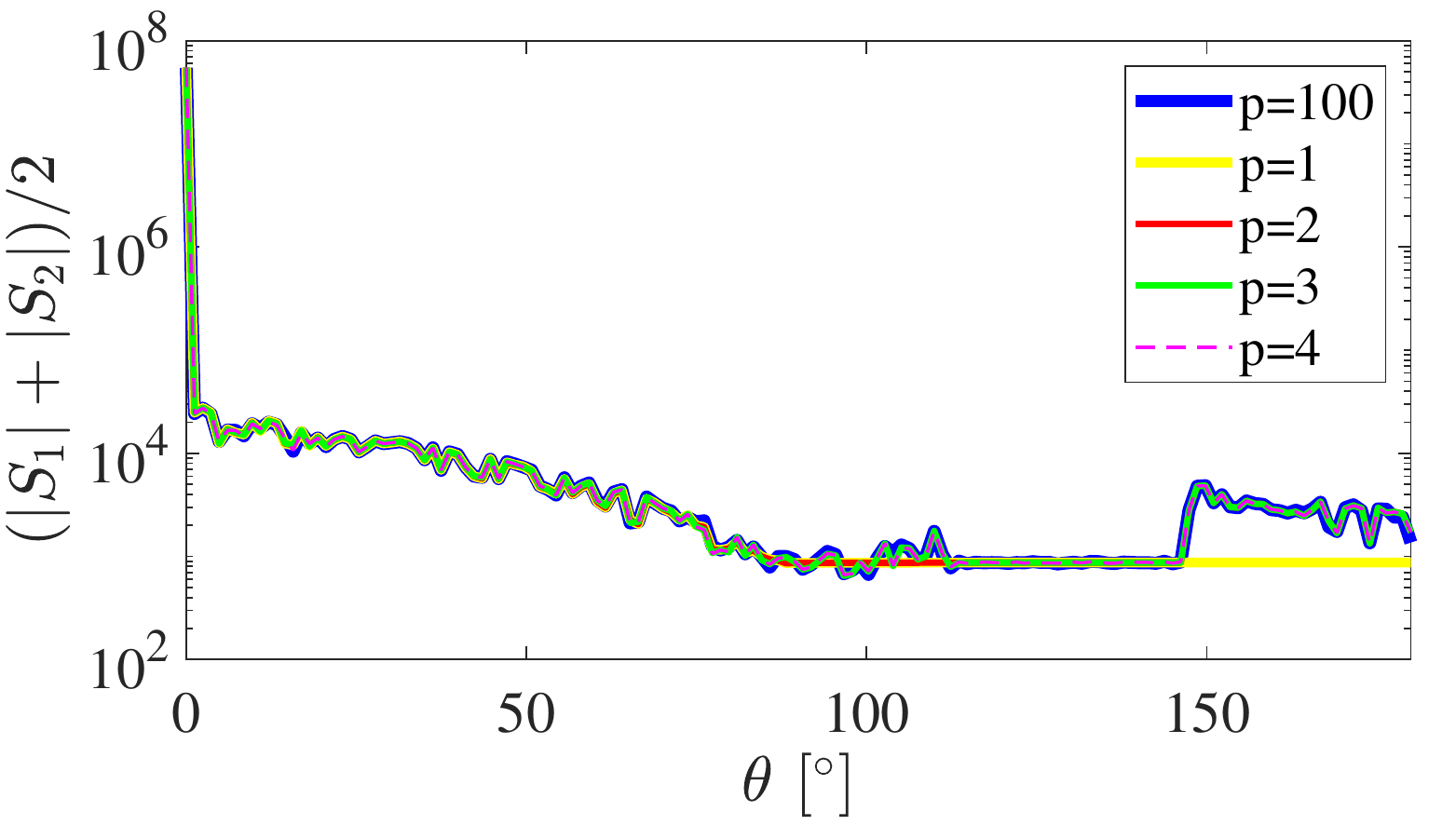}
\end{overpic}
\end{minipage}
}
\hspace{-0.12in}
\subfigure[$r=1000~\mathrm{{\mu}m}$, $\lambda = 0.7~\mathrm{{\mu}m}$]{
\begin{minipage}[b]{0.32\linewidth}
\begin{overpic}[width=1.0\linewidth]{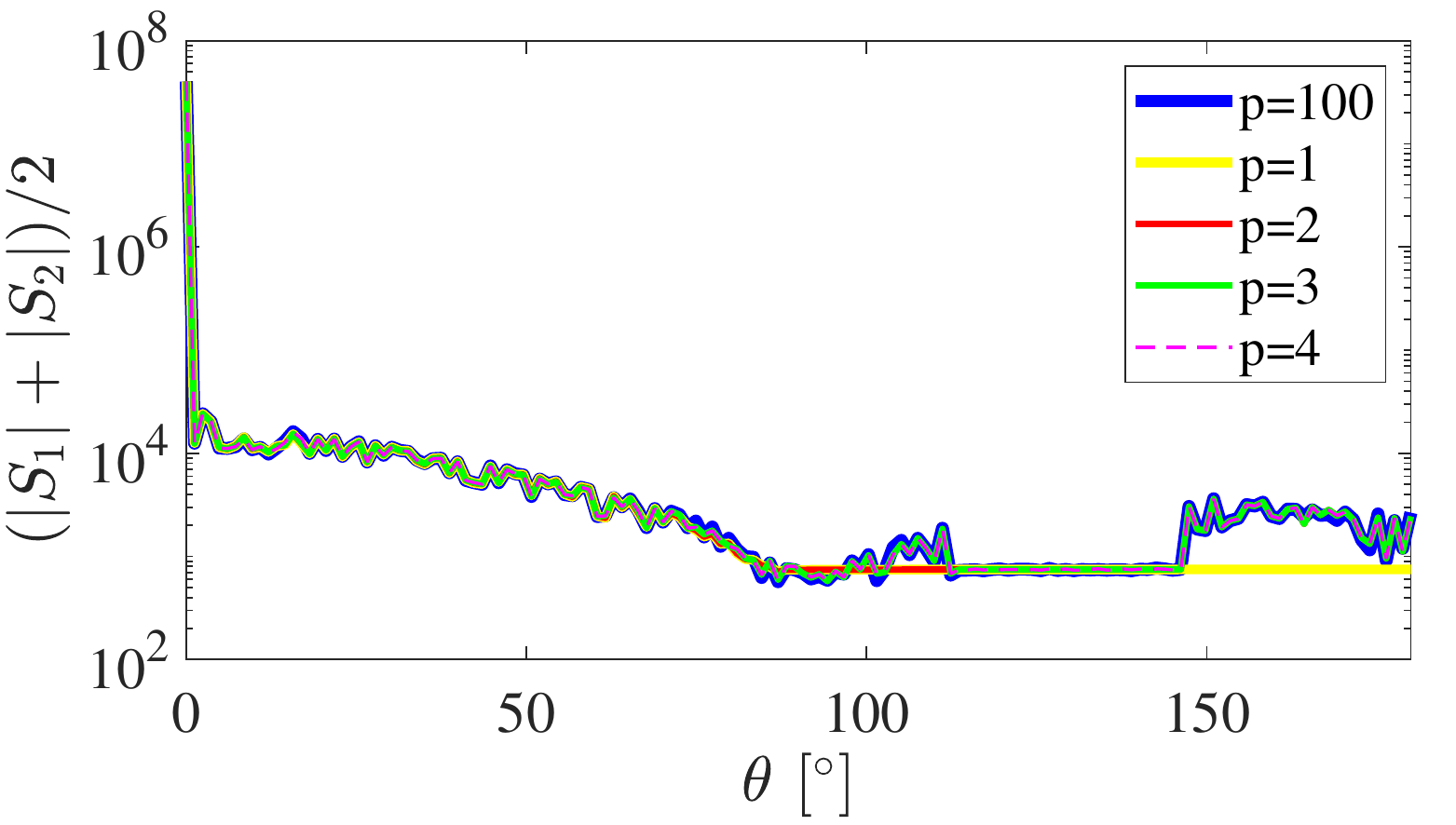}
\end{overpic}
\end{minipage}
}
\caption{\label{fig:goa_p_2} Visual comparisons of $(|S_1|+|S_2|)/2$ with increasing values of $p$ in GOA. Here, we show different combinations of particle size $r$ and wavelength $\lambda$, while the relative refractive index of the particle $\eta$ is set to 1.40.}
\end{figure*}

In GOA, the parameter $p$ is the number of chords that each ray makes inside the particle. The ray is internal reflected $p-1$ times before leaving the particle. Since higher-order rays ($p>3$) have negligible light intensities as compared with other lower-order rays ($p\leq 3$), they can be removed in the computation of the scattering amplitude functions $S_1$ and $S_2$. To validate this, we plot $(|S_1|+|S_2|)/2$ with increasing values of $p$ in Fig. \ref{fig:goa_p} for $\eta=1.33$ and in Fig. \ref{fig:goa_p_2} for $p=1.40$. As see, when $p$ is small (i.e., $p=1$), the simulated $(|S_1|+|S_2|)/2$ curves have remarkable differences compared with the ground truths generated with a very high order (i.e., $p=100$). However, the $(|S_1|+|S_2|)/2$ curves with $p=3$ and $p=4$ are almost identical, and closely match the ground truths. This implies that $p=3$ is sufficient in computing $S_1$ and $S_2$ with GOA.

\begin{figure}[h]
\centering
\subfigure[$\eta=1.33$]{
\begin{minipage}{0.48\linewidth}
\begin{overpic}[width=1.0\linewidth]{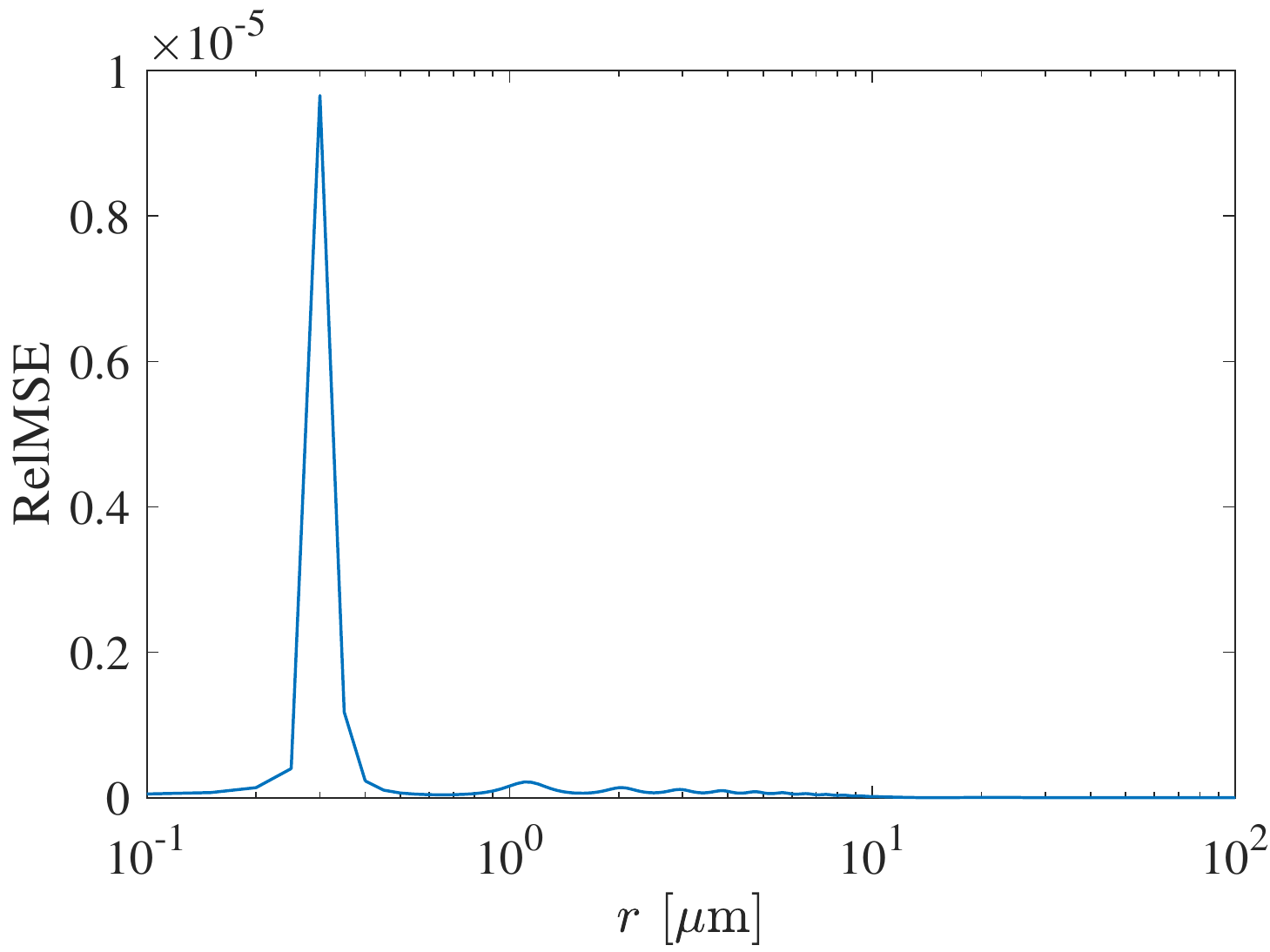}
\end{overpic}
\end{minipage}
}
\hspace{-0.08in}
\subfigure[$\eta=1.40$]{
\begin{minipage}{0.48\linewidth}
\begin{overpic}[width=1.0\linewidth]{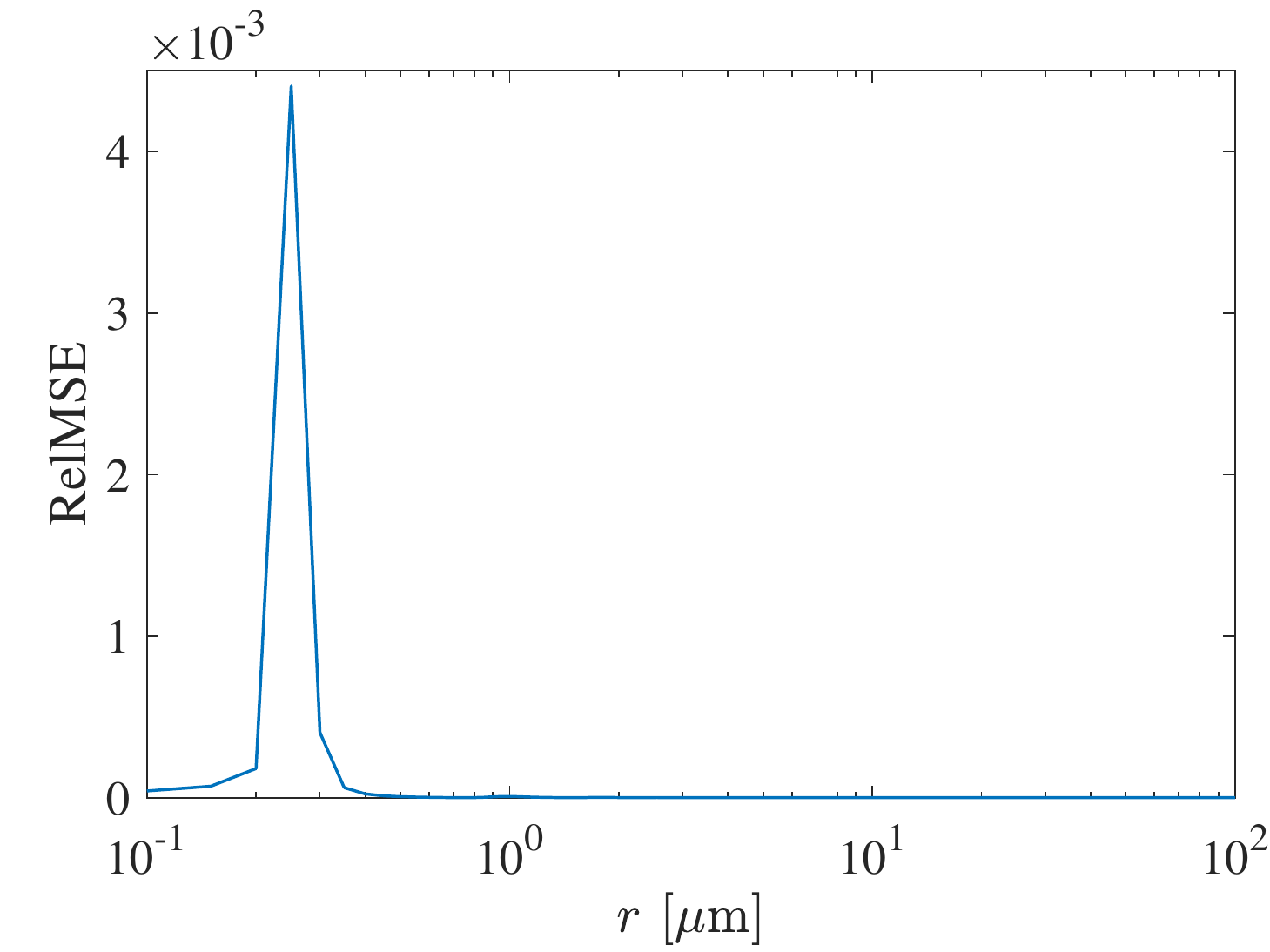}
\end{overpic}
\end{minipage}
}
\caption{\label{fig:ct_mse} Variation of $C_\mathrm{t}$'s RelMSE as a function of the radius $r$.}
\end{figure}

For evaluating the extinction cross section $C_\mathrm{t}$, we can further reduce $p$ to 1. This simplification will lower the computational cost while introducing negligible error, as verified in Fig. \ref{fig:ct_mse}. Here, we adopt the Relative Mean Squared Error (RelMSE) between $p=3$ and $p=1$:
\begin{equation}
  \mathrm{RelMSE}\{C_\mathrm{t}\} = \frac{\left(C_\mathrm{t}^{p=3}-C_\mathrm{t}^{p=1}\right)^2}{\left(C_\mathrm{t}^{p=3}\right)^2}
  \tag{38}
\end{equation}
to measure the error of $C_\mathrm{t}$. As seen, the RelMSE of $C_\mathrm{t}$ is very low, especially for $r>1~\mathrm{{\mu}m}$.

\begin{figure*}[h]
\centering
\subfigure[$r=1~\mathrm{{\mu}m}$, $\lambda = 0.5~\mathrm{{\mu}m}$]{
\begin{minipage}[b]{0.32\linewidth}
\begin{overpic}[width=1.0\linewidth]{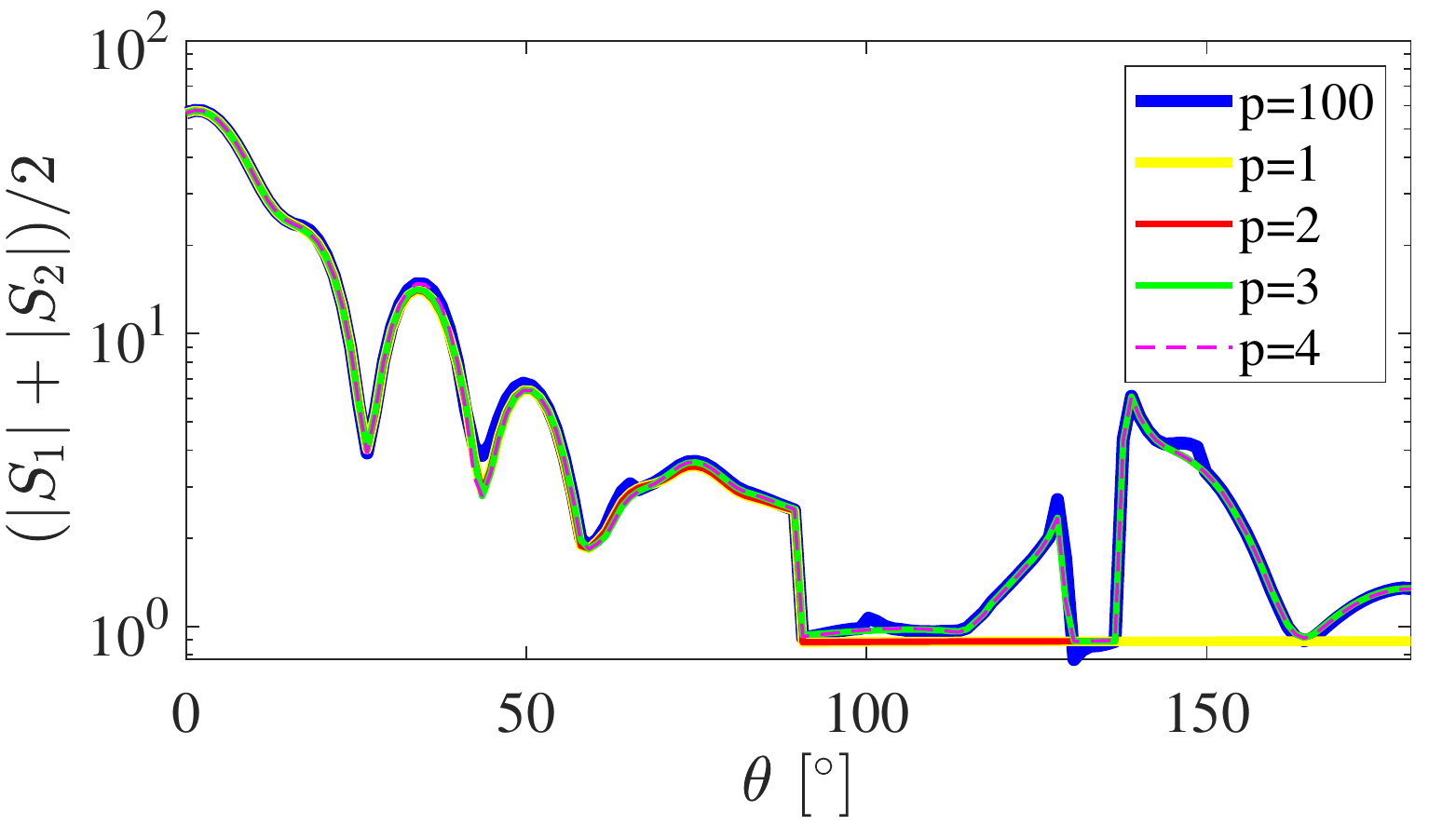}
\end{overpic}
\end{minipage}
}
\hspace{-0.12in}
\subfigure[$r=1~\mathrm{{\mu}m}$, $\lambda = 0.6~\mathrm{{\mu}m}$]{
\begin{minipage}[b]{0.32\linewidth}
\begin{overpic}[width=1.0\linewidth]{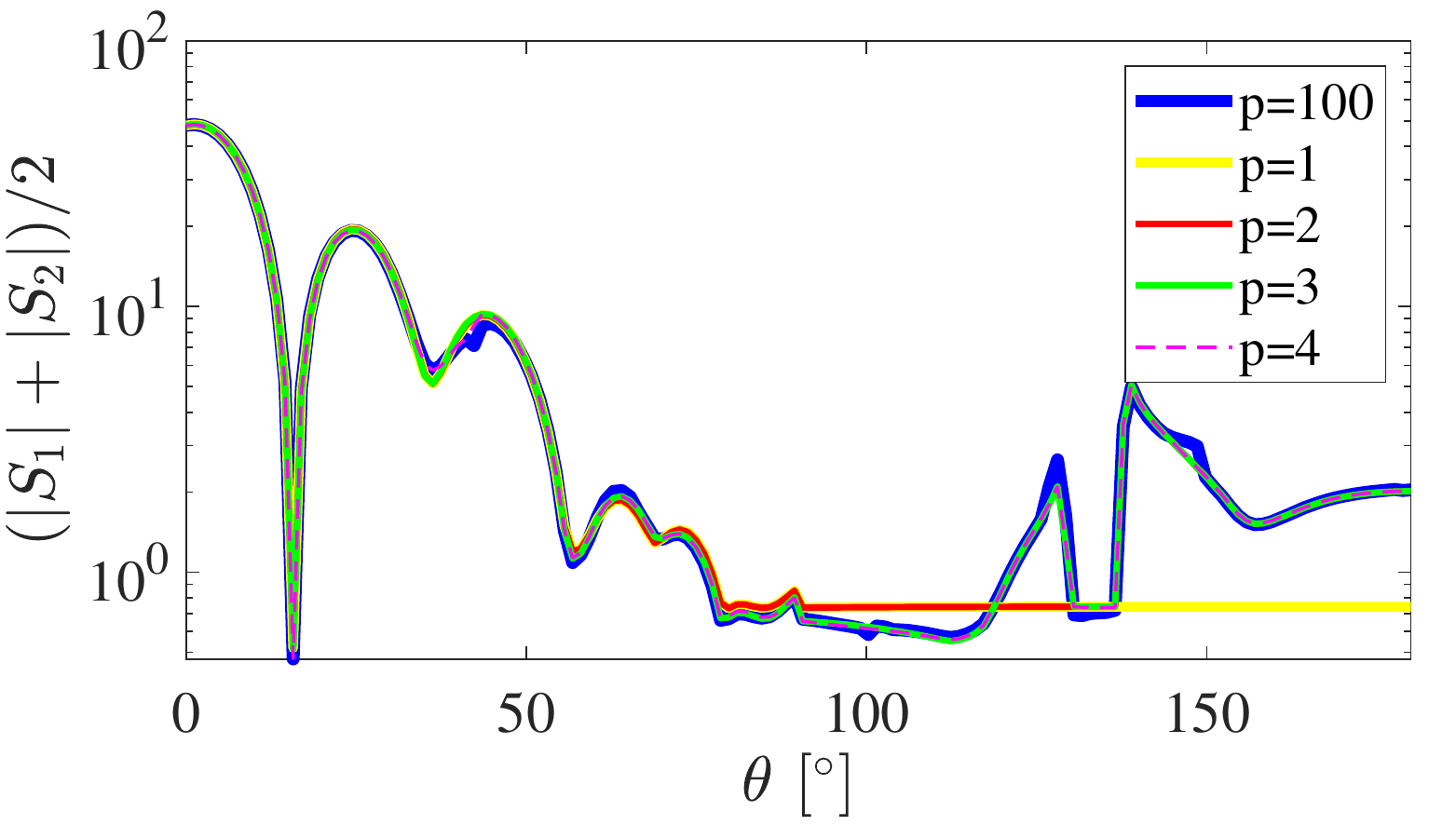}
\end{overpic}
\end{minipage}
}
\hspace{-0.12in}
\subfigure[$r=1~\mathrm{{\mu}m}$, $\lambda = 0.7~\mathrm{{\mu}m}$]{
\begin{minipage}[b]{0.32\linewidth}
\begin{overpic}[width=1.0\linewidth]{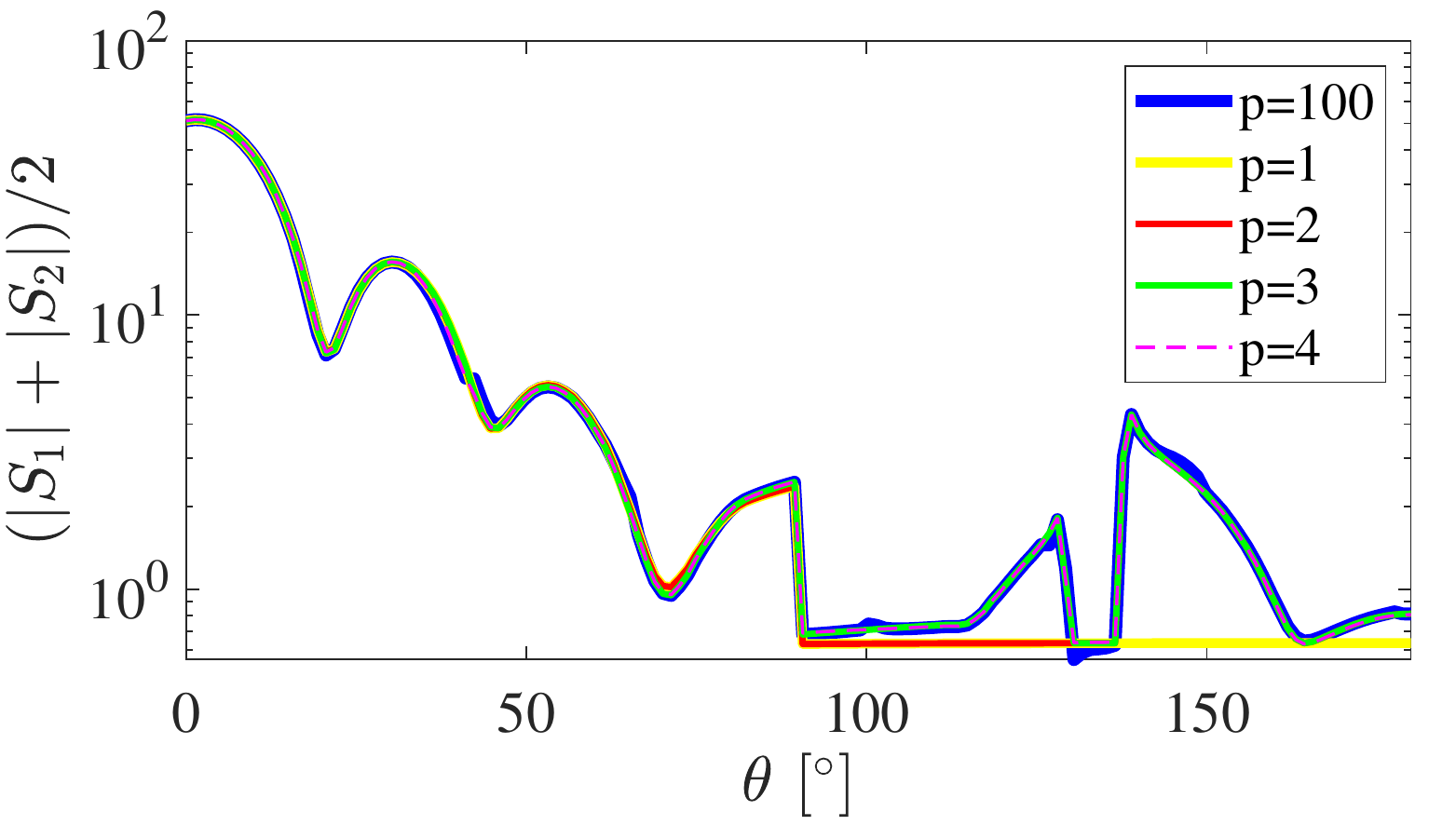}
\end{overpic}
\end{minipage}
}
\subfigure[$r=10~\mathrm{{\mu}m}$, $\lambda = 0.5~\mathrm{{\mu}m}$]{
\begin{minipage}[b]{0.32\linewidth}
\begin{overpic}[width=1.0\linewidth]{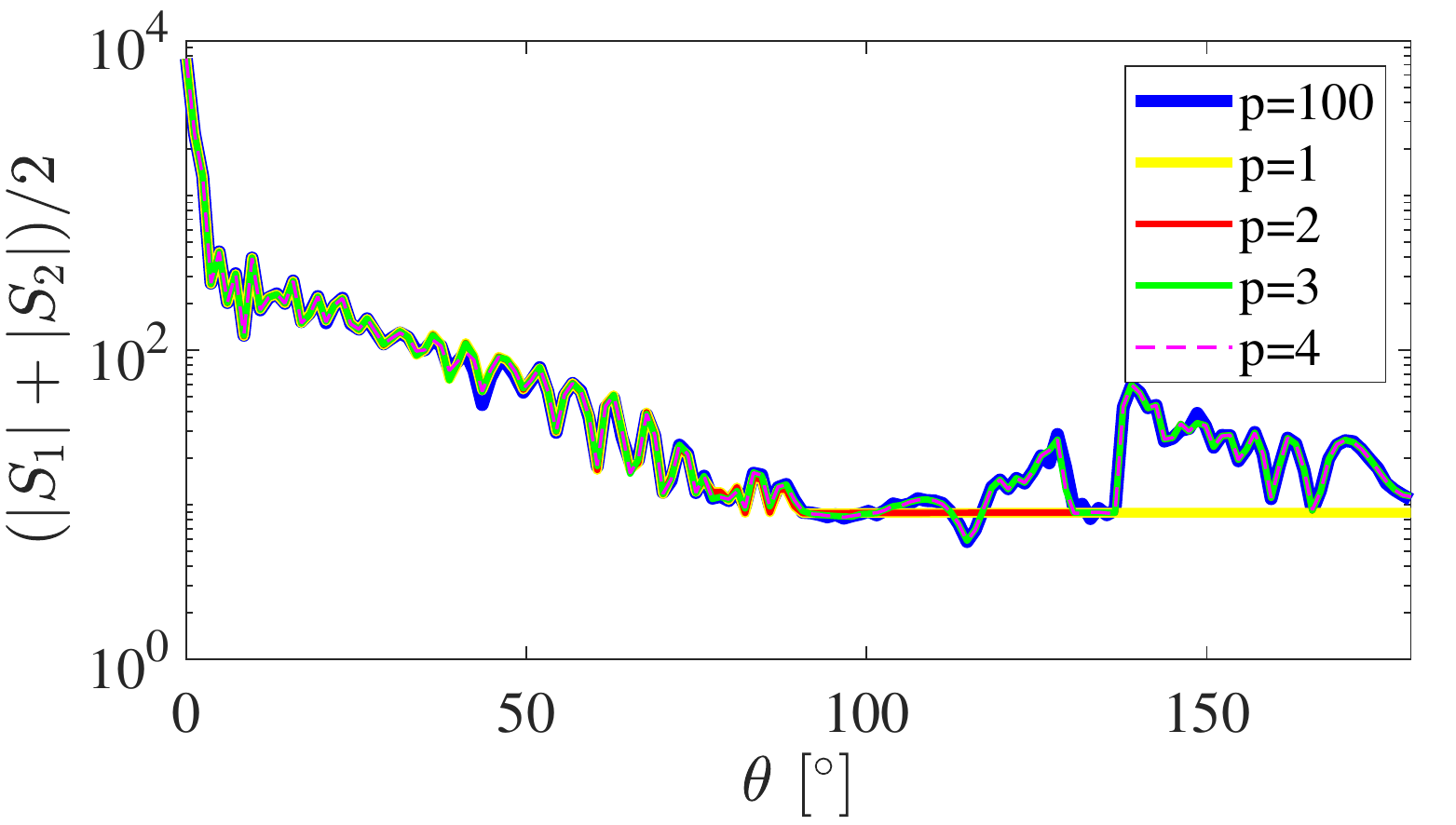}
\end{overpic}
\end{minipage}
}
\hspace{-0.12in}
\subfigure[$r=10~\mathrm{{\mu}m}$, $\lambda = 0.6~\mathrm{{\mu}m}$]{
\begin{minipage}[b]{0.32\linewidth}
\begin{overpic}[width=1.0\linewidth]{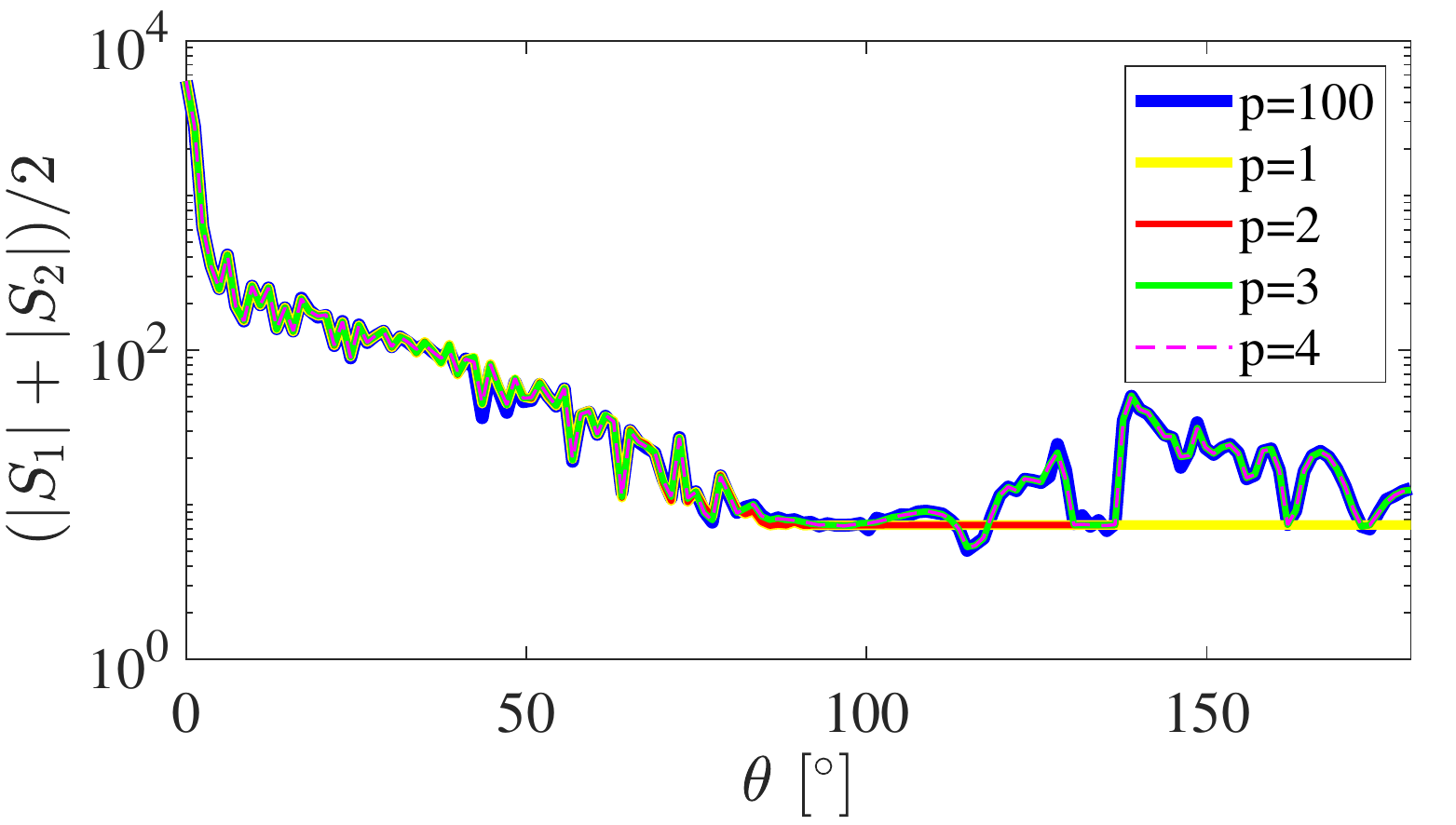}
\end{overpic}
\end{minipage}
}
\hspace{-0.12in}
\subfigure[$r=10~\mathrm{{\mu}m}$, $\lambda = 0.7~\mathrm{{\mu}m}$]{
\begin{minipage}[b]{0.32\linewidth}
\begin{overpic}[width=1.0\linewidth]{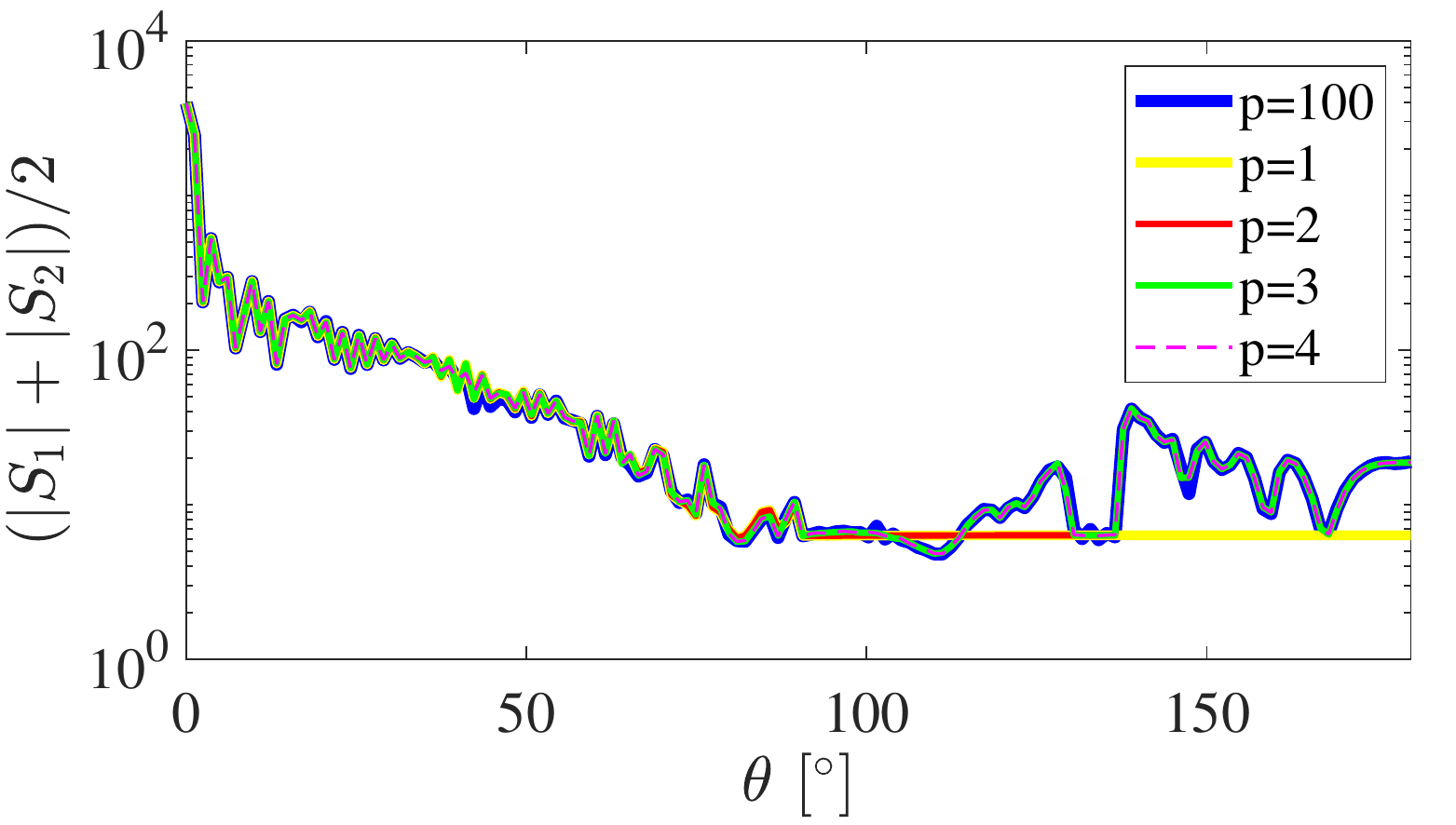}
\end{overpic}
\end{minipage}
}
\subfigure[$r=100~\mathrm{{\mu}m}$, $\lambda = 0.5~\mathrm{{\mu}m}$]{
\begin{minipage}[b]{0.32\linewidth}
\begin{overpic}[width=1.0\linewidth]{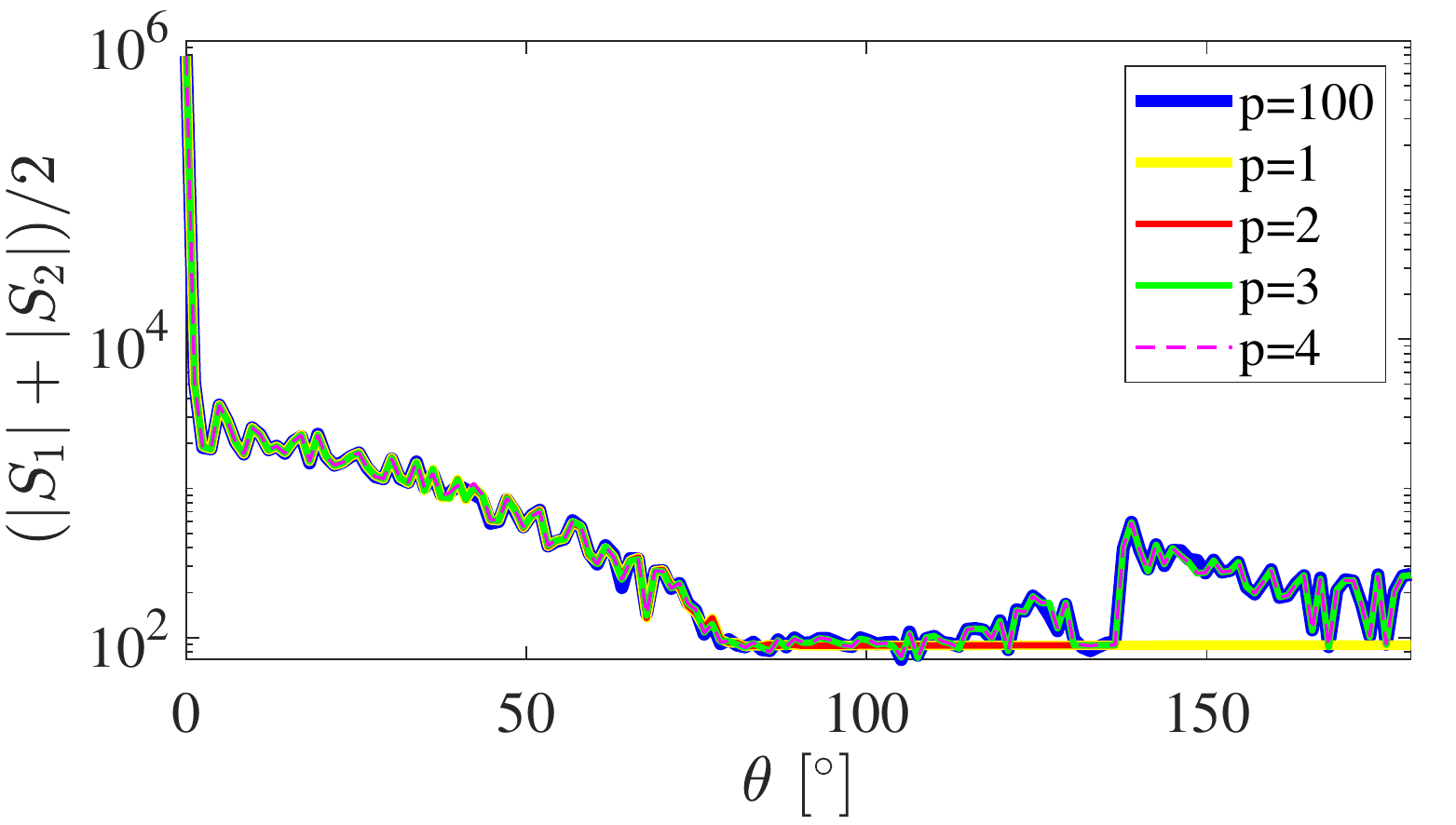}
\end{overpic}
\end{minipage}
}
\hspace{-0.12in}
\subfigure[$r=100~\mathrm{{\mu}m}$, $\lambda = 0.6~\mathrm{{\mu}m}$]{
\begin{minipage}[b]{0.32\linewidth}
\begin{overpic}[width=1.0\linewidth]{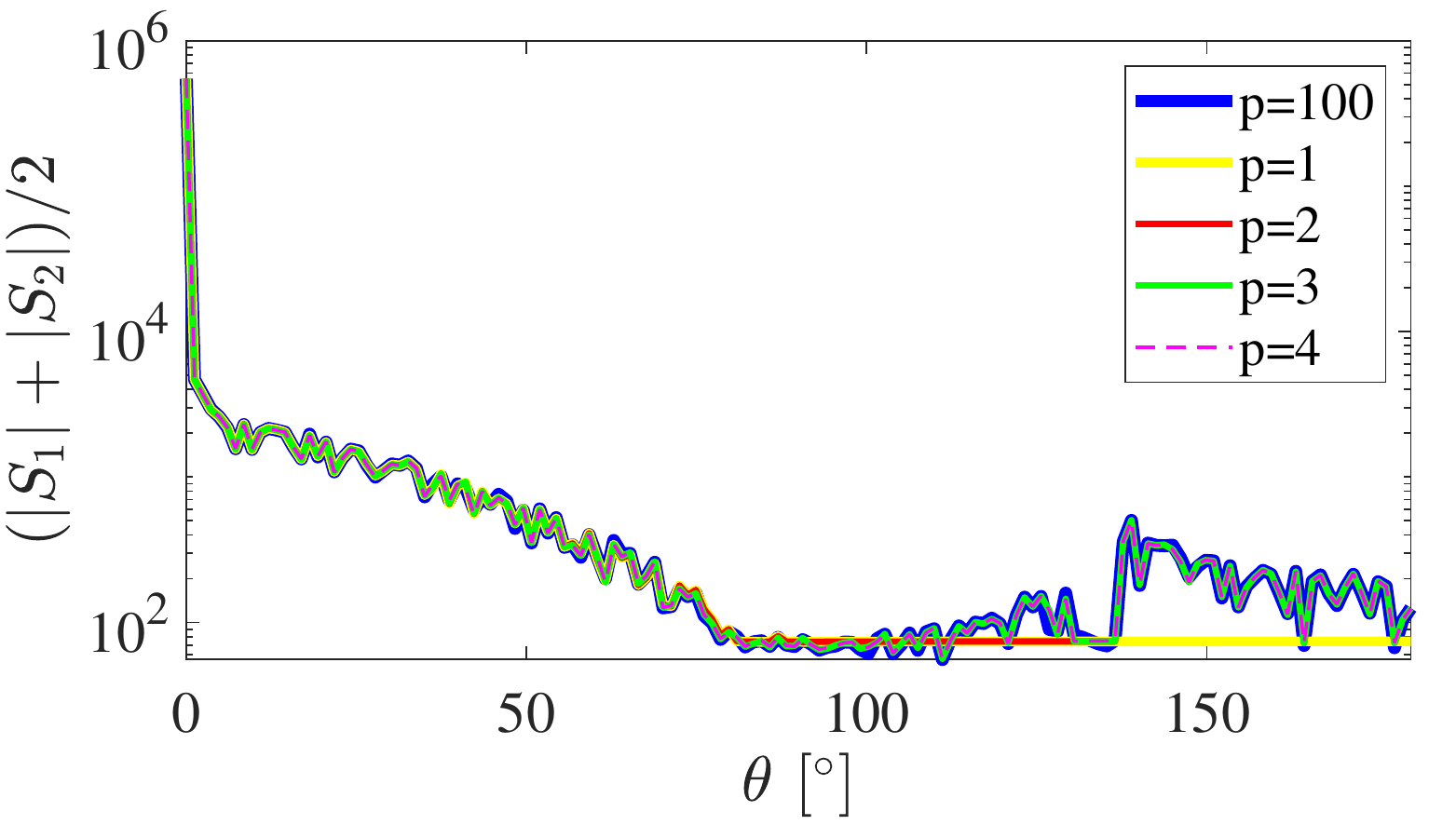}
\end{overpic}
\end{minipage}
}
\hspace{-0.12in}
\subfigure[$r=100~\mathrm{{\mu}m}$, $\lambda = 0.7~\mathrm{{\mu}m}$]{
\begin{minipage}[b]{0.32\linewidth}
\begin{overpic}[width=1.0\linewidth]{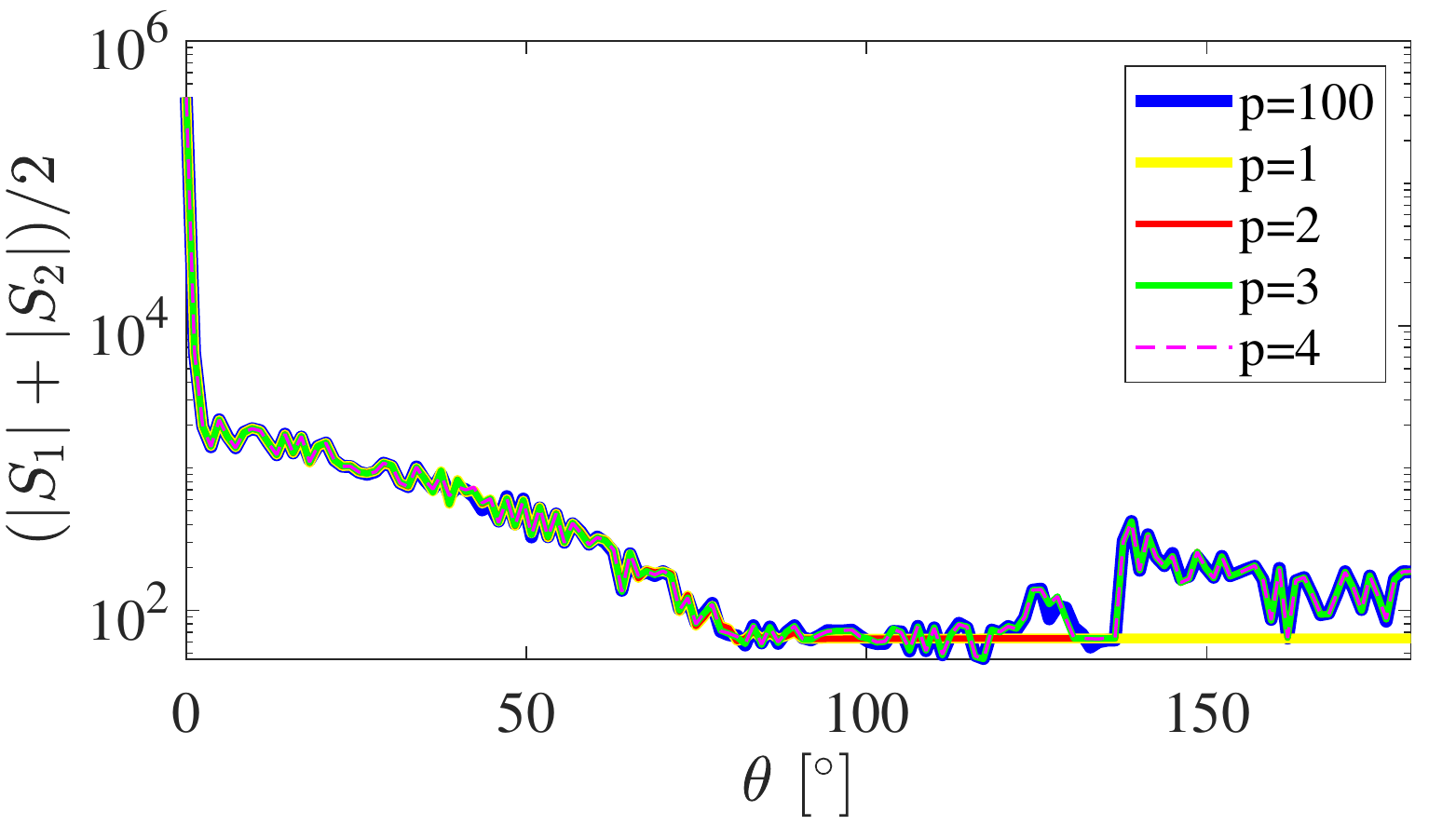}
\end{overpic}
\end{minipage}
}
\subfigure[$r=1000~\mathrm{{\mu}m}$, $\lambda = 0.5~\mathrm{{\mu}m}$]{
\begin{minipage}[b]{0.32\linewidth}
\begin{overpic}[width=1.0\linewidth]{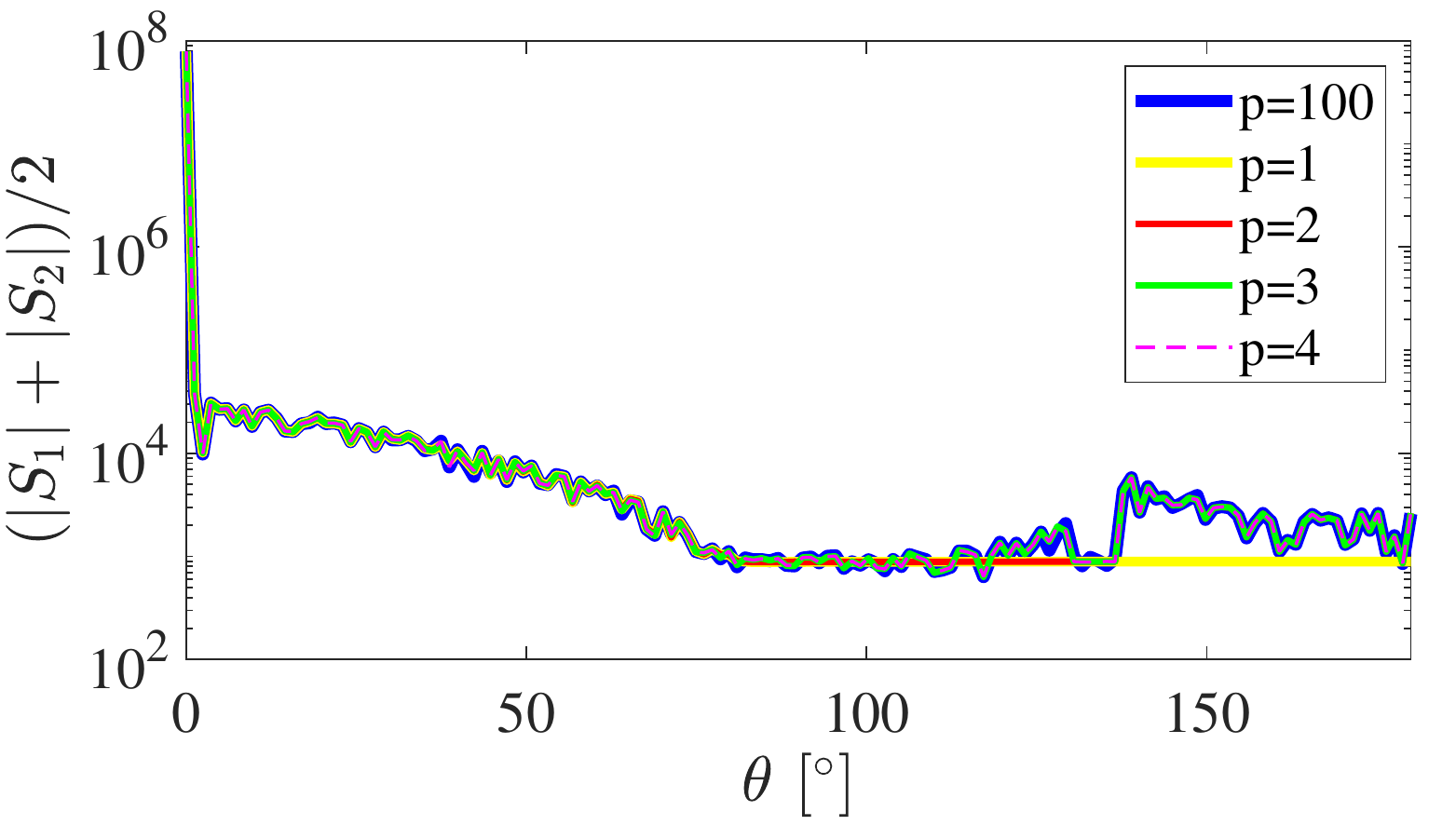}
\end{overpic}
\end{minipage}
}
\hspace{-0.12in}
\subfigure[$r=1000~\mathrm{{\mu}m}$, $\lambda = 0.6~\mathrm{{\mu}m}$]{
\begin{minipage}[b]{0.32\linewidth}
\begin{overpic}[width=1.0\linewidth]{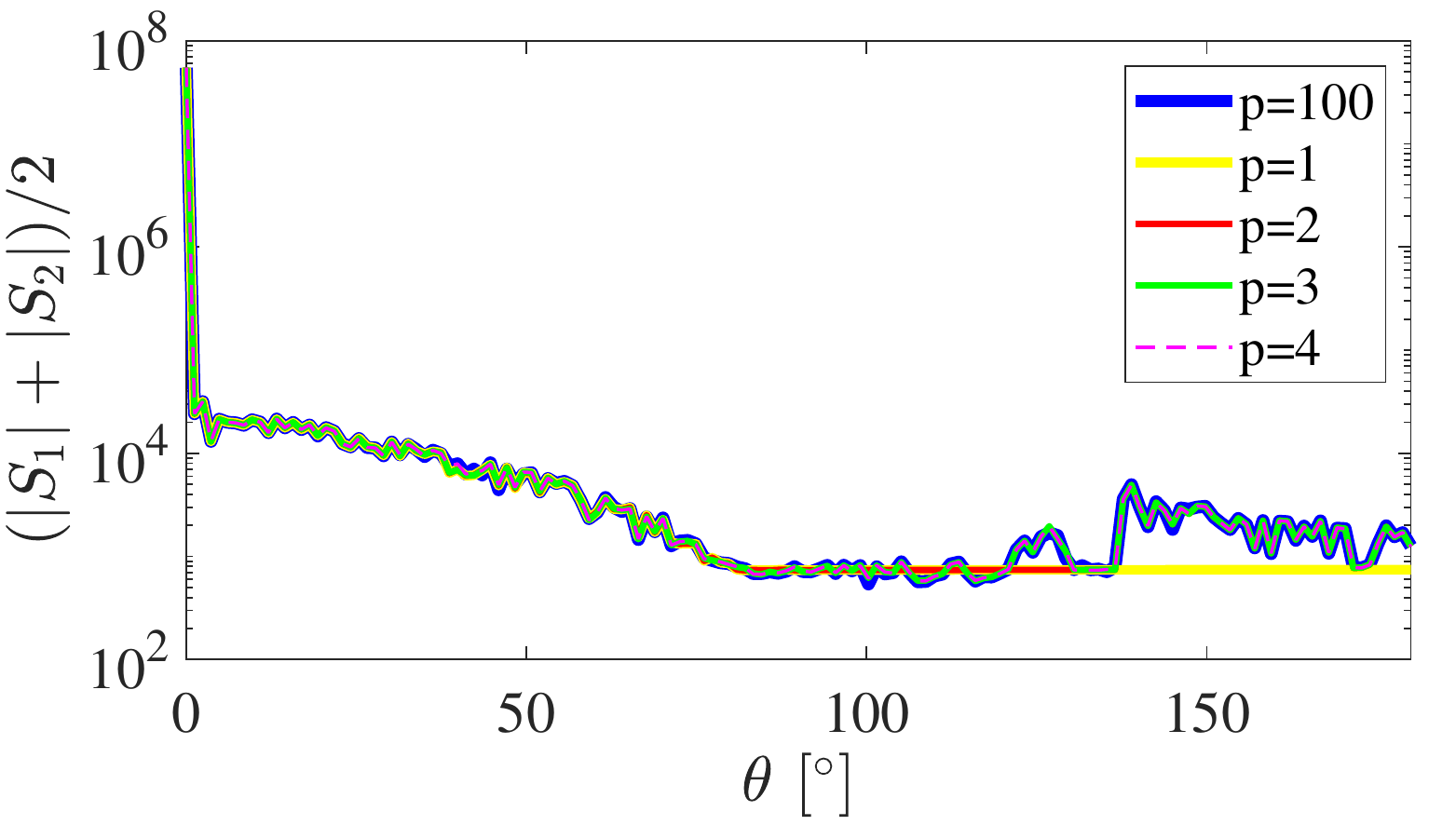}
\end{overpic}
\end{minipage}
}
\hspace{-0.12in}
\subfigure[$r=1000~\mathrm{{\mu}m}$, $\lambda = 0.7~\mathrm{{\mu}m}$]{
\begin{minipage}[b]{0.32\linewidth}
\begin{overpic}[width=1.0\linewidth]{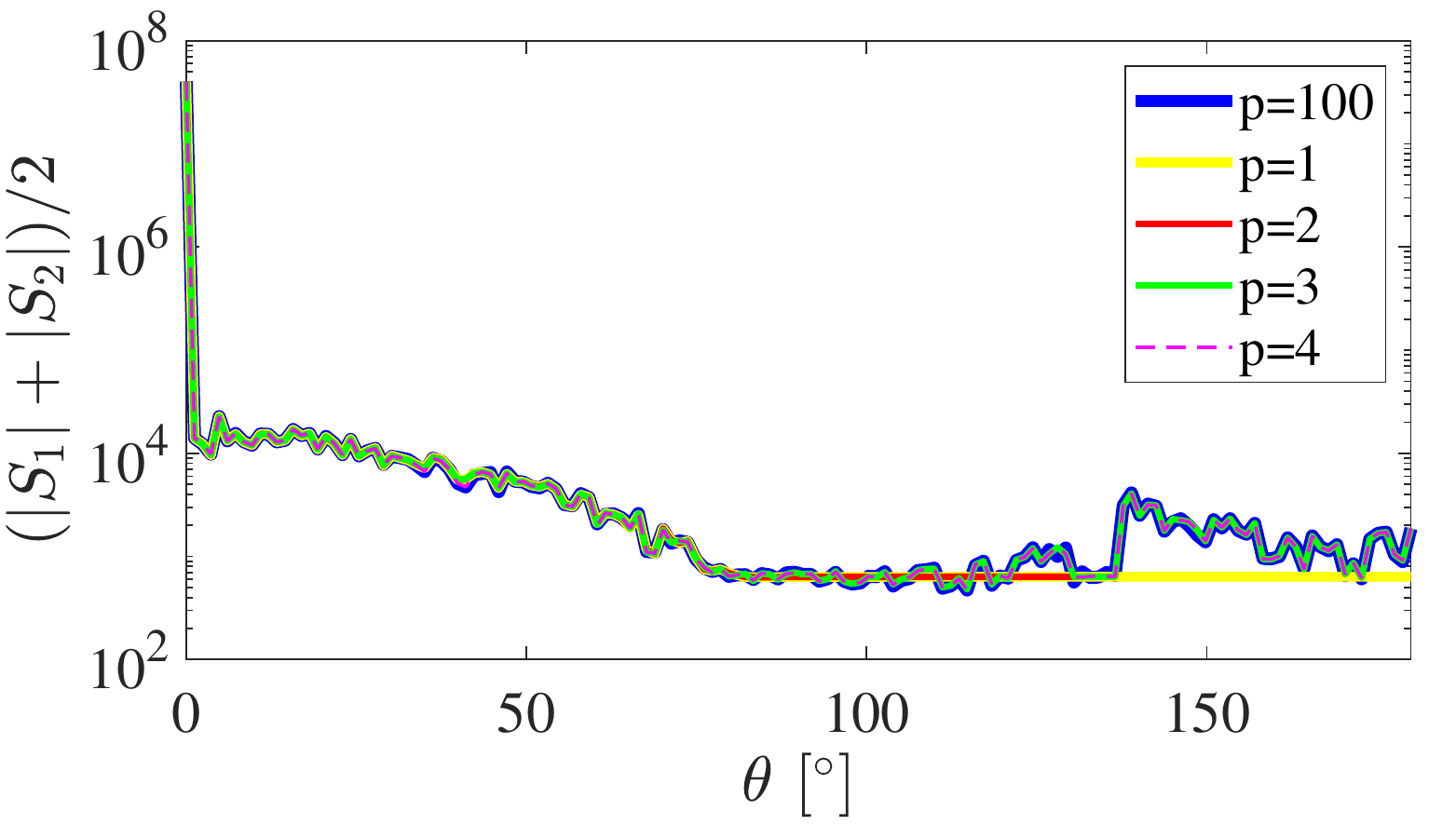}
\end{overpic}
\end{minipage}
}
\caption{\label{fig:goa_p} Visual comparisons of $(|S_1|+|S_2|)/2$ with increasing values of $p$ in GOA. Here, we show different combinations of particle size $r$ and wavelength $\lambda$, while the relative refractive index of the particle $\eta$ is set to 1.33.}
\end{figure*}

\section{More Comparisons between GOA and Lorenz-Mie Theory}
\begin{figure*}[t]
\centering
\subfigure{
\begin{minipage}[b]{0.19\linewidth}
\begin{overpic}[width=1.0\linewidth, trim={70px, 200px, 70px, 200px}, clip]{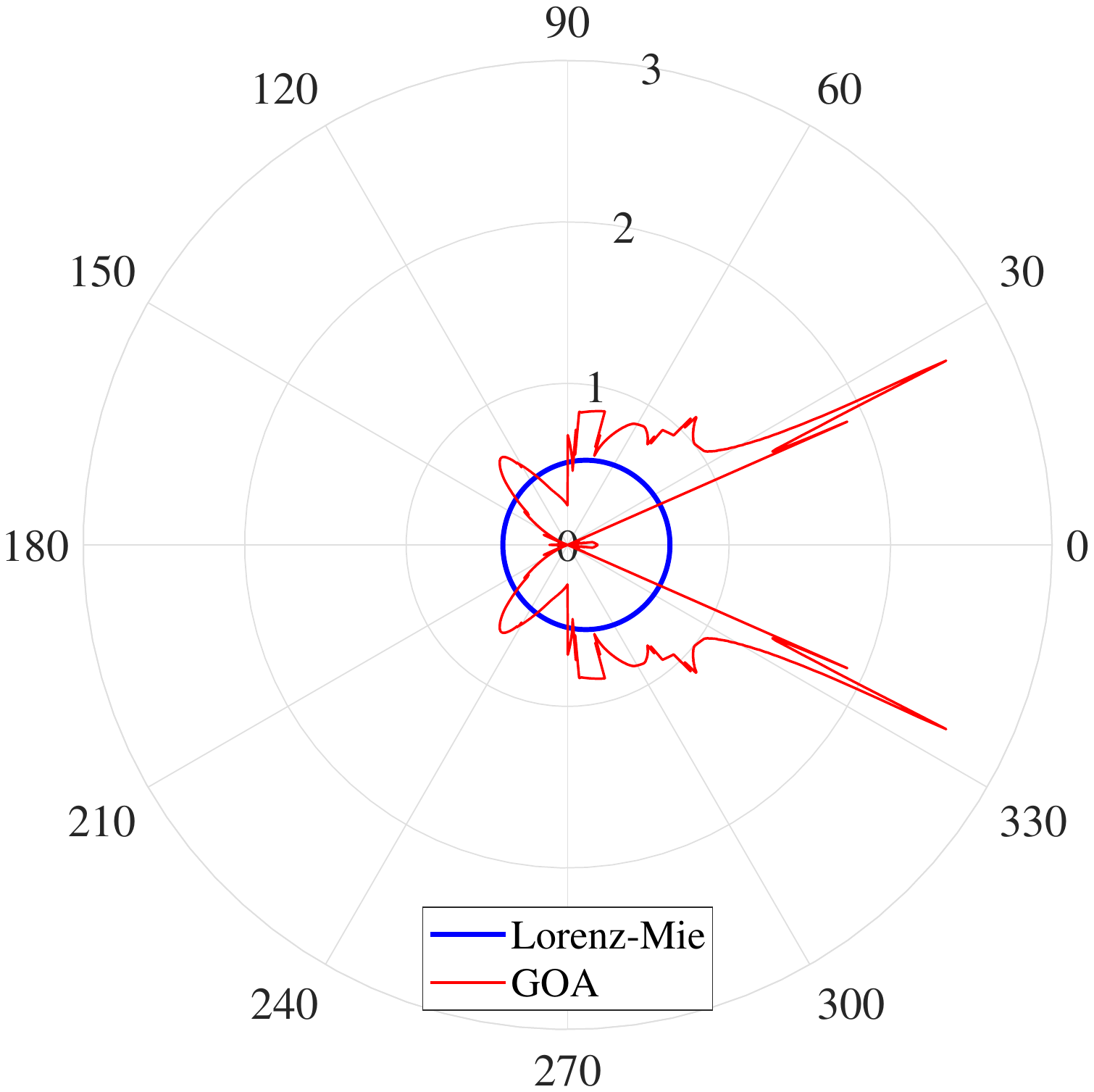}
\end{overpic}
\end{minipage}
}
\hspace{-0.1in}
\subfigure{
\begin{minipage}[b]{0.19\linewidth}
\begin{overpic}[width=1.0\linewidth, trim={70px, 200px, 70px, 200px}, clip]{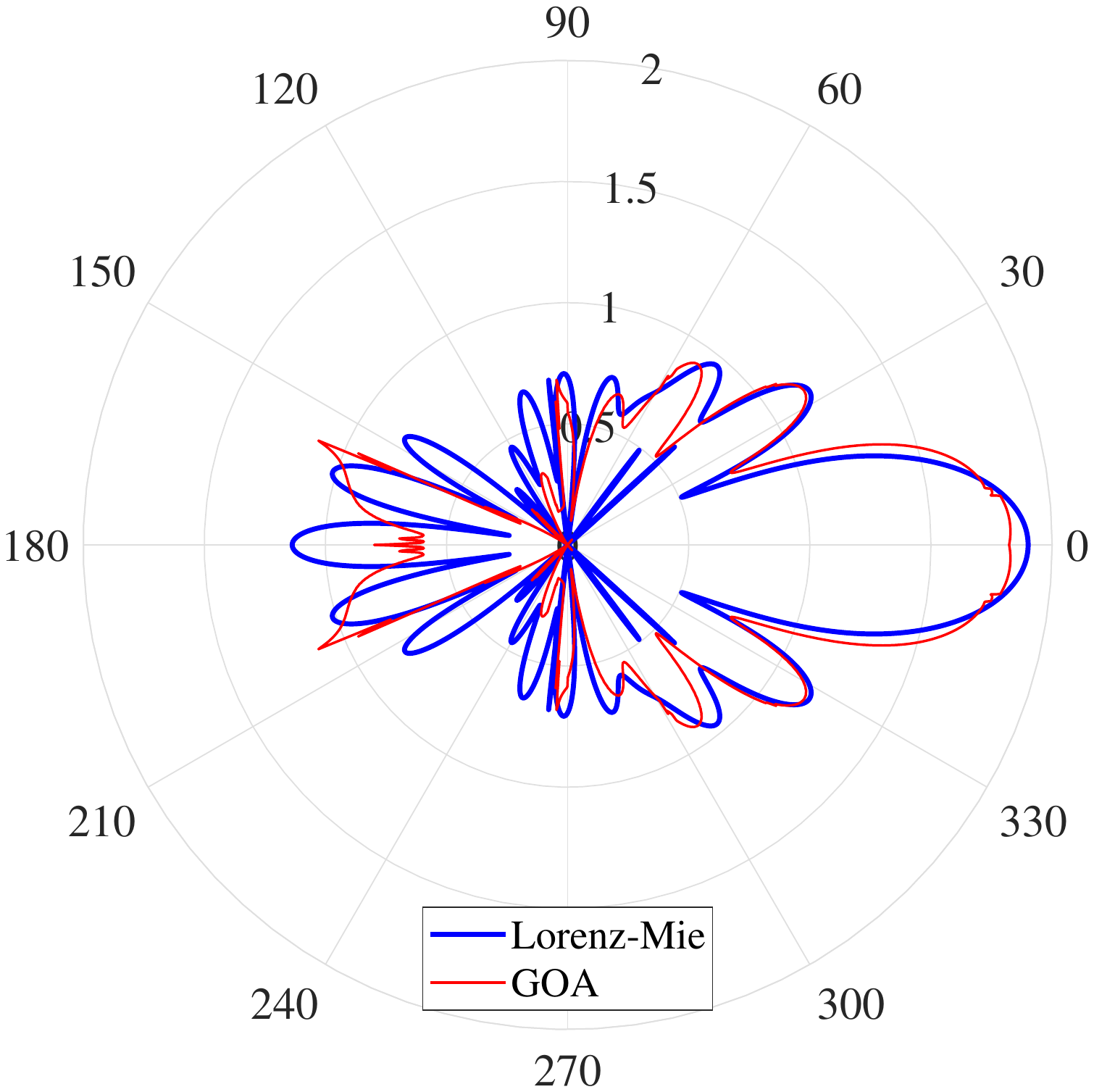}
\end{overpic}
\end{minipage}
}
\hspace{-0.1in}
\subfigure{
\begin{minipage}[b]{0.19\linewidth}
\begin{overpic}[width=1.0\linewidth, trim={70px, 200px, 70px, 200px}, clip]{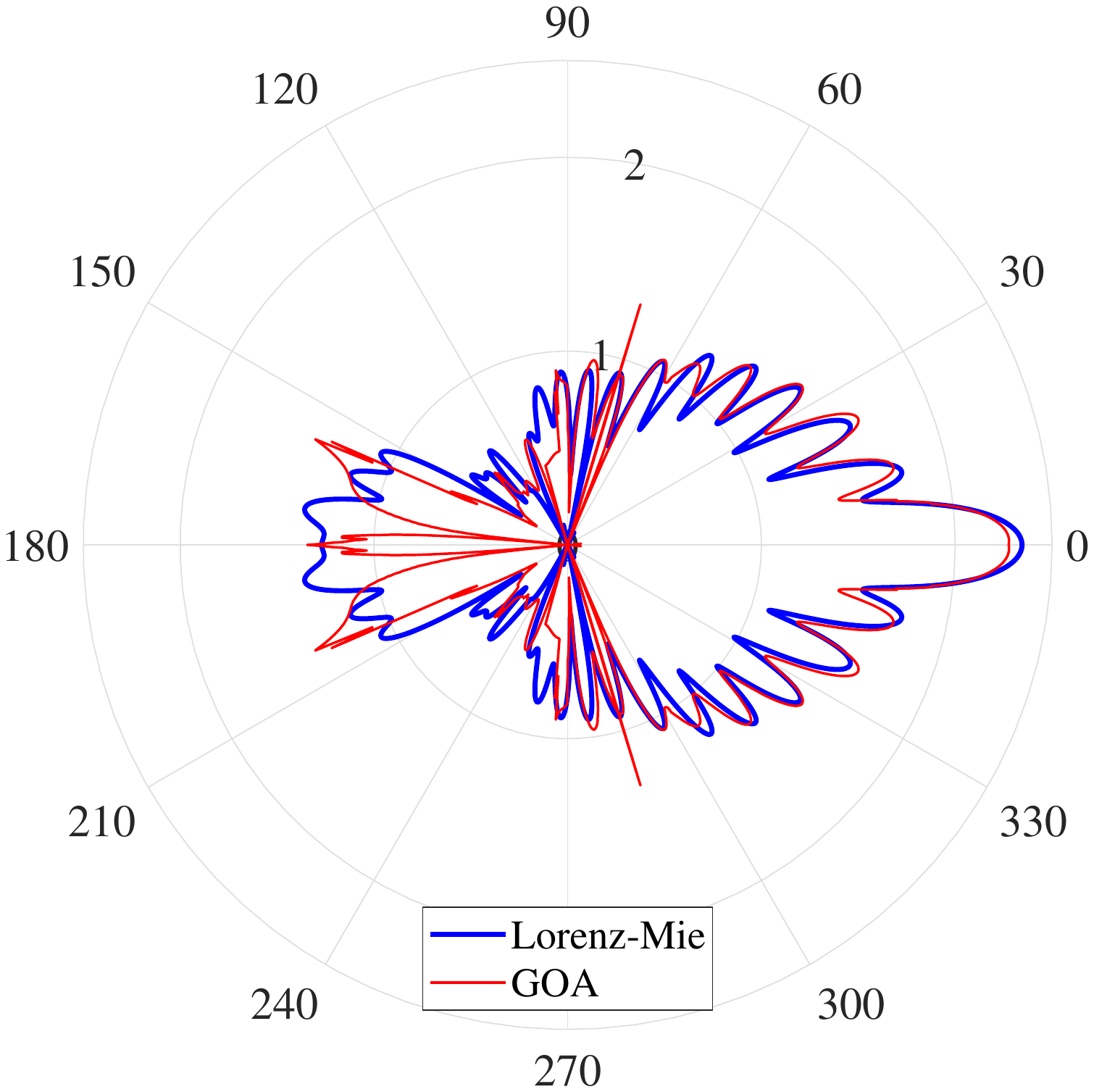}
\end{overpic}
\end{minipage}
}
\hspace{-0.1in}
\subfigure{
\begin{minipage}[b]{0.19\linewidth}
\begin{overpic}[width=1.0\linewidth, trim={70px, 200px, 70px, 200px}, clip]{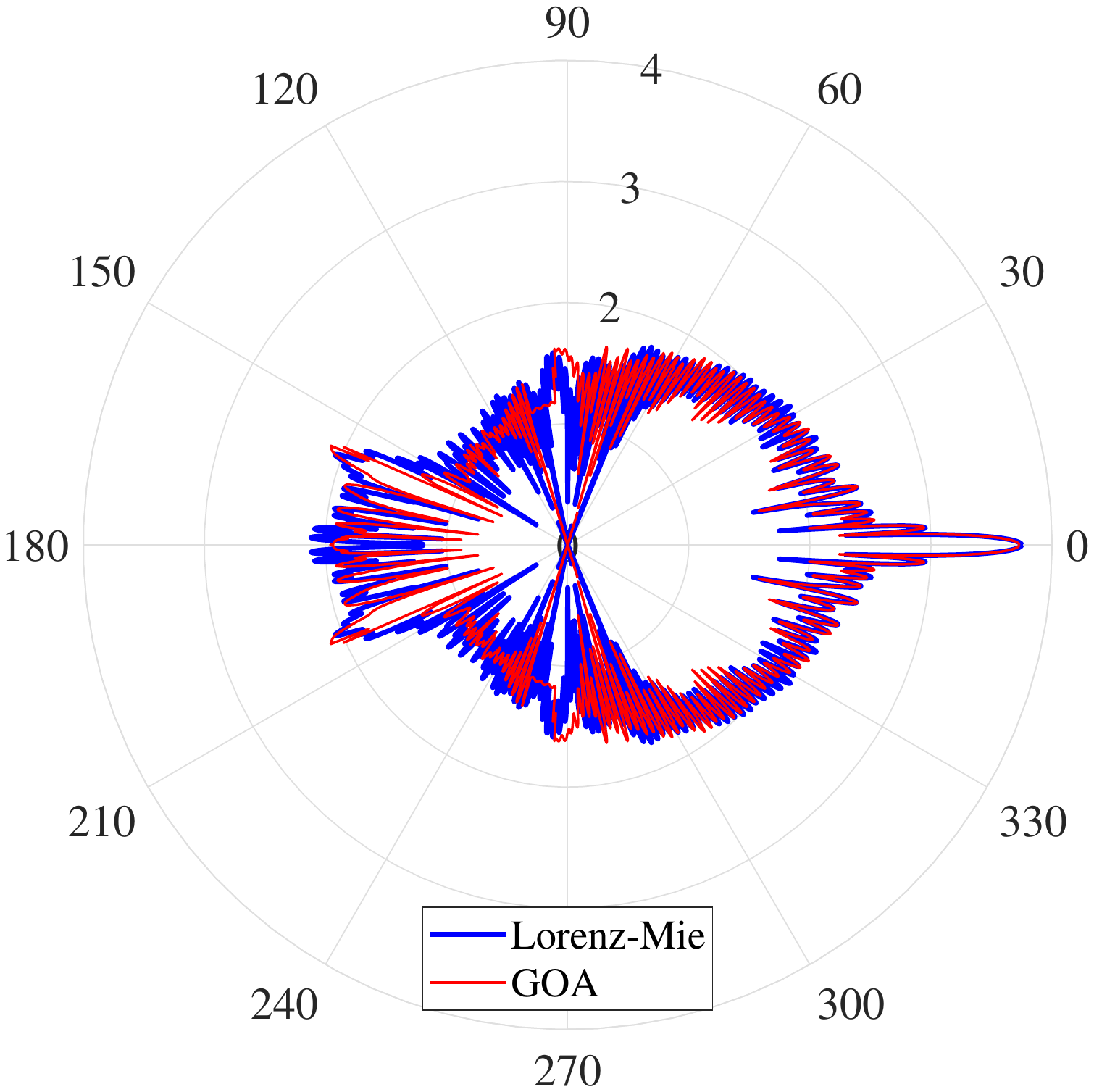}
\end{overpic}
\end{minipage}
}
\hspace{-0.1in}
\subfigure{
\begin{minipage}[b]{0.19\linewidth}
\begin{overpic}[width=1.0\linewidth, trim={70px, 200px, 70px, 200px}, clip]{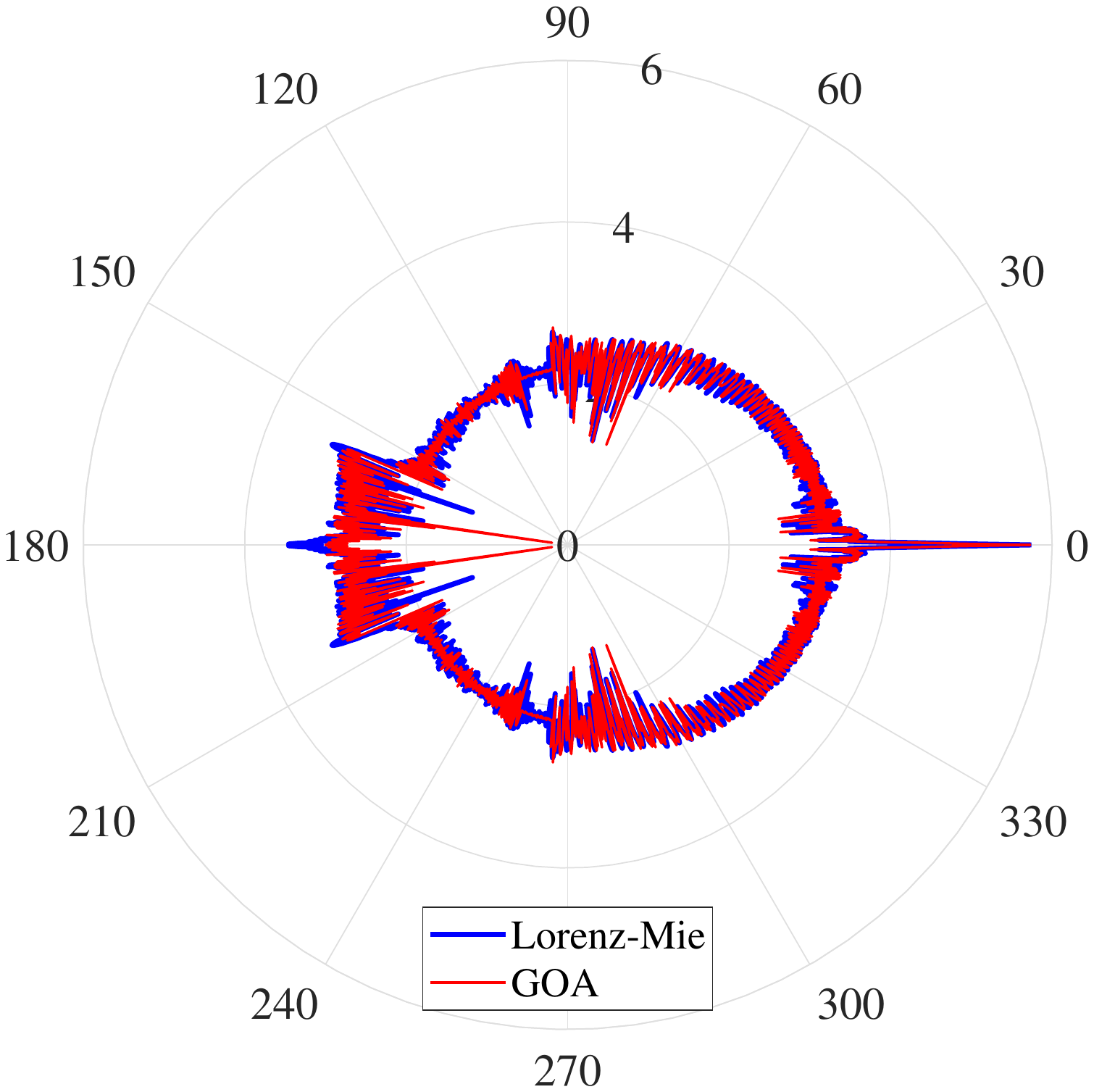}
\end{overpic}
\end{minipage}
}
\setcounter{subfigure}{0}
\subfigure[$r=0.1~\mathrm{{\mu}m}$]{
\begin{minipage}[b]{0.19\linewidth}
\begin{overpic}[width=1.0\linewidth, trim={70px, 200px, 70px, 200px}, clip]{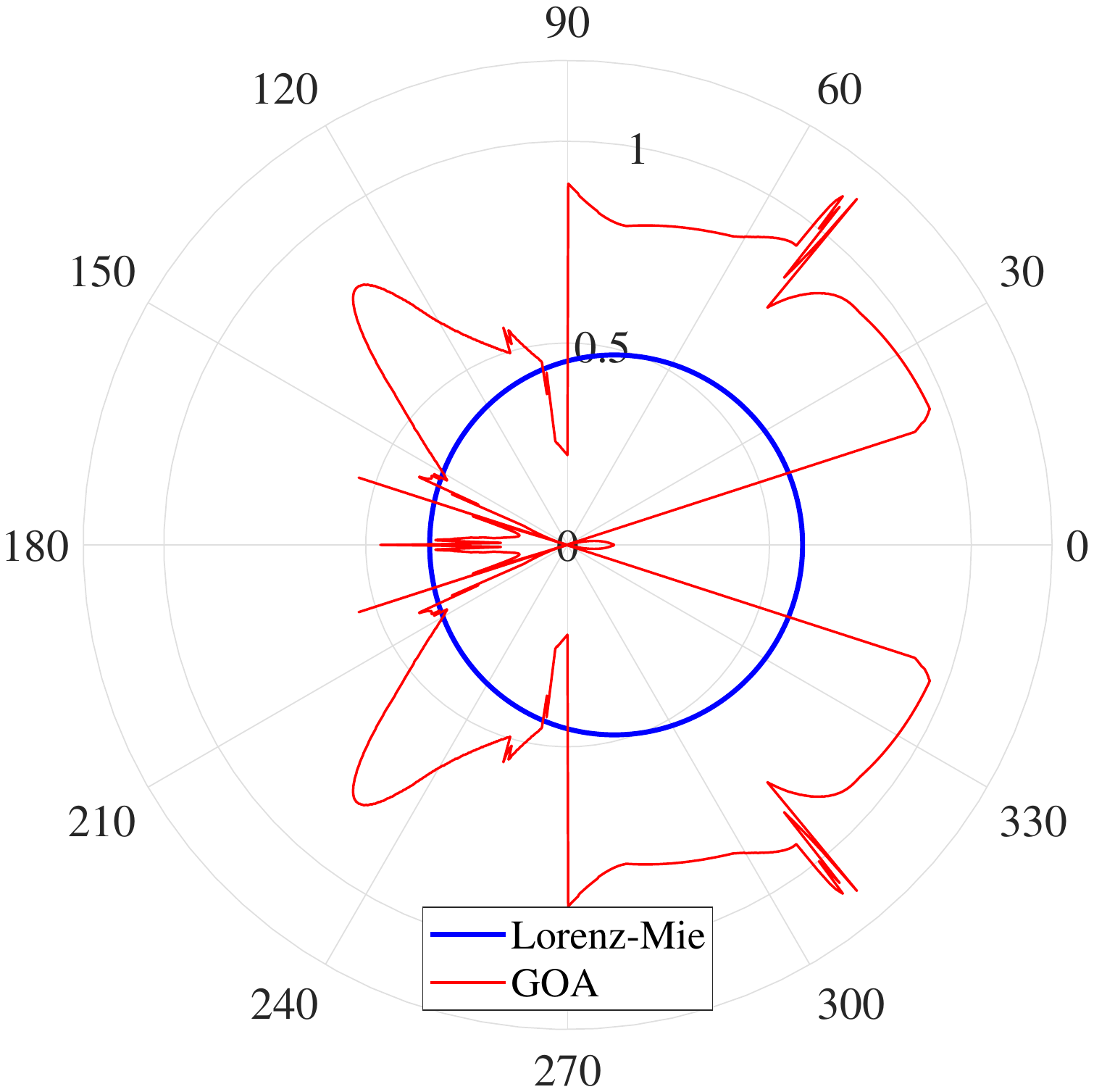}
\end{overpic}
\end{minipage}
}
\hspace{-0.1in}
\subfigure[$r=1~\mathrm{{\mu}m}$]{
\begin{minipage}[b]{0.19\linewidth}
\begin{overpic}[width=1.0\linewidth, trim={70px, 200px, 70px, 200px}, clip]{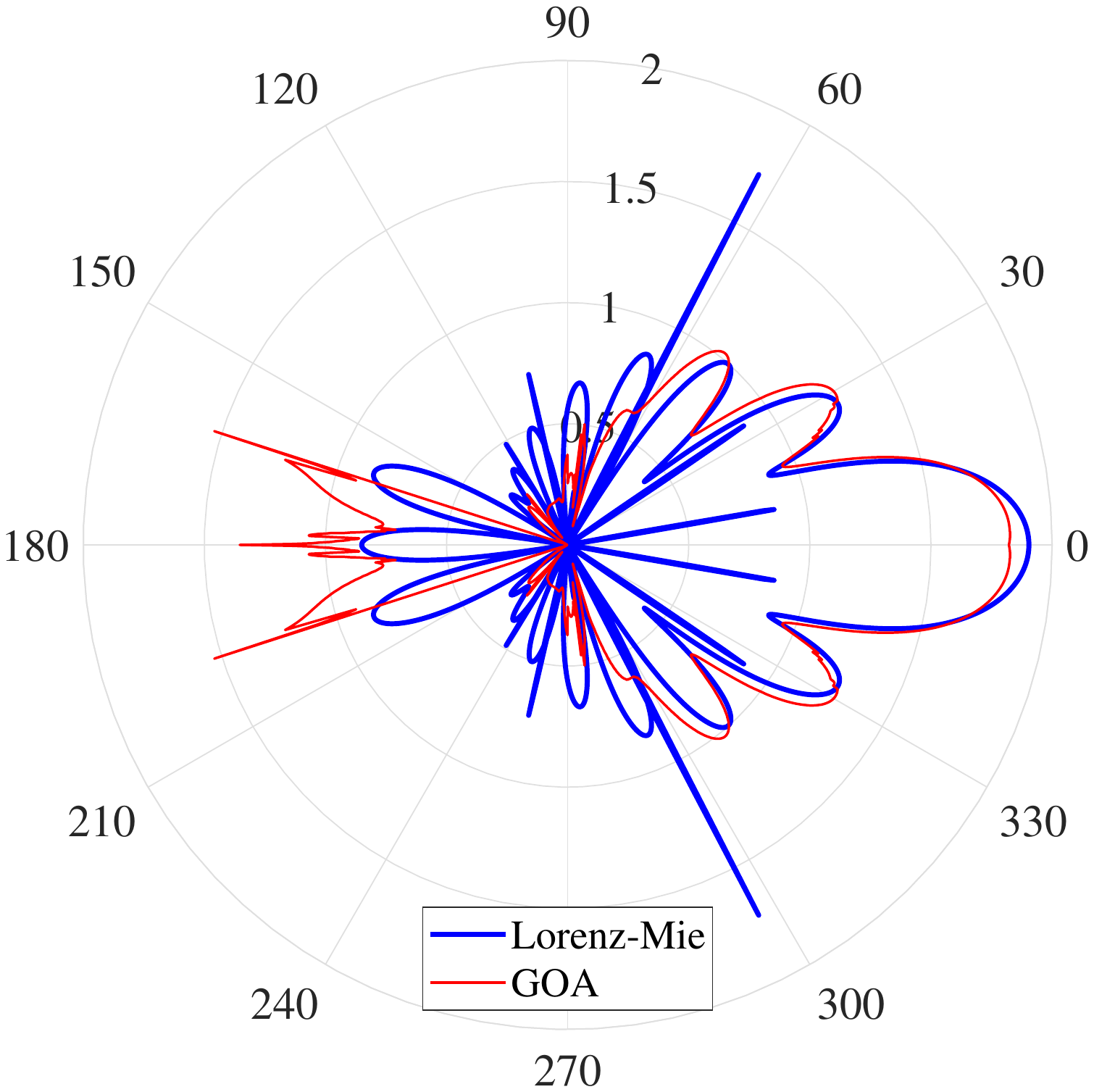}
\end{overpic}
\end{minipage}
}
\hspace{-0.1in}
\subfigure[$r=2~\mathrm{{\mu}m}$]{
\begin{minipage}[b]{0.19\linewidth}
\begin{overpic}[width=1.0\linewidth, trim={70px, 200px, 70px, 200px}, clip]{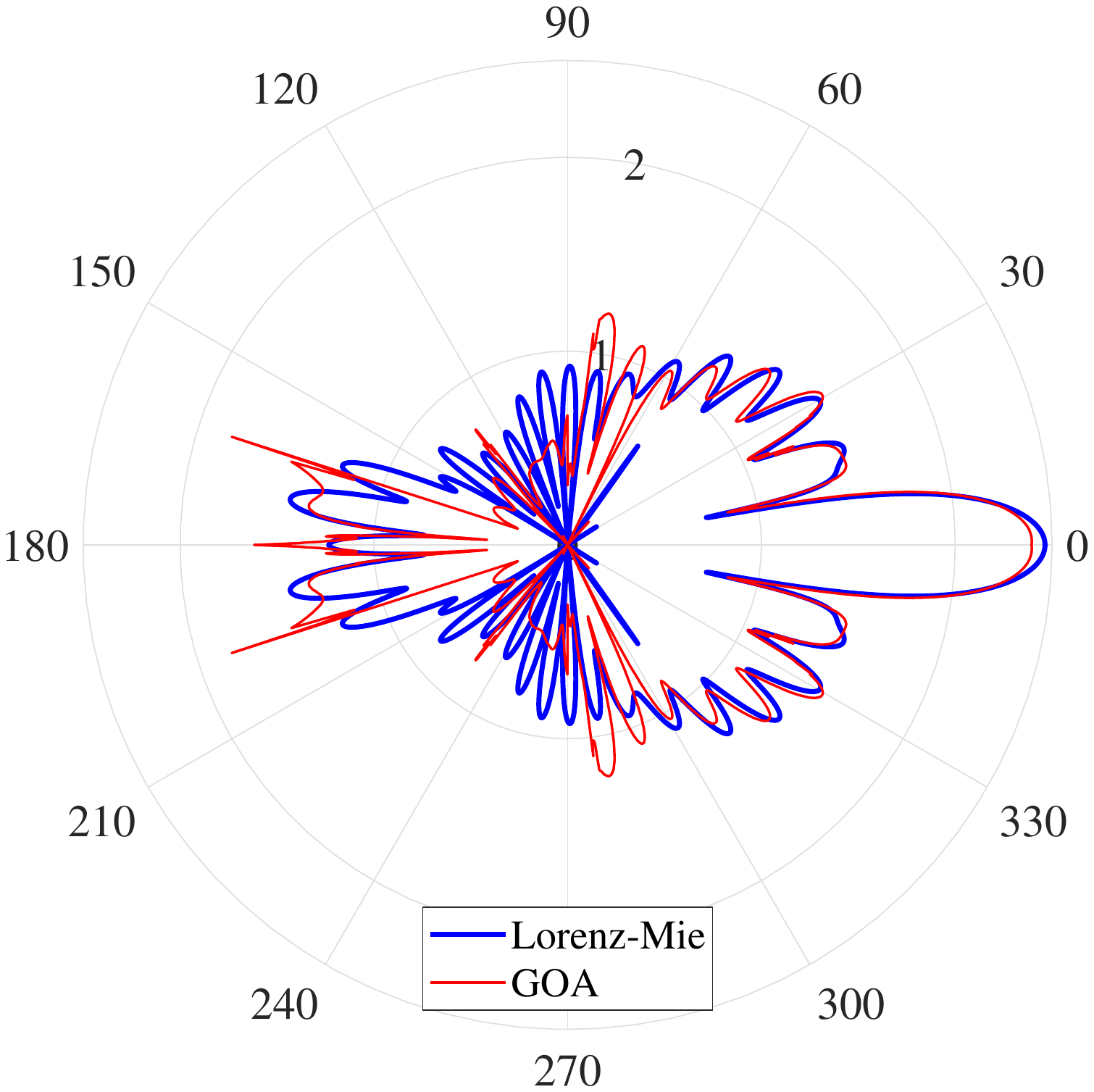}
\end{overpic}
\end{minipage}
}
\hspace{-0.1in}
\subfigure[$r=10~\mathrm{{\mu}m}$]{
\begin{minipage}[b]{0.19\linewidth}
\begin{overpic}[width=1.0\linewidth, trim={70px, 200px, 70px, 200px}, clip]{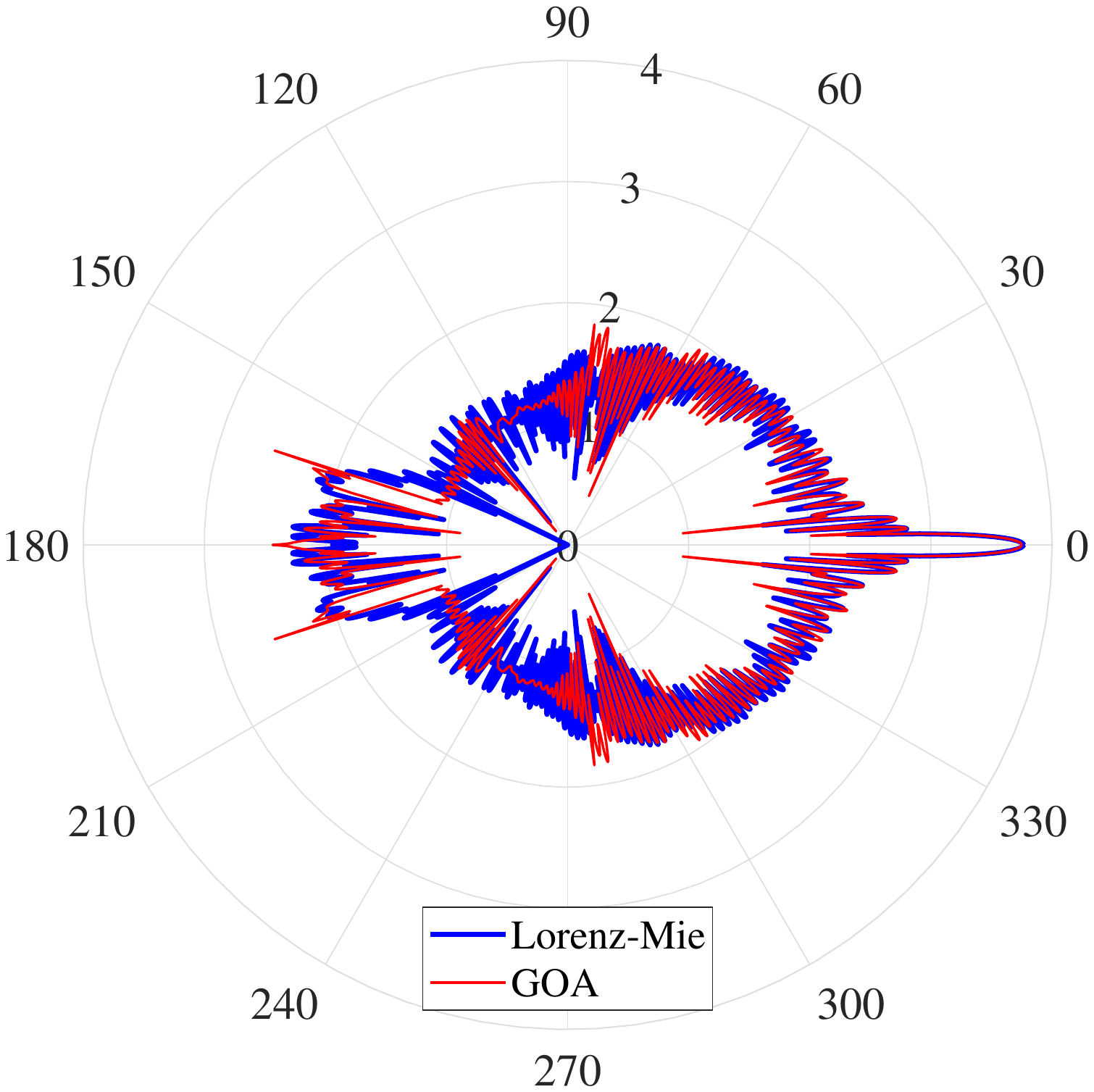}
\end{overpic}
\end{minipage}
}
\hspace{-0.1in}
\subfigure[$r=100~\mathrm{{\mu}m}$]{
\begin{minipage}[b]{0.19\linewidth}
\begin{overpic}[width=1.0\linewidth, trim={70px, 200px, 70px, 200px}, clip]{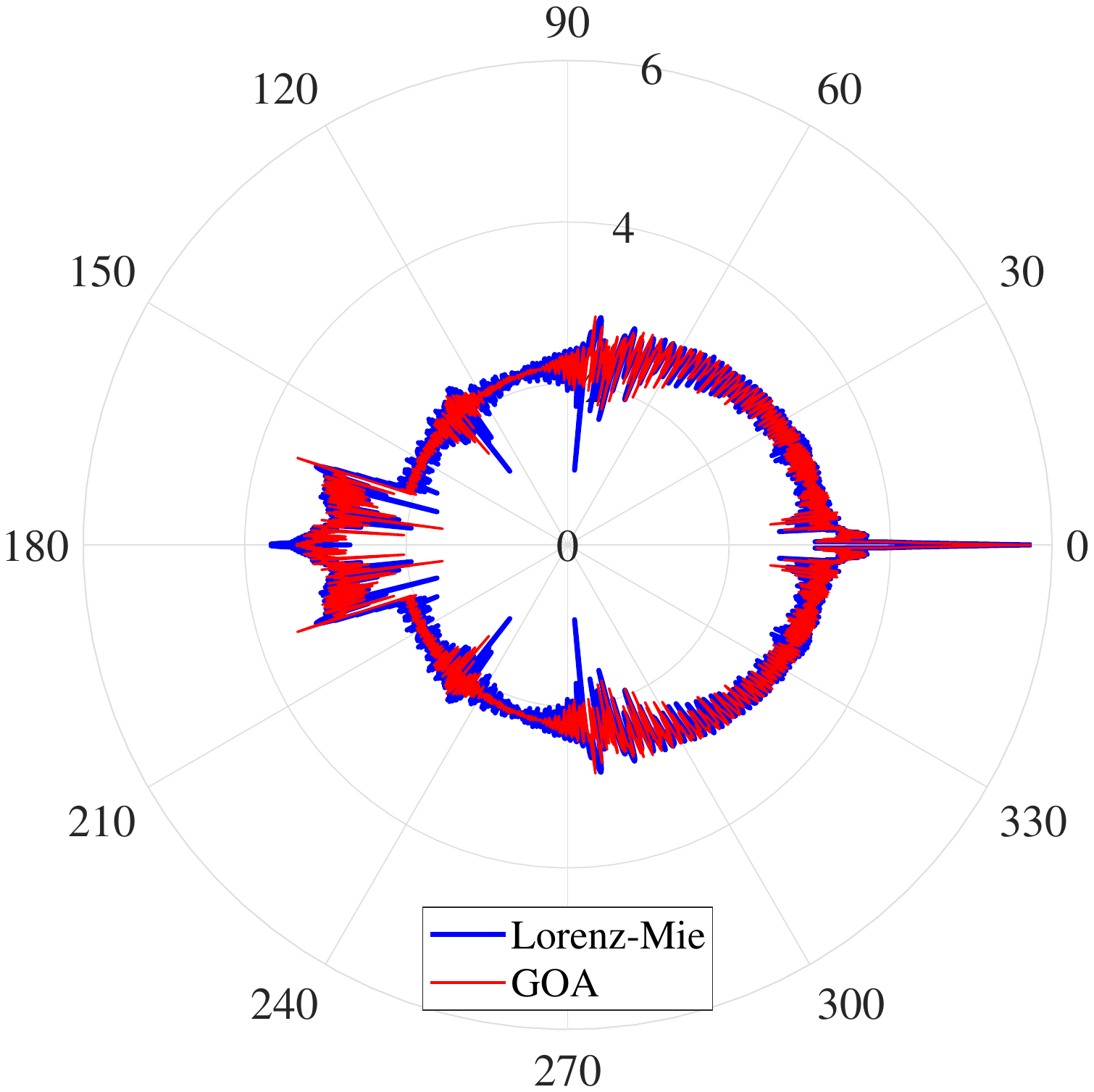}
\end{overpic}
\end{minipage}
}
\caption{\label{fig:goa_mie_1} Visual comparisons of $\log |S_1|$ by Lorenz-Mie calculations (blue curves) with those by GOA (red curves). First row: $\eta = 1.49$ and $\lambda = 0.6~\mathrm{{\mu}m}$. Second row: $\eta = 1.56$ and $\lambda = 0.6~\mathrm{{\mu}m}$. The particle radius is set to $r=0.1, 1, 2, 10$ and $100~\mathrm{{\mu}m}$, respectively.}
\end{figure*}

\begin{figure*}[t]
\centering
\subfigure{
\begin{minipage}[b]{0.19\linewidth}
\begin{overpic}[width=1.0\linewidth, trim={70px, 200px, 70px, 200px}, clip]{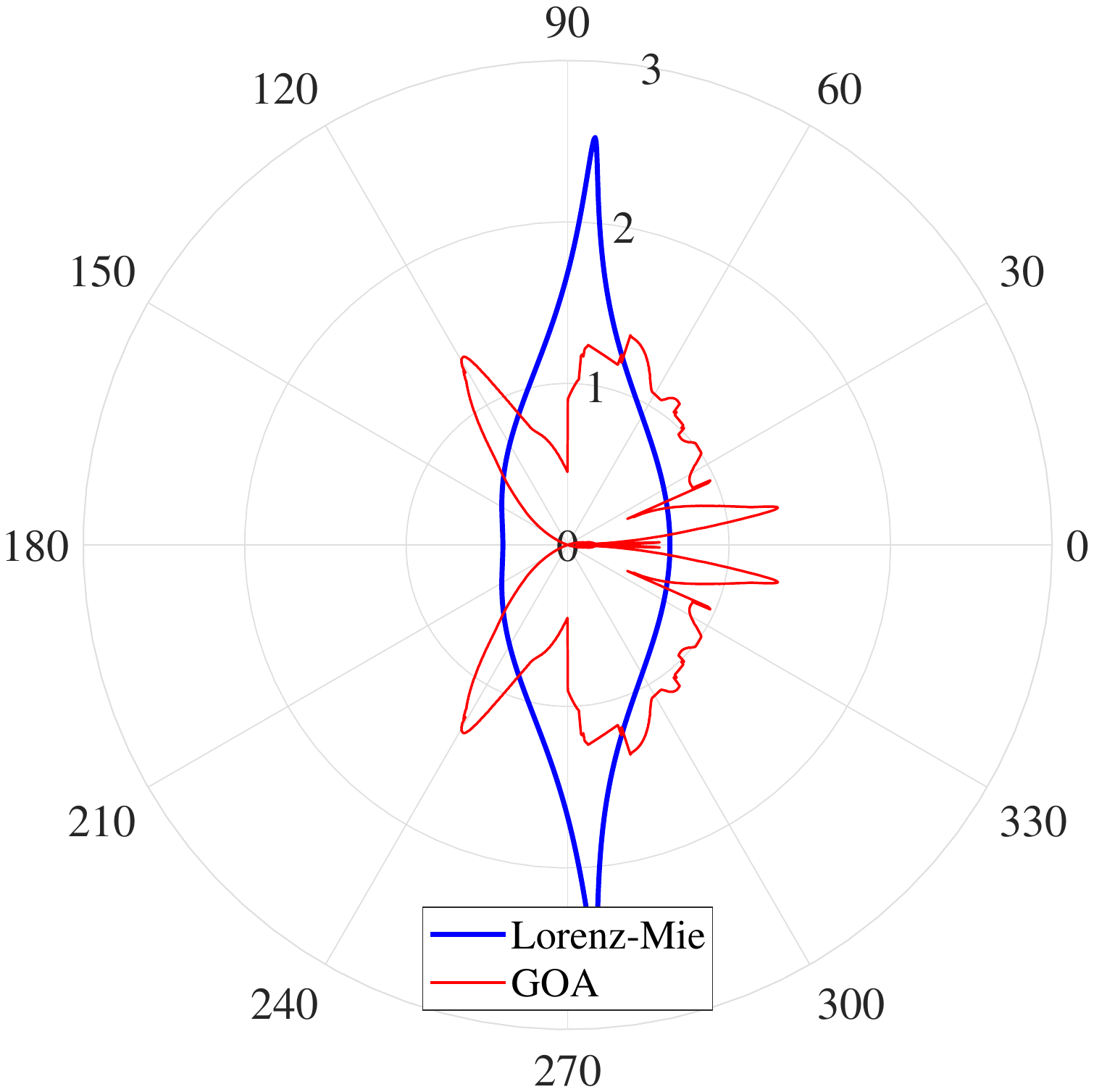}
\end{overpic}
\end{minipage}
}
\hspace{-0.1in}
\subfigure{
\begin{minipage}[b]{0.19\linewidth}
\begin{overpic}[width=1.0\linewidth, trim={70px, 200px, 70px, 200px}, clip]{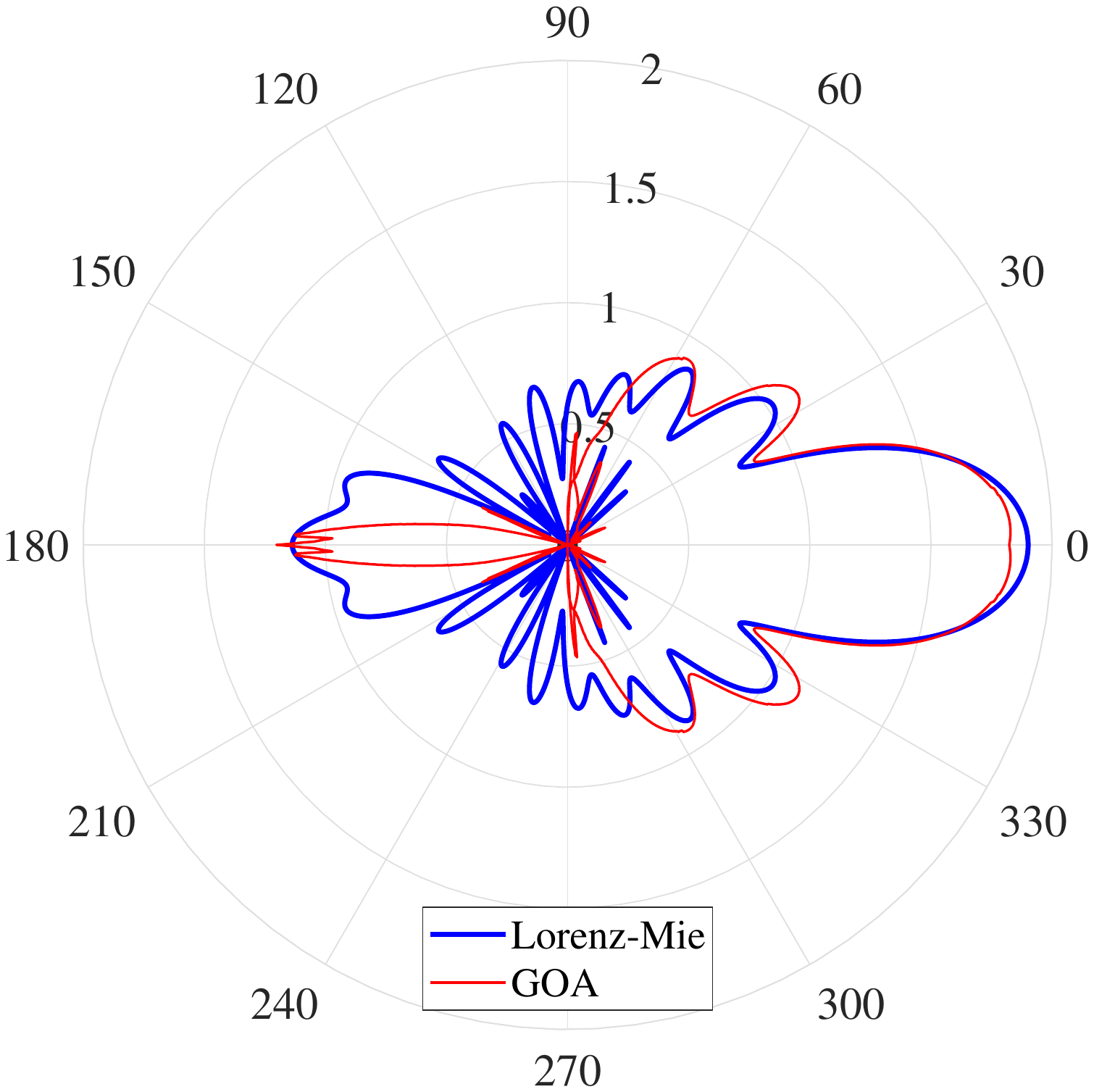}
\end{overpic}
\end{minipage}
}
\hspace{-0.1in}
\subfigure{
\begin{minipage}[b]{0.19\linewidth}
\begin{overpic}[width=1.0\linewidth, trim={70px, 200px, 70px, 200px}, clip]{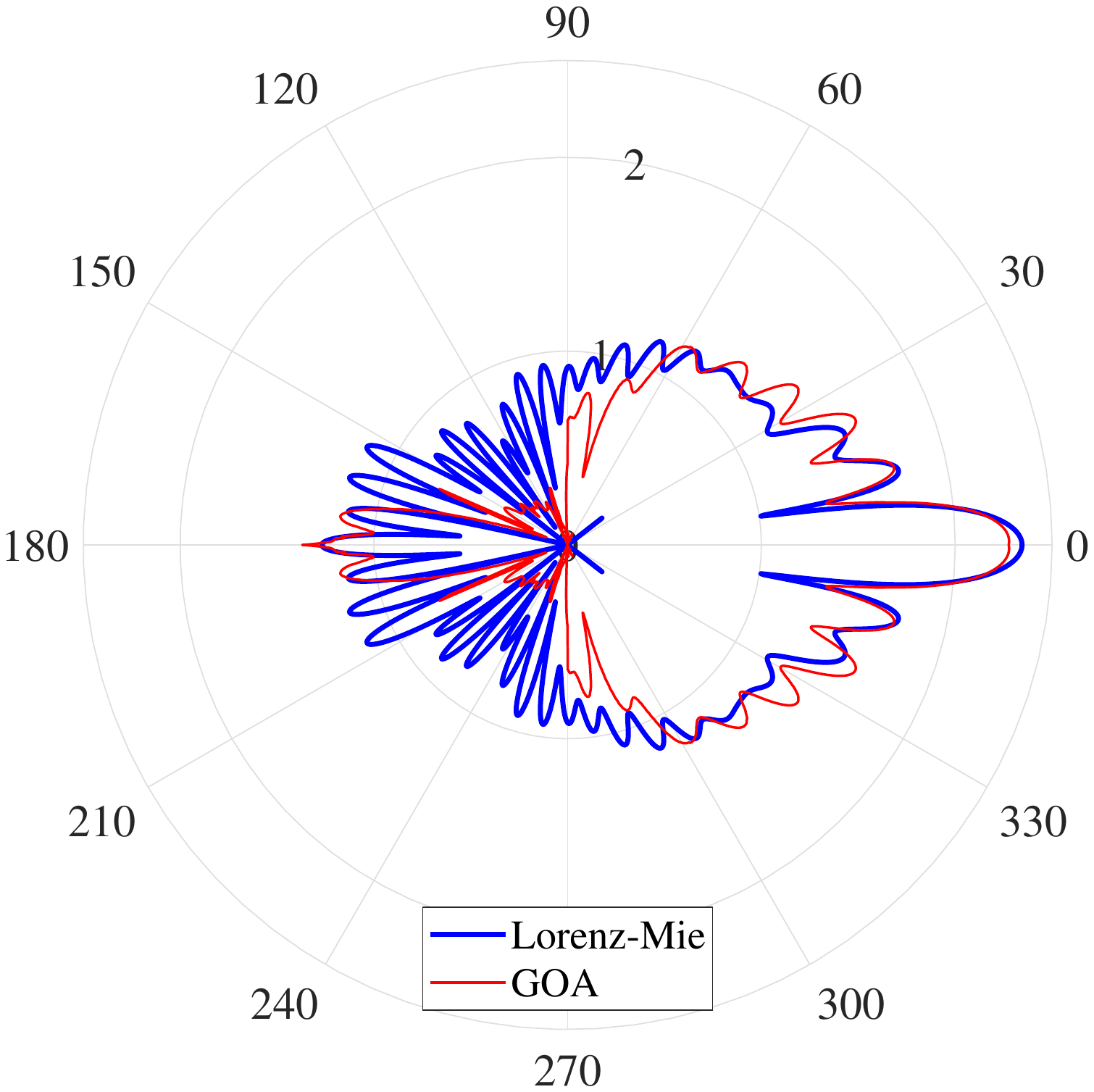}
\end{overpic}
\end{minipage}
}
\hspace{-0.1in}
\subfigure{
\begin{minipage}[b]{0.19\linewidth}
\begin{overpic}[width=1.0\linewidth, trim={70px, 200px, 70px, 200px}, clip]{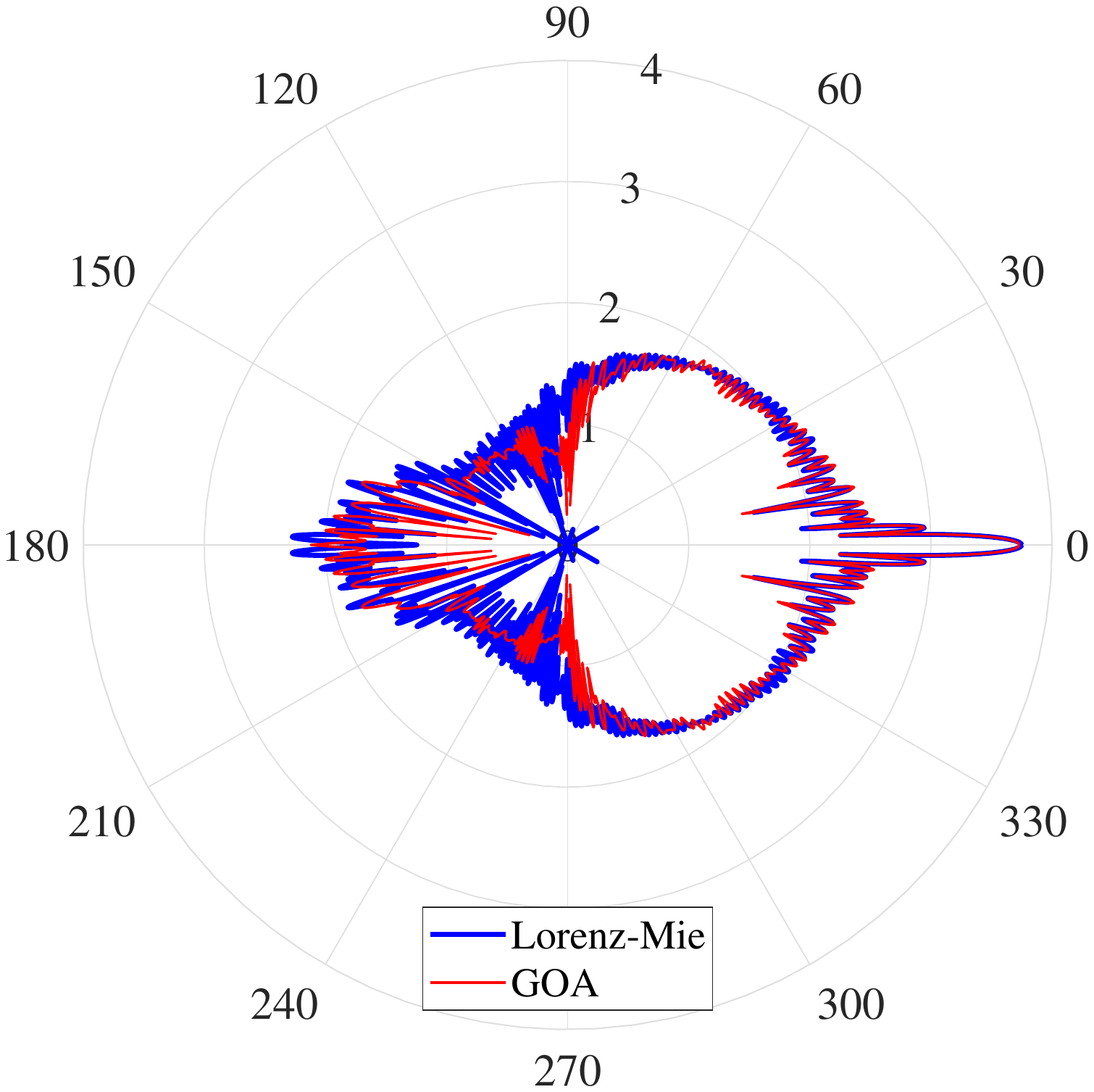}
\end{overpic}
\end{minipage}
}
\hspace{-0.1in}
\subfigure{
\begin{minipage}[b]{0.19\linewidth}
\begin{overpic}[width=1.0\linewidth, trim={70px, 200px, 70px, 200px}, clip]{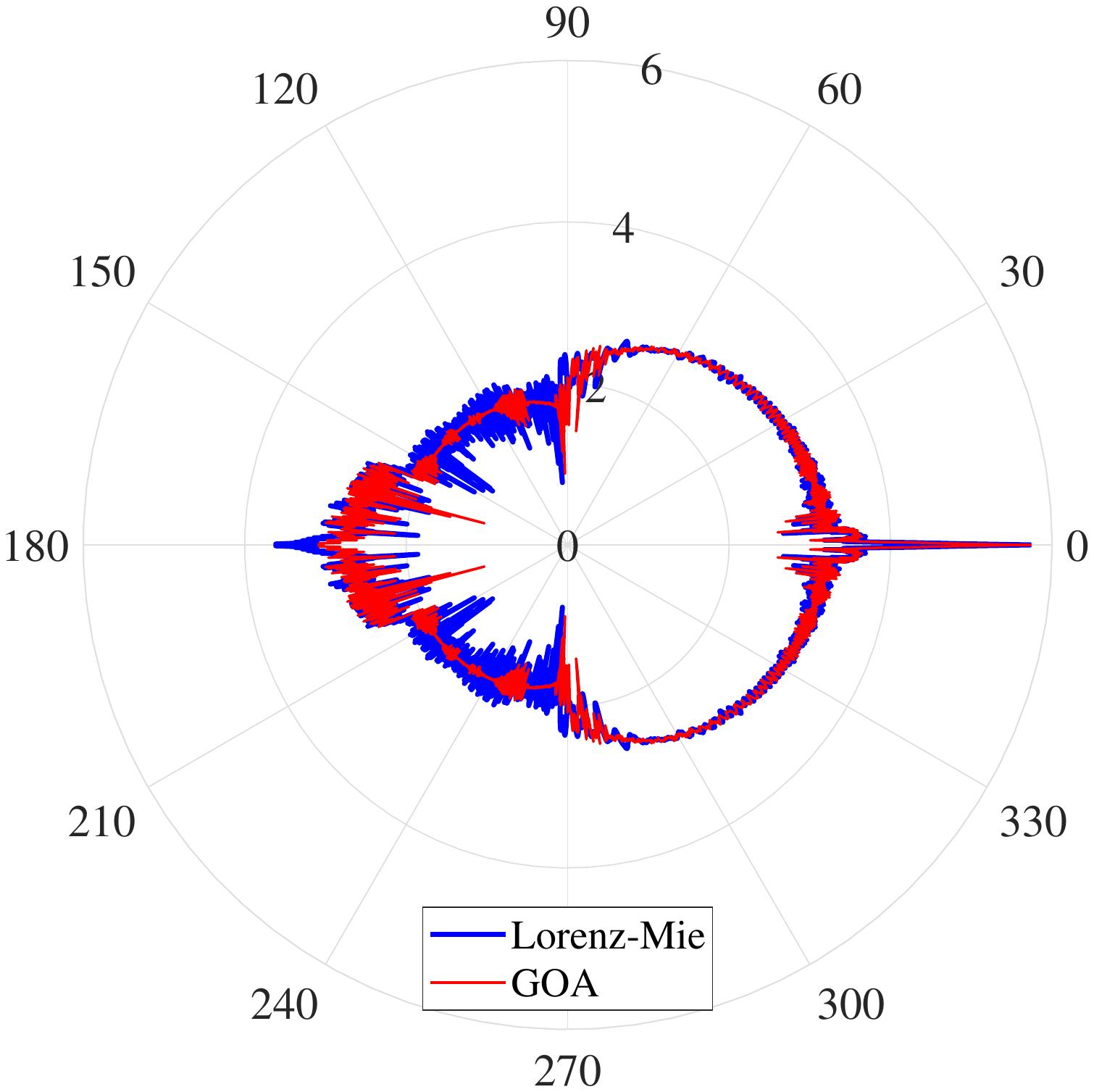}
\end{overpic}
\end{minipage}
}
\setcounter{subfigure}{0}
\subfigure[$r=0.1~\mathrm{{\mu}m}$]{
\begin{minipage}[b]{0.19\linewidth}
\begin{overpic}[width=1.0\linewidth, trim={70px, 200px, 70px, 200px}, clip]{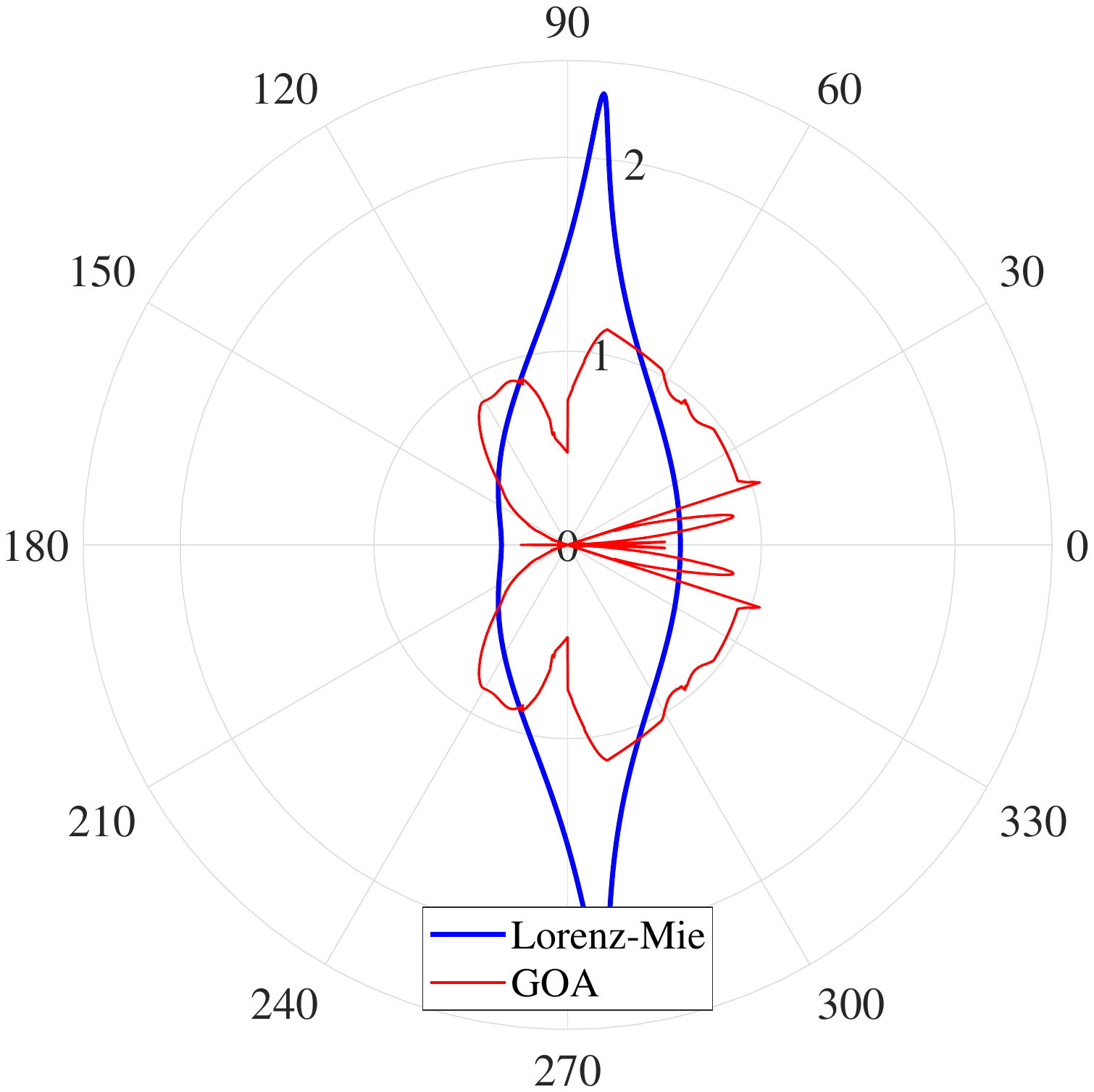}
\end{overpic}
\end{minipage}
}
\hspace{-0.1in}
\subfigure[$r=1~\mathrm{{\mu}m}$]{
\begin{minipage}[b]{0.19\linewidth}
\begin{overpic}[width=1.0\linewidth, trim={70px, 200px, 70px, 200px}, clip]{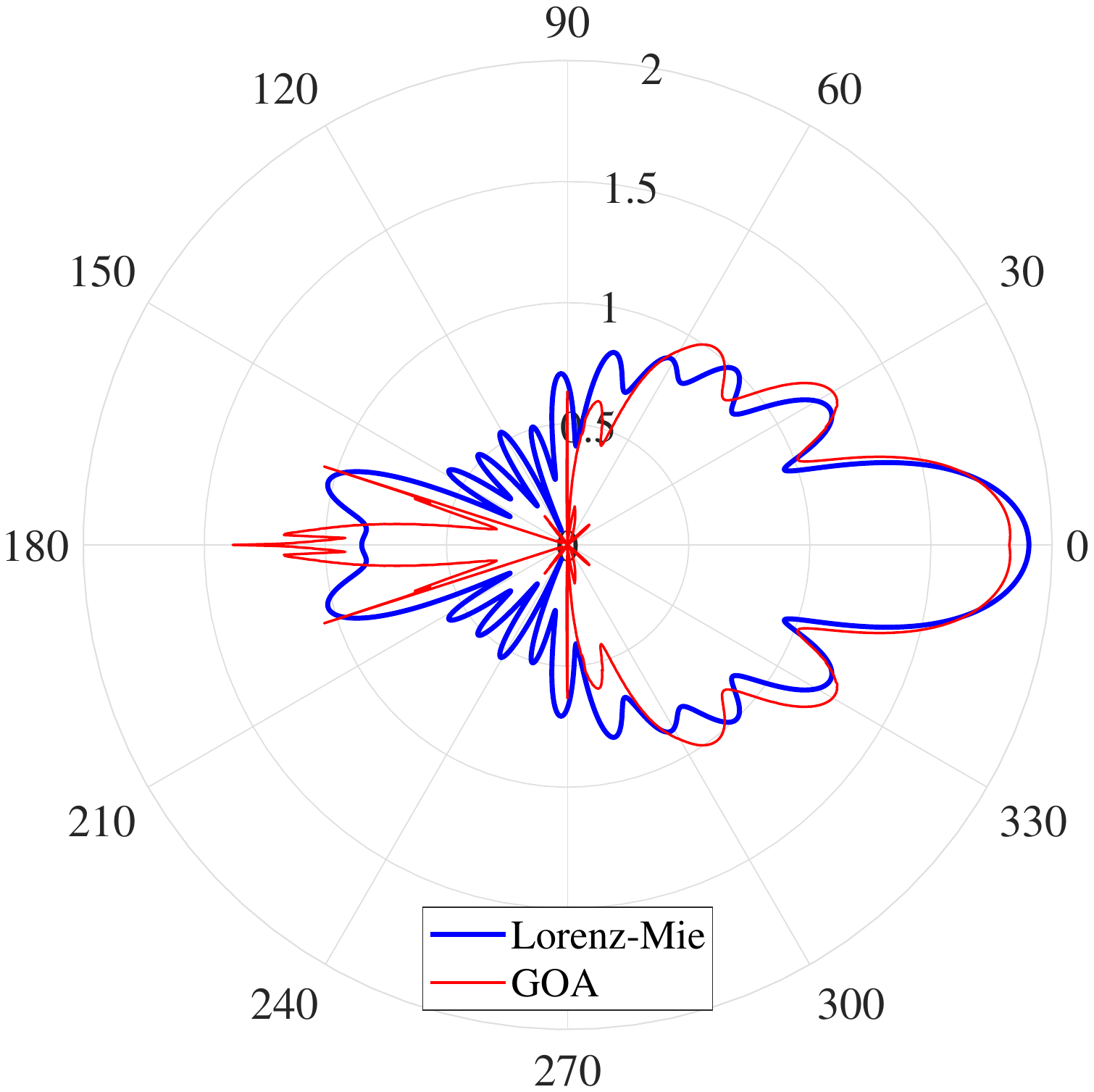}
\end{overpic}
\end{minipage}
}
\hspace{-0.1in}
\subfigure[$r=2~\mathrm{{\mu}m}$]{
\begin{minipage}[b]{0.19\linewidth}
\begin{overpic}[width=1.0\linewidth, trim={70px, 200px, 70px, 200px}, clip]{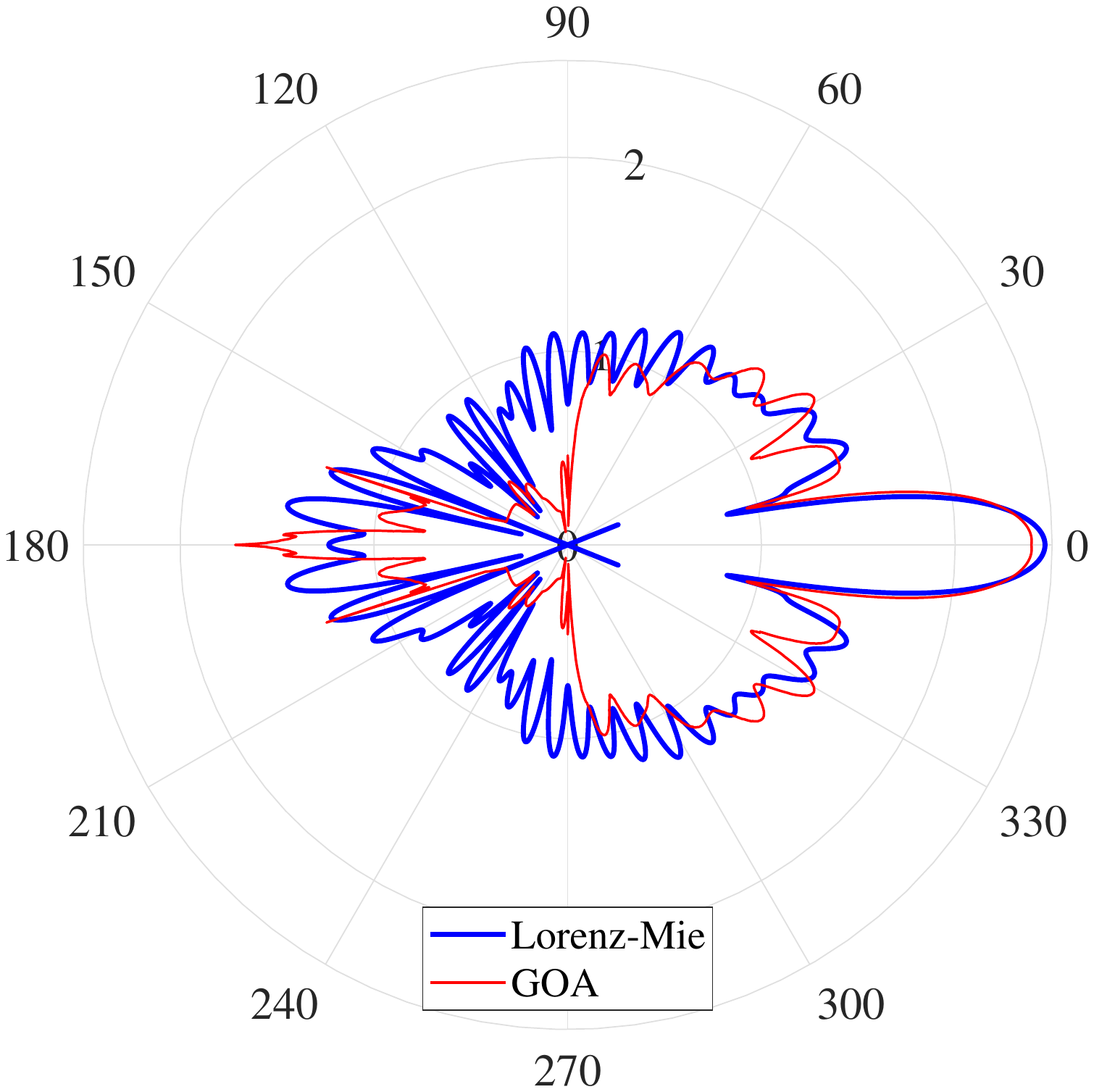}
\end{overpic}
\end{minipage}
}
\hspace{-0.1in}
\subfigure[$r=10~\mathrm{{\mu}m}$]{
\begin{minipage}[b]{0.19\linewidth}
\begin{overpic}[width=1.0\linewidth, trim={70px, 200px, 70px, 200px}, clip]{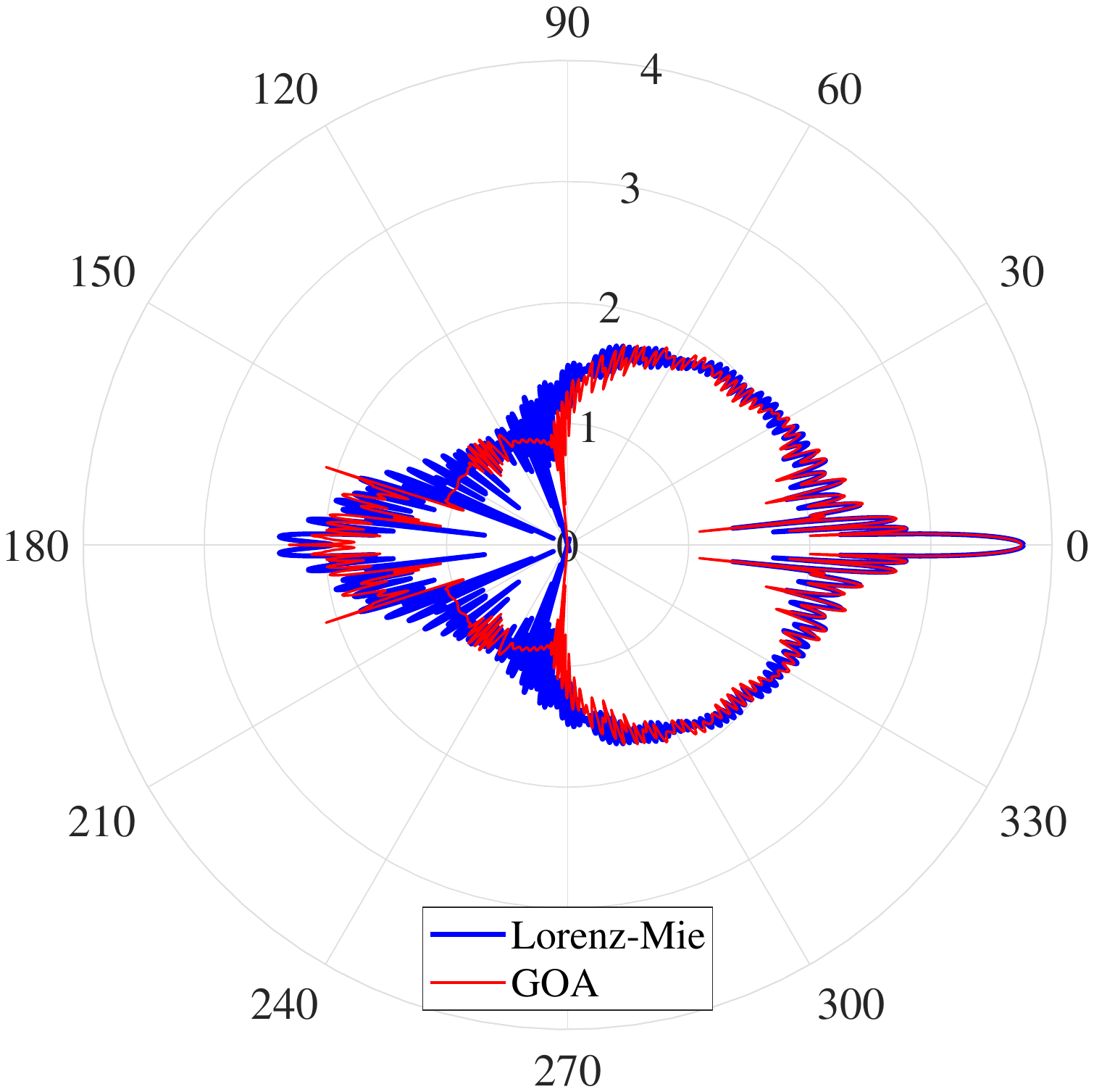}
\end{overpic}
\end{minipage}
}
\hspace{-0.1in}
\subfigure[$r=100~\mathrm{{\mu}m}$]{
\begin{minipage}[b]{0.19\linewidth}
\begin{overpic}[width=1.0\linewidth, trim={70px, 200px, 70px, 200px}, clip]{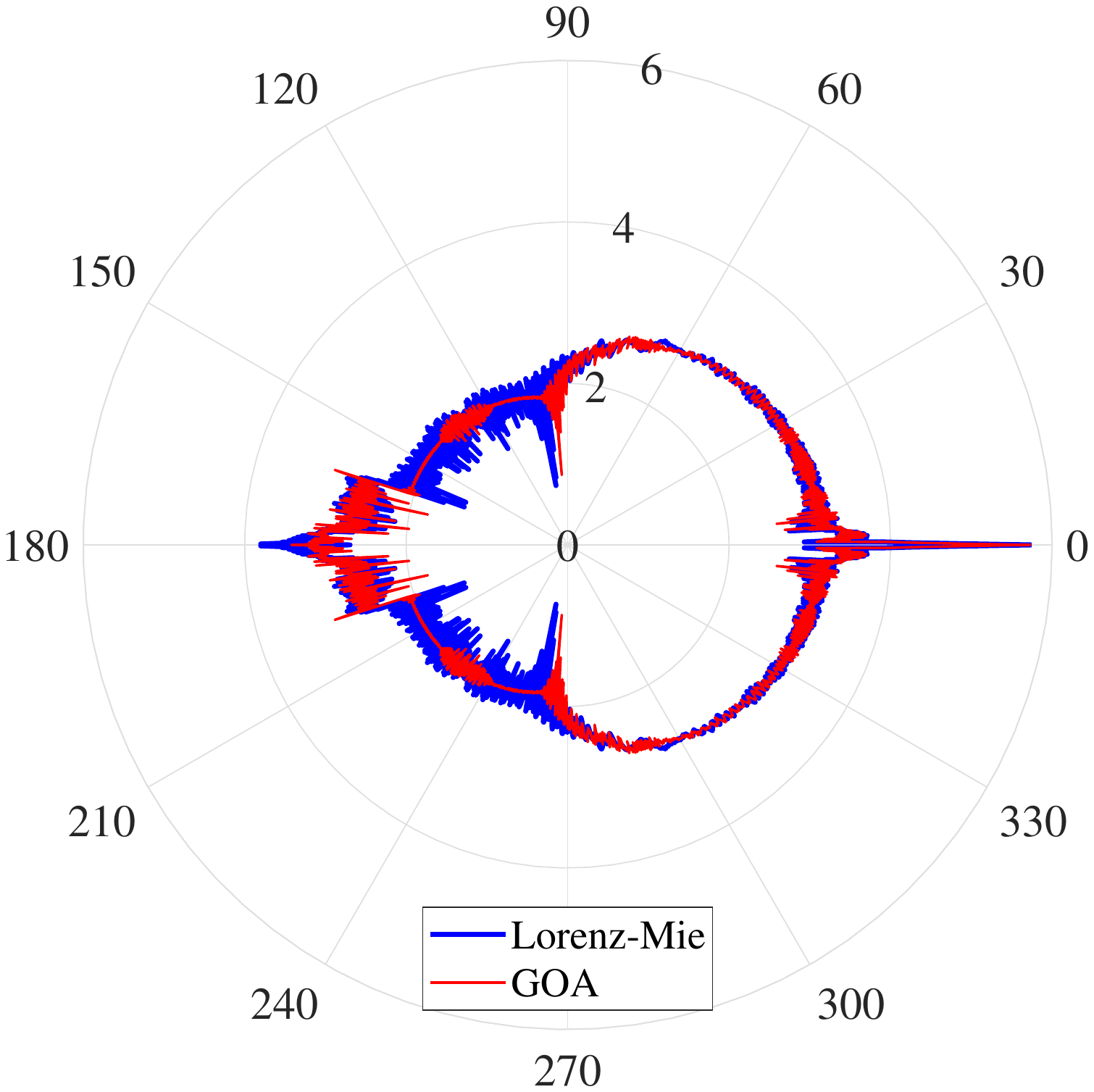}
\end{overpic}
\end{minipage}
}
\caption{\label{fig:goa_mie_2} Visual comparisons of $\log |S_2|$ by Lorenz-Mie calculations (blue curves) with those by GOA (red curves). First row: $\eta = 1.49$ and $\lambda = 0.6~\mathrm{{\mu}m}$. Second row: $\eta = 1.56$ and $\lambda = 0.6~\mathrm{{\mu}m}$. The particle radius is set to $r=0.1, 1, 2, 10$ and $100~\mathrm{{\mu}m}$, respectively.}
\end{figure*}

This section provides more visual comparisons between GOA and Lorenz-Mie theory. In Fig. \ref{fig:goa_mie_1}, we visualize the curves of $\log S_1(\theta)$ generated by GOA (red curves) and Lorenz-Mie theory (blue curves), respectively. Here, we test two different relative refractive indexes: $\eta=1.49$ and $\eta=1.56$. The particle radius ranges from $0.1~\mathrm{{\mu}m}$ to $100~\mathrm{{\mu}m}$. Again, close agreements are found when $r>1~\mathrm{{\mu}m}$ with some differences existing mainly on the backward peaks. When $r=0.1~\mathrm{{\mu}m}$, large errors occur in any direction, indicating that GOA does not work properly in this case.
Similar conclusions are reached when comparing GOA and Lorenz-Mie theory for the generation of $\log S_2(\theta)$ curves in Fig. \ref{fig:goa_mie_2}.

\begin{figure*}[t]
\centering
\subfigure{
\begin{minipage}[b]{0.19\linewidth}
\begin{overpic}[width=1.0\linewidth]{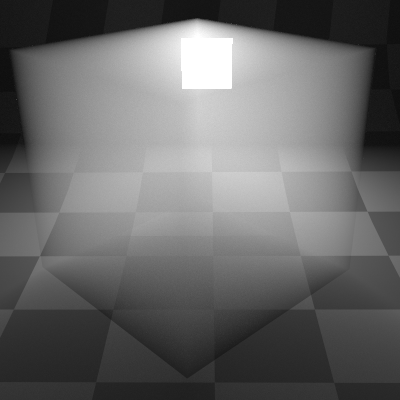}
\end{overpic}
\end{minipage}
}
\hspace{-0.1in}
\subfigure{
\begin{minipage}[b]{0.19\linewidth}
\begin{overpic}[width=1.0\linewidth]{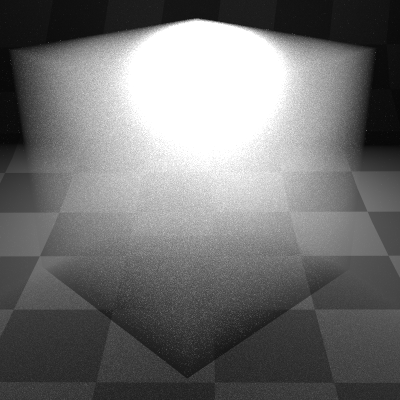}
\end{overpic}
\end{minipage}
}
\hspace{-0.1in}
\subfigure{
\begin{minipage}[b]{0.19\linewidth}
\begin{overpic}[width=1.0\linewidth]{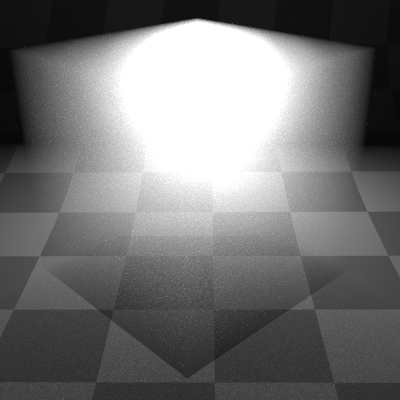}
\end{overpic}
\end{minipage}
}
\hspace{-0.1in}
\subfigure{
\begin{minipage}[b]{0.19\linewidth}
\begin{overpic}[width=1.0\linewidth]{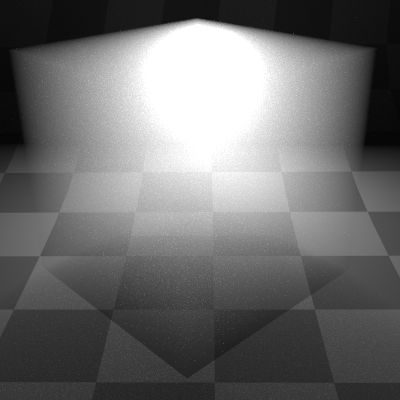}
\end{overpic}
\end{minipage}
}
\hspace{-0.1in}
\subfigure{
\begin{minipage}[b]{0.19\linewidth}
\begin{overpic}[width=1.0\linewidth]{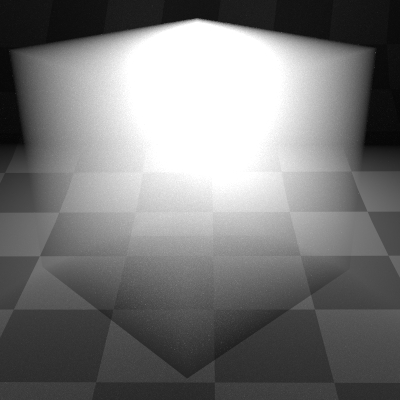}
\end{overpic}
\end{minipage}
}
\setcounter{subfigure}{0}
\subfigure[$r=0.1~\mathrm{{\mu}m}$, $N=10^{14}$]{
\begin{minipage}[b]{0.19\linewidth}
\begin{overpic}[width=1.0\linewidth]{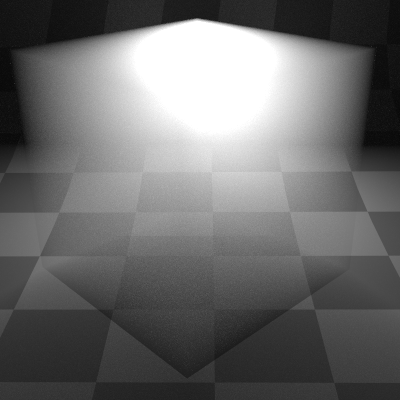}
\end{overpic}
\end{minipage}
}
\hspace{-0.1in}
\subfigure[$r=1~\mathrm{{\mu}m}$, $N=10^{11}$]{
\begin{minipage}[b]{0.19\linewidth}
\begin{overpic}[width=1.0\linewidth]{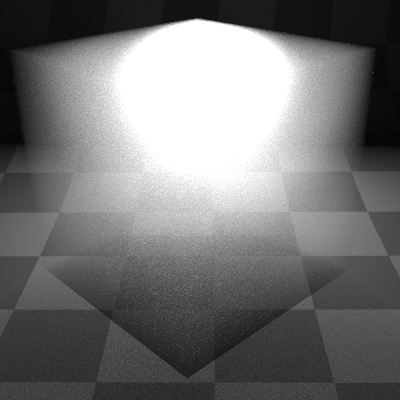}
\end{overpic}
\end{minipage}
}
\hspace{-0.1in}
\subfigure[$r=2~\mathrm{{\mu}m}$, $N=2\cdot10^{10}$]{
\begin{minipage}[b]{0.19\linewidth}
\begin{overpic}[width=1.0\linewidth]{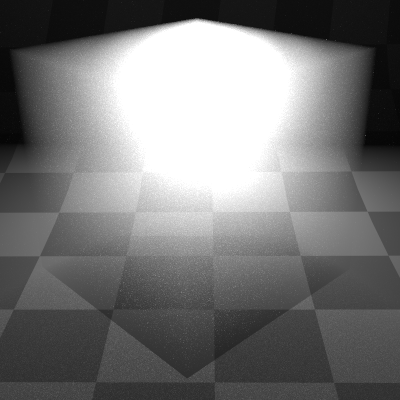}
\end{overpic}
\end{minipage}
}
\hspace{-0.1in}
\subfigure[$r=10~\mathrm{{\mu}m}$, $N=10^{9}$]{
\begin{minipage}[b]{0.19\linewidth}
\begin{overpic}[width=1.0\linewidth]{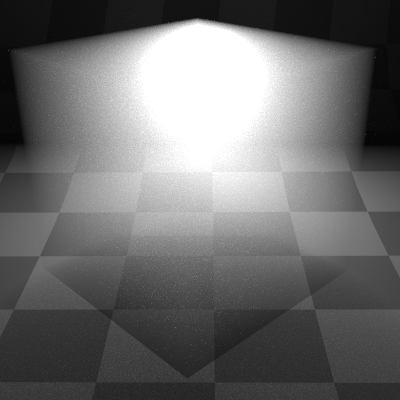}
\end{overpic}
\end{minipage}
}
\hspace{-0.1in}
\subfigure[$r=100~\mathrm{{\mu}m}$, $N=10^{7}$]{
\begin{minipage}[b]{0.19\linewidth}
\begin{overpic}[width=1.0\linewidth]{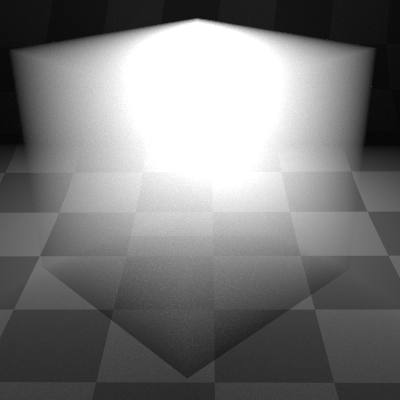}
\end{overpic}
\end{minipage}
}
\caption{\label{fig:goa_mie_cube_1} Rendering a smooth cubic medium with optical quantities derived from Lorenz-Mie theory (top row) and GOA (bottom row), respectively. Here, the relative refractive index $\eta$ is set to $1.49$. }
\end{figure*}

\begin{figure*}[t]
\centering
\subfigure{
\begin{minipage}[b]{0.19\linewidth}
\begin{overpic}[width=1.0\linewidth]{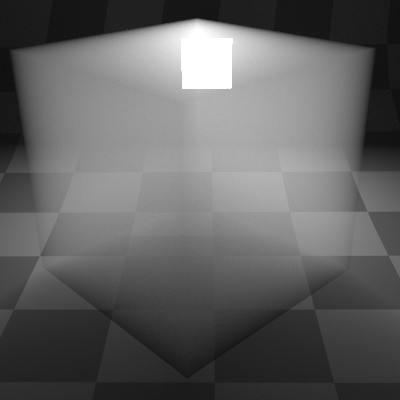}
\end{overpic}
\end{minipage}
}
\hspace{-0.1in}
\subfigure{
\begin{minipage}[b]{0.19\linewidth}
\begin{overpic}[width=1.0\linewidth]{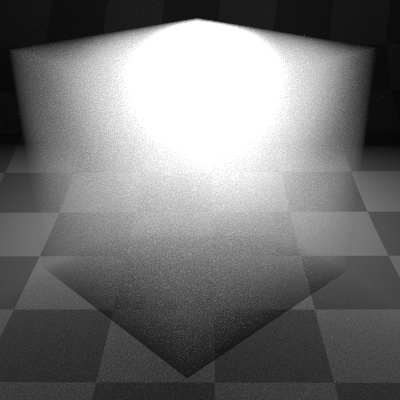}
\end{overpic}
\end{minipage}
}
\hspace{-0.1in}
\subfigure{
\begin{minipage}[b]{0.19\linewidth}
\begin{overpic}[width=1.0\linewidth]{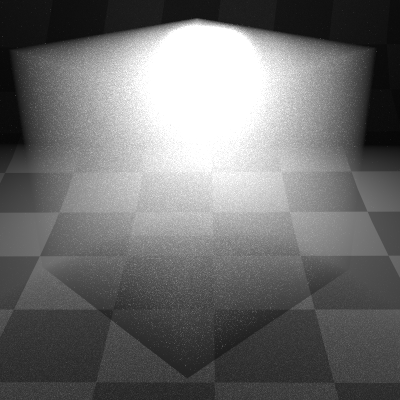}
\end{overpic}
\end{minipage}
}
\hspace{-0.1in}
\subfigure{
\begin{minipage}[b]{0.19\linewidth}
\begin{overpic}[width=1.0\linewidth]{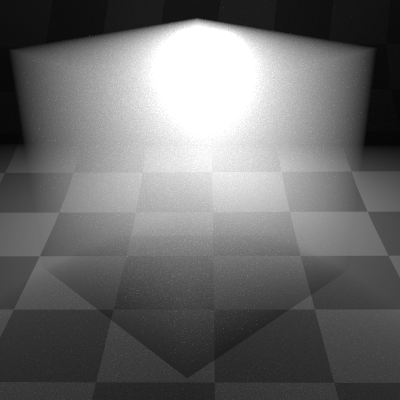}
\end{overpic}
\end{minipage}
}
\hspace{-0.1in}
\subfigure{
\begin{minipage}[b]{0.19\linewidth}
\begin{overpic}[width=1.0\linewidth]{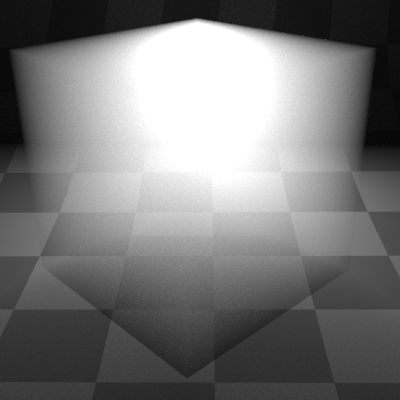}
\end{overpic}
\end{minipage}
}
\setcounter{subfigure}{0}
\subfigure[$r=0.1~\mathrm{{\mu}m}$, $N=10^{14}$]{
\begin{minipage}[b]{0.19\linewidth}
\begin{overpic}[width=1.0\linewidth]{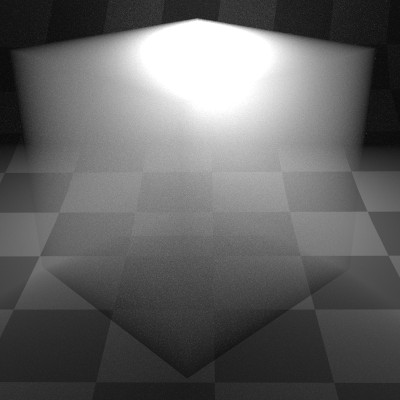}
\end{overpic}
\end{minipage}
}
\hspace{-0.1in}
\subfigure[$r=1~\mathrm{{\mu}m}$, $N=10^{11}$]{
\begin{minipage}[b]{0.19\linewidth}
\begin{overpic}[width=1.0\linewidth]{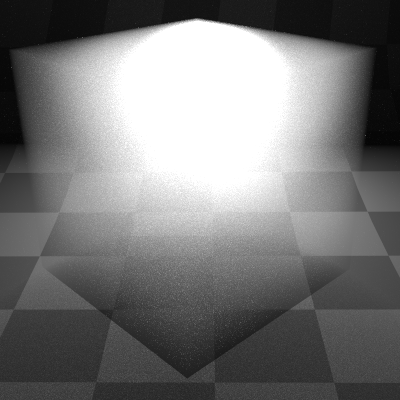}
\end{overpic}
\end{minipage}
}
\hspace{-0.1in}
\subfigure[$r=2~\mathrm{{\mu}m}$, $N=2\cdot10^{10}$]{
\begin{minipage}[b]{0.19\linewidth}
\begin{overpic}[width=1.0\linewidth]{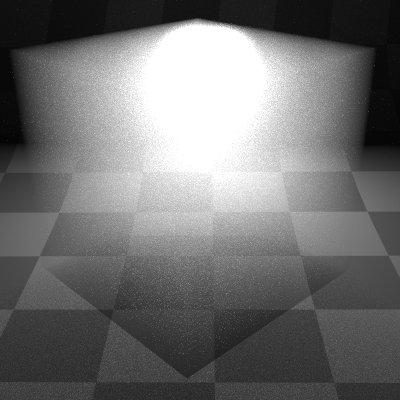}
\end{overpic}
\end{minipage}
}
\hspace{-0.1in}
\subfigure[$r=10~\mathrm{{\mu}m}$, $N=10^{9}$]{
\begin{minipage}[b]{0.19\linewidth}
\begin{overpic}[width=1.0\linewidth]{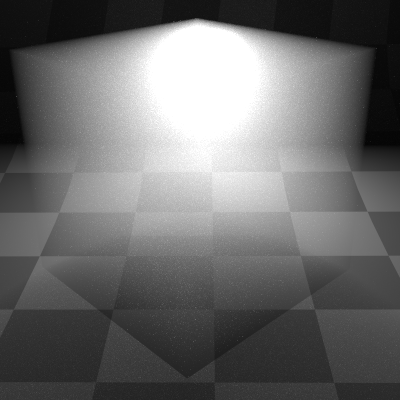}
\end{overpic}
\end{minipage}
}
\hspace{-0.1in}
\subfigure[$r=100~\mathrm{{\mu}m}$, $N=10^{7}$]{
\begin{minipage}[b]{0.19\linewidth}
\begin{overpic}[width=1.0\linewidth]{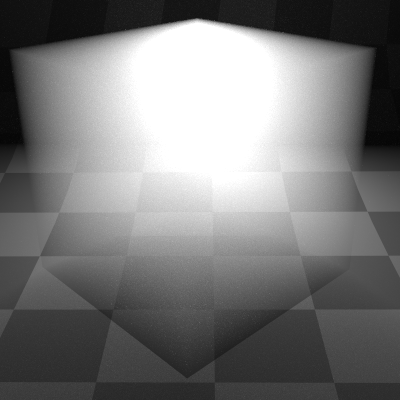}
\end{overpic}
\end{minipage}
}
\caption{\label{fig:goa_mie_cube_2} Rendering a smooth cubic medium with optical quantities derived from Lorenz-Mie theory (top row) and GOA (bottom row), respectively. Here, the relative refractive index $\eta$ is set to $1.56$.}
\end{figure*}

Although there are some mismatches between GOA and Lorenz-Mie theory in the case of $r=2~\mathrm{{\mu}m}$, the influence on the appearance of rendered media is subtle. To see this, we render a smooth cubic medium in Fig. \ref{fig:goa_mie_cube_1} and Fig. \ref{fig:goa_mie_cube_2} with different scene configurations. The medium is assumed to comprise monodisperse particles. The extinction coefficient and the phase function are respectively determined by Lorenz-Mie theory and GOA in a preprocessing stage, according to the particle radius $r$ and the particle number $N$. However, for $r=0.1~\mathrm{{\mu}m}$ we use the same extinction coefficient derived from Lorenz-Mie theory in both cases since GOA yields a negative value. This guarantees the fairness of comparison. Nonetheless, quite different appearances are achieved by Lorenz-Mie theory and GOA when $r=0.1~\mathrm{{\mu}m}$ due to the large discrepancy in $S_1(\theta)$ and $S_2(\theta)$. The difference of translucent appearance becomes less noticeable when $r$ goes up to $2~\mathrm{{\mu}m}$ and shrinks further as $r$ increases.

%




\end{document}